\documentclass{article}

\usepackage{arxiv}

\usepackage[utf8]{inputenc} % allow utf-8 input
\usepackage[T1]{fontenc}    % use 8-bit T1 fonts
\usepackage{hyperref}       % hyperlinks
\usepackage{url}            % simple URL typesetting
\usepackage{booktabs}       % professional-quality tables
\usepackage{amsfonts}       % blackboard math symbols
\usepackage{nicefrac}       % compact symbols for 1/2, etc.
\usepackage{microtype}      % microtypography
\usepackage{lipsum}		% Can be removed after putting your text content
\usepackage{graphicx}
\usepackage{natbib}
\usepackage{doi}
\usepackage{natbib}

\usepackage{textcomp}
\usepackage{graphicx}
\usepackage{xcolor}
\usepackage[T1]{fontenc}
\usepackage{amssymb}
\usepackage{amsmath}
\usepackage{multirow}
\usepackage{longtable,enumitem}
\usepackage{pifont} % http://ctan.org/pkg/pifont
\usepackage{xcolor}
\usepackage{comment}
\usepackage{gensymb}
\usepackage{setspace}
\newcommand{\cmark}{\ding{51}}%
\newcommand{\xmark}{\ding{55}}%
\usepackage{adjustbox}

\title{A Survey of Sound Source Localization with Deep Learning Methods}

\author{ \hspace{1mm}Pierre-Amaury Grumiaux\thanks{During the writing of this paper, Pierre-Amaury Grumiaux was also at Orange Labs,  4 Rue du Clos Courtel, F-35510 Cesson-S\'evign\'e, France, and at Univ.~Grenoble Alpes, Grenoble-INP, CNRS, GIPSA-lab, 11 Rue des Math\'ematiques, F-38400 Saint-Martin-d'H\`eres, France.} \\
	Nantes Universit{\'e}, {\'E}cole Centrale Nantes, CNRS, LS2N\\
	2 chemin de la Houssini\`ere\\
	F-44332 Nantes, France \\
	\texttt{pierreamaury.grumiaux@gmail.com}
	\And
	\hspace{1mm}Sr\dj{}an Kiti\'c \\
	Orange Labs\\
	4 Rue du Clos Courtel\\
	35510 Cesson-S\'evign\'e, France \\
	\texttt{srdan.kitic@orange.com}
	\And
	\hspace{1mm}Laurent Girin \\
	Univ.~Grenoble Alpes, Grenoble-INP, GIPSA-lab\\
	11 Rue des Math\'ematiques\\
	38400 Saint-Martin-d'H\`eres, France \\
	\texttt{laurent.girin@grenoble-inp.fr}
	\And
	\hspace{1mm}Alexandre Gu{\'e}rin \\
	Orange Labs\\
	4 Rue du Clos Courtel\\
	35510 Cesson-S\'evign\'e, France \\
	\texttt{alexandre.guerin@orange.com}
}

\date{}

\renewcommand{\headeright}{}
\renewcommand{\undertitle}{}
\renewcommand{\shorttitle}{A Survey of DL-based SSL}

\hypersetup{
pdftitle={A Survey of DL-based SSL},
pdfsubject={eess.AS},
pdfauthor={P.A.~Grumiaux, S.~Kiti\'c, L.~Girin, A.~Gu\'erin},
pdfkeywords={},
}

\begin{document}
\maketitle

\begin{abstract}
	This article is a survey of deep learning methods for single and multiple sound source localization, with a focus on sound source localization in indoor environments, where reverberation and diffuse noise are present. We provide an extensive topography of the neural network-based sound source localization literature in this context, organized according to the neural network architecture, the type of input features, the output strategy (classification or regression), the types of data used for model training and evaluation, and the model training strategy. 
%This way, an interested reader can easily comprehend the vast panorama of deep learning-based sound source localization methods. 
Tables summarizing the literature survey are provided at the end of the paper, allowing a quick search of methods with a given set of target characteristics. 
\end{abstract}

% keywords can be removed
%\keywords{}

\section{Introduction}

Sound source localization (SSL) is the problem of estimating the position of one or several sound sources relative to some arbitrary reference position, which is generally the position of the recording microphone array, based on the recorded multichannel acoustic signals. In most practical cases, SSL is simplified to the estimation of the sources' direction of arrival (DoA), \emph{i.e.}, it focuses on the estimation of azimuth and elevation angles, without estimating the distance to the microphone array (therefore, unless otherwise specified, in this article we use the terms ``SSL'' and ``DoA estimation'' interchangeably). SSL has numerous practical applications -- for instance, in source separation, e.g., \citep{chazan_multi-microphone_2019}, automatic speech recognition (ASR), e.g., \citep{lee_dnn-based_2016}, speech enhancement, e.g., \citep{xenaki_sound_2018}, human-robot interaction, e.g., \citep{li_reverberant_2016}, noise control, e.g., \citep{chiariotti2019acoustic}, and room acoustic analysis, e.g., \citep{amengual2017spatial}. As detailed in the following, in this paper, we focus on sound sources in the audible range (typically speech and audio signals) in indoor (office or domestic) environments.

Although SSL is a longstanding and widely researched topic  \citep{gerzon1992general, dibiase_robust_2001, argentieri2015survey, cobos2017survey, benesty_microphone_2008, knapp_generalized_1976, brandstein2001microphone, nehorai1994acoustic, hickling1993finding}, it remains a very challenging problem to date. Traditional SSL methods are based on signal/channel models and signal processing (SP) techniques. Although they have shown notable advances in the domain over the years, they are known to perform poorly in difficult yet common scenarios where noise, reverberation, and  several simultaneously emitting sound sources may be present \citep{blandin2012multi, evers2020locata}. In the last decade, the potential of data-driven deep learning (DL) techniques for addressing such difficult scenarios has received an increasing interest. As a result, an increasing number of SSL systems based on deep neural networks (DNNs) have been proposed in the recent years. Most of the reported works have indicated the superiority of DNN-based SSL methods over conventional (i.e., SP-based) SSL methods. For example, \citet{chakrabarty_broadband_2017} showed that, in low signal-to-noise ratio conditions, using a CNN led to a two-fold increase in overall DoA classification accuracy compared to using the conventional method called steered response power with phase transform (SRP-PHAT) (see Section \ref{sec:traditional}). In \citet{perotin_crnn-based_2018}, the authors were able to obtain a $25\%$ increase of DoA classification accuracy when using a convolutional recurrent neural network (CRNN) over a method based on independent component analysis (ICA). Finally \citet{adavanne_direction_2018} proved that employing a CRNN can reduce the average angular error by $50\%$ in reverberant conditions compared to the conventional MUSIC algorithm (see Section \ref{sec:traditional}). 

This kind of results has further motivated the expansion of scientific papers on DL applied to SSL. 
%In the last three years (2019 to 2021), we have witnessed a threefold increase in the number of corresponding publications. 
In the meantime, there has been no comprehensive survey of the existing approaches, which would be very useful for researchers and practitioners in the domain. Although we can find reviews mostly focused on conventional methods, e.g., \citep{argentieri2015survey, cobos2017survey, evers2020locata, gannot2019introduction}, to the best of our knowledge only a very few have explicitly targeted SSL with DL methods. \citet{Ahmad2021survey} presented a short survey of several existing DL models and datasets for SSL before proposing a DL architecture of their own. \citet{bianco2019machine} and \citet{purwins2019deep} presented an interesting overview of machine learning applied to various problems in audio and acoustics. Nevertheless, only a short portion of each of these two reviews is dedicated to SSL with DNNs.

\subsection{Aim of the paper}

The goal of this paper is to fill this gap, and to provide a thorough survey of the SSL literature using DL techniques. More precisely, we examined and review 156 papers published from 2011 to 2021. We classify and discuss the different approaches in terms of characteristics of the employed methods and addressed configurations (e.g., single-source vs.~multi-source localization setup or neural network architecture; the exact list is given in Section~\ref{subsec:intro-outline}). In other words, we present a taxonomy of the DL-based SSL literature published in the last decade. At the end of the paper, we present a summary of this survey in the form of four tables (one for the period 2011--2018, and one for each of the years 2019, 2020 and 2021). All of the methods that we reviewed are reported in these tables with a summary of their characteristics presented in different columns. This enables the reader to rapidly select the subset of methods having a given set of characteristics, if they are interested in that particular type of method.

Note that in this survey paper, we do not aim to evaluate and compare the performance of the different systems. Due to the large number of DNN-based SSL papers and the diversity of configurations, such a contribution would be very difficult and cumbersome (albeit very useful), especially because the discussed systems are often trained and evaluated on different datasets. As we will see later, listing and commenting on these different datasets is, however, part of our survey effort. Note also that we do not consider SSL systems that exploit other modalities in addition to sound, e.g., audio-visual systems  \citep{ban_accounting_2018, wu2021binaural, masuyama2020self}. Finally, we do consider DL-based methods for joint sound event localization and detection (SELD), which is a combination of sound event detection (SED; here detection actually means classification) and SSL, and in that case, we focus on the localization task. In particular, we include in the review the SELD methods presented to the DCASE Challenge (and/or to the corresponding DCASE Workshop) in 2019, 2020 and 2021 (see the DCASE website at \url{https://dcase.community/}). One of the task of this challenge is precisely dedicated to SELD, which has contributed to make the DL-based SSL (and SED) problem a popular research topic over the recent years.

\subsection{General principle of DL-based SSL}
\label{subsec:intro-general-principle}

\begin{figure}[!t]
\centering
\includegraphics[width=0.8\linewidth]{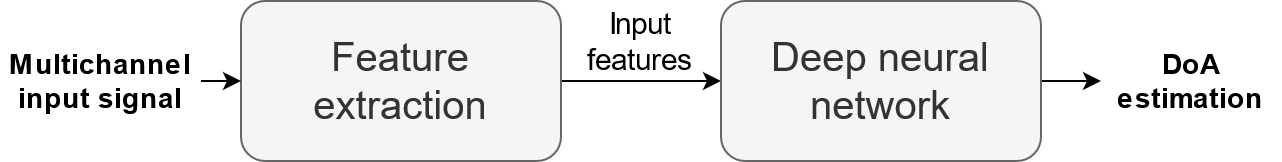}
\caption{General pipeline of a DL-based SSL system.}
\label{fig:DL_SSL_pipeline}
\end{figure}

The general principle of DL-based SSL methods and systems can be schematized with a simple pipeline, as illustrated in Fig.~\ref{fig:DL_SSL_pipeline}. A multichannel input signal recorded with a microphone array is processed by a feature extraction module to provide input features. These input features are fed into a DNN, which delivers an estimate of the source location or DoA. As discussed later in the paper, a recent trend is to skip the feature extraction module to directly feed the network with multichannel raw data. In any case, the two fundamental reasons behind the design of such SSL are detailed below. 

First, multichannel signals recorded with an array of $I$ microphones distributed in space contain information about the location of the source(s). Indeed, when the microphones are close to each other compared to their distance to the source(s), the microphone signal waveforms, although appearing similar from a distance, exhibit more or less notable and complex differences in terms of delay and amplitude, depending on the experimental setup. These interchannel differences are due to distinct propagation paths from the source to the different microphones, for both the direct path (line of sight between source and microphone) and the numerous reflections that compose the reverberation in an indoor environment. In other words, a source signal $s_j(t)$ is convolved with different room impulse responses (RIRs) $a_{i,j}(t)$, which depend on the source position, microphone position and directivity ($i$ denotes the microphone index in the array), and acoustic  environment configuration (e.g., room shape):
\begin{align}
x_{i}(t) = a_{i,j}(t) \star s_j(t) + n_i(t) = \sum_{\tau=0}^{T-1} a_{i,j}(\tau)s_j(t-\tau) + n_i(t),
\label{eq:sensor-signal-time-domain}
\end{align}
where $x_{i}(n)$ denotes the resulting recorded signal at microphone $i$, $n_i(t)$ is the noise signal at microphone $i$ (diffuse, ``background'' noise and possibly some sensor noise), and $\star$ denotes the convolution (note that we work with digital signals and $t$ and $\tau$ are discrete time indexes; $T$ is the effective length of the RIR). Therefore, the recorded signal contains information on the relative source-to-microphone array position. 
The microphone signals are often expressed in the time-frequency (TF) domain, using the short-term Fourier transform (STFT), where the convolution in Eq.~\eqref{eq:sensor-signal-time-domain} is assumed to transform into a product between the STFT of the source signal $S_j(f,n)$ and the acoustic transfer function (ATF) $A_{i,j}(f)$, which is the (discrete) Fourier transform of the corresponding RIR and is thus encoding the source spatial information ($f$ denotes the frequency bin, and $n$ is the STFT frame index) \citep{gannot2017consolidated, vincent2018audio}:
\begin{equation}
 X_{i}(f,n) = A_{i,j}(f)S_j(f,n) + N_{i}(f,n).
 \label{eq:sensor-signal-STFT-domain}
\end{equation}
% The above equation uses the fact that the convolution between the source signal and an RIR (see Eq.~\eqref{eq:sensor-signal-time-domain}) is approximated in the STFT domain by the product between $S(f,n)$ and the ATF. This approximation is referred to as the narrow-band assumption in the speech/audio source localization/separation literature \citep{gannot2017consolidated, vincent2018audio}, in which it is widely used even if the fact that the STFT analysis window is generally shorter than the RIRs length may question its accuracy. 
When several, say $J$, sources are present, the recorded signal is the sum of their contribution (plus the noise):
\begin{equation}
    x_{i}(t) = \sum_{j=1}^J a_{i,j}(t) \star s_j(t) + n_i(t).
\label{eq:microphone-signal-sum}
\end{equation}
This latter equation is often reformulated in the TF domain in matrix form:
\begin{equation}
\mathbf{X}(f,n) = \mathbf{A}(f)\mathbf{S}(f,n) + \mathbf{N}(f,n),
\label{eq:microphone-signal-sum-STFT}
\end{equation}
where $\mathbf{X}(f,n) = [X_{1}(f,n), ..., X_{I}(f,n)]^\top$ is the microphone signal vector, $\mathbf{A}(f)$ is the matrix gathering the ATFs, $\mathbf{S}(f,n)= [S_{1}(f,n), ..., S_{J}(f,n)]^\top$ is the source signal vector, and $\mathbf{N}(f,n)= [N_{1}(f,n), ..., N_{I}(f,n)]^\top$ is the noise vector. In that multi-source case, the difficulty of the SSL problem is that the contributions of the different sources generally overlap in time. SSL then requires to proceed to some kind of source clustering, which is generally easier to proceed in the frequency or time-frequency domain due to the natural sparsity of audio sources in that domain \citep{rickard_approximate_2002}. In this paper, we do not describe the foundations of source-to-microphone propagation in more detail. They can be found in several references on general acoustics, e.g.,  \citep{jacobsen2013fundamentals,rossing2007springer}, room acoustics, e.g.,  \citep{kuttruff2016room}, array signal processing, e.g., \citep{brandstein2001microphone,benesty_microphone_2008,jarrett2017theory,rafaely_fundamentals_2019}, speech enhancement and audio source separation, e.g.,  \citep{gannot2017consolidated,vincent2018audio}, and many papers on conventional SSL. 

The second reason for designing DNN-based SSL systems is that even if the relationship between the information contained in the multichannel signal and the location of the source(s) is generally complex (especially in a multisource reverberant and noisy configuration, see Eqs.~\eqref{eq:microphone-signal-sum} and \eqref{eq:microphone-signal-sum-STFT}), DNNs are powerful models that are able to automatically identify and exploit this relationship, given that they are provided with a sufficiently large number of representative training examples. This ability of data-driven DL methods to replace conventional methods based on a signal/channel model and SP techniques ---or at least a part of them, since the feature extractor module can be based on conventional processing--- makes them attractive for addressing problems such as SSL. 
%An appealing property of DL-based methods is their capacity to deal with real-world data, whereas conventional methods often suffer from oversimplistic assumptions compared to the complexity of real-world acoustics. The major drawback of the DNN-based approaches is their lack of generality.
While some conventional methods can adapt to the observed signals, e.g., \cite{dvorkind2005time,li2016estimation,laufer2020data, li_reverberant_2016}, they are all intrinsically based on certain (more or less plausible) modeling assumptions, which can limit their effectiveness when exposed to the complexity of real-world acoustics. Deep learning models do not \emph{explicitly} impose any such assumptions, and instead they efficiently adapt to the presented training data. This is, however, also the major drawback of the DNN-based approaches, as they are less generic than traditional methods. A deep model designed for and trained in a given configuration (e.g., a given microphone array geometry) will not provide satisfying localization results if the setup changes \citep{liu_direction--arrival_2018,moing_data-efficient_2021}, unless some relevant adaptation method can be used, which is still an open problem in DL in general. 
%In this paper, we do not consider this aspect, and we do not intend to further detail the pros and cons of DL-based methods vs.~conventional methods. Our goal is rather to present, in a soundly organized manner, a representative (if not exhaustive) panorama of DL-based SSL methods published in the last decade. 

\subsection{Outline of the paper}
\label{subsec:intro-outline}

The remainder of the paper is organized as follows. In Section~\ref{sec:context}, we specify the context and scope of the survey in terms of the considered acoustic environment and sound source configurations. In Section~\ref{sec:traditional}, we briefly present the most common conventional SSL methods, for two reasons: first, they are often used as a baseline for the evaluation of DL-based methods; and second, we will see that several types of features extracted by conventional methods can be used in DL-based methods. Section~\ref{sec:NN-archi} aims to classify the different neural network architectures used for SSL. Section~\ref{sec:inputFeatures} presents the various types of input features used for SSL with neural networks. In Section~\ref{sec:output}, we explain the two output strategies employed in DL-based SSL: classification and regression. We then discuss in Section~\ref{sec:data} the datasets used for training and evaluating the models. In Section~\ref{sec:learningStrategies}, learning paradigms such as supervised or semi-supervised learning are discussed from the SSL perspective. Section~\ref{sec:conclusion} provides the two summary tables and concludes the paper. Note that, due to the large number of acronyms used in this survey paper, we provide a list of these acronyms in Table~\ref{tab:acronyms}.

\begin{table}[h!]
\caption{Table of acronyms.}
\label{tab:acronyms}
    \begin{tabular}{|lr|}
    \hline
        \textbf{ACCDOA}     & activity-coupled Cartesian direction of arrival \\
        \textbf{AE}         & autoencoder \\
        \textbf{ATF}        & acoustic transfer function \\
        \textbf{ASR} & automatic speech recognition \\
        \textbf{BGRU}       & bidirectional gated recurrent unit \\
        \textbf{BIR}        & binaural impulse response \\
        \textbf{BRIR}       & binaural room impulse response \\
        \textbf{CC}         & cross-correlation \\
        \textbf{CNN}        & convolutional neural network \\
        \textbf{CRNN}       & convolutional recurrent neural network \\
        \textbf{CPS}        & cross power spectrum \\
        \textbf{DCASE}      & Detection and Classification \\
            & of Acoustic Scenes and Events \\
        \textbf{DIRHA}      & distant-speech interaction \\
                            & for robust home applications \\
        \textbf{DL}         & deep learning \\
        \textbf{DNN}        & deep neural network \\
        \textbf{DoA}        & direction of arrival \\
        \textbf{DP-RTF}     & direct-path relative transfer function \\
        \textbf{DRR} & direct-to-reverberant ratio \\
        \textbf{EM}         & expectation maximization \\
        \textbf{ESPRIT}     & Estimation of Signal Parameters \\
                            & via Rotational Invariance Techniques \\
        \textbf{EVD}        & eigenvalue decomposition \\
        \textbf{FFNN}       & feed-forward neural network \\
        \textbf{FOA}        & first-order Ambisonics \\
        \textbf{GAN}        & generative adversarial network \\
        \textbf{GCC}        & generalized cross-correlation \\
        \textbf{GLU}        & gated linear unit \\
        \textbf{GMM}        & Gaussian mixture models \\
        \textbf{GMR}        & Gaussian mixture regression \\
        \textbf{GPU}        & graphical processing unit \\
        \textbf{GRU}        & gated recurrent unit \\
        \textbf{HATS}       & head-and-torso simulator \\
        \textbf{HOA}        & higher-order Ambisonics \\
    \hline
    \end{tabular}
    \begin{tabular}{|lr|}
    \hline
        \textbf{HRTF}       & head-related transfer function \\
        \textbf{ICA}        & independent component analysis \\
        \textbf{ILD}        & interaural level difference \\
        \textbf{IPD}        & interaural phase difference \\
        \textbf{ITD}        & interaural time difference \\
        \textbf{ISM}        & image source method \\
        \textbf{LSTM}       & long short-term memory \\
        \textbf{MHSA}        & multihead self-attention \\
        \textbf{MLP}        & multiLayer Perceptron \\
        \textbf{MOT}        & multi-object tracking \\       \textbf{MUSIC}      & MUltiple SIgnal Classification \\
        \textbf{NLP}        & natural language processing \\
        \textbf{NoS}        & number of sources \\
        \textbf{PHAT}       & PHAse Transform \\
        \textbf{RIR}        & room impulse response \\
        \textbf{RNN}        & recurrent neural network \\
        \textbf{RTF}        & relative transfer function \\
        \textbf{SA}         & self-attention \\
        \textbf{SCM} & spatial covariance matrix \\
        \textbf{SED}        & sound event detection \\
        \textbf{SELD}       & sound event localization and detection \\
        \textbf{SH} & spherical harmonics \\
        \textbf{SMIR}       & spherical microphone impulse response \\
       \textbf{SMN}       & sequence matching network \\ 
        \textbf{SP}        & signal processing \\
        \textbf{SPS}        & spatial pseudo-spectrum \\
        \textbf{SRP}        & steered power response \\
        \textbf{SSL}        & sound source localization \\
        \textbf{STFT}       & short-term Fourier transform \\
        \textbf{TCN}        & temporal convolutional network \\
        \textbf{TDoA}       & time difference of arrival \\
        \textbf{TF}         & time-frequency \\
        \textbf{VAD}        & voice activity detection \\
        \textbf{VAE}        & variational autoencoder \\
        \textbf{WDO}        & W-disjoint orthogonality \\
        \hline
    \end{tabular}
\end{table}

%%% ENVIRONMENT %%%
\section{Acoustic environment and sound source configurations}
\label{sec:context}

SSL has been applied in different configurations, depending on the application. In this section, we specify the scope of our survey in terms of acoustic environment (noisy, reverberant, or even multi-room) and the nature of the considered sound sources (their type, number, and static/mobile status). 

\subsection{Acoustic environments}
\label{subsec:acoustic-environments}

In this paper, we focus on SSL in an indoor environment, \emph{i.e.}, when the microphone array and the sound source(s) are present in a closed room, generally of moderate size, typically an office room or a domestic environment. This implies reverberation: in addition to the direct source-to-microphone propagation path, the recorded sound contains many other multi-path components of the same source. All of these components form the room impulse response (RIR), which is defined for each source position and microphone array position (including orientation) and for a given room configuration.

In a general manner, the presence of reverberation is seen as a notable perturbation that makes SSL more difficult compared to the simpler (but somewhat unrealistic) \textit{anechoic} case, which assumes the absence of reverberation, as is obtained in the \textit{free field} propagation setup. Another important adverse factor to take into account in SSL is noise. On the one hand, noise can come from interfering sound sources in the surrounding environment: TV, background music, pets, street noise passing through open or closed windows, etc. Often, noise is considered as diffuse, \emph{i.e.}, it does not originate from a clear direction. On the other hand, the imperfections of the recording devices are another source of noise that are generally considered as artifacts.

Early works on using neural networks for DoA estimation most often considered direct-path propagation only (the anechoic setting), e.g., \citep{rastogi_array_1987, goryn_neural_1988, jha_bearing_1988, jha_bearing_1989, jha_direction_1991, falong_ml_1993, yang_complex-valued_1994, southall_direction_1995, el_zooghby_neural_2000}, though a model of the acoustical environment was used to generate simulated data to train the neural network of \citet{datum1996artificial}. Most of these works are from the pre-deep-learning era, using ``shallow'' neural networks with only one or two hidden layers \cite{goodfellow_deep_2016}. We do not detail these works in our survey, although we acknowledge them as pioneering contributions to the neural network-based DoA estimation problem. A few more recent works based on more ``modern'' neural network architectures also focused on anechoic propagation only, or did not consider sound sources in the audible bandwidth \citep{liu_direction--arrival_2018,unlersen_direction_2016, bialer_performance_2019, elbir_deepmusic_2020, choi_convolutional_2020}. 
% We do not detail these papers either, since we focus on SSL in real-world reverberant environments.

\subsection{Source types}\label{sec:sourcetypes}

In the SSL literature, a great proportion of systems focuses on localizing speech sources because of their importance in related tasks such as speech enhancement or speech recognition.
Examples of speaker localization systems can be found in papers by \citet{chakrabarty_multi-speaker_2019, grumiaux_saladnet_2021, he_neural_2021, hao_spectral_2020}. In such systems, the neural networks are trained to estimate the DoA of speech sources so that they are somewhat specialized in this type of source. Other systems, in particular those participating in the DCASE Challenge, consider a variety of sound source types \citep{politis_overview_2020}. Depending on the challenge task and its corresponding dataset, these methods are capable of localizing alarms, crying babies, crashes, barking dogs, female/male screams, female/male speech, footsteps, knockings on doors, ringings, phones, and piano sounds. Note that the localization of such sources, even if they overlap in time, is not necessarily a more difficult problem than localization of several overlapping speakers, since the former usually have distinct spectral characteristics that neural models may exploit for better detection and localization.

\subsection{Number of sources}
\label{sec:numberOfSources}

The number of sources (NoS) in a recorded mixture signal is an important parameter for SSL. In the SSL literature, the NoS might be considered as known (as a working hypothesis). Alternatively, it can be estimated along with the source location, in which case the SSL problem is a combination of detection and localization. Examples of conventional (non-deep) SSL works including NoS estimation can be found in papers by \citet{arberet2009robust,landschoot2019model}.

Many DNN-based works have considered only one source to localize, as it is the simplest scenario to address, e.g., \citep{perotin_crnn-based_2018, bologni_acoustic_2021, liu_deep_2021}. We refer to this scenario as \textit{single-source} SSL. In this case, the networks are trained and evaluated on datasets with only at most one active source (a source is said to be active when emitting sound and inactive otherwise). In terms of NoS, we thus have here either 1 or 0 active source. The activity of the source in the processed signal, which generally contains background noise, can be artificially controlled, \emph{i.e.}, the knowledge of source activity is a working hypothesis. This is a reasonable approach at training time when using synthetic data, but it is quite unrealistic at test time on real-world data. Alternatively, the source activity can be estimated, which is a more realistic approach at test time. In the latter case, there are two ways of dealing with the source activity detection problem. 
The first is to employ a source detection algorithm beforehand and then apply the SSL method only on the signal portions with an active source. For example, a voice activity detection (VAD) technique has been used in the SSL systems of \citet{kim_voice_2018, chang_temporal_2018, sehgal_convolutional_2018, li_voice_2016}. The other way is to detect the activity of the source at the same time as the localization algorithm. For example, an additional neuron was added by \citet{yalta_sound_2017} to the output layer of their DNN, which outputted $1$ when no source was active (in that case, all other localization neurons were trained to output $0$), and $0$ otherwise.

Multi-source localization is a much more difficult problem than single-source SSL. Current state-of-the-art DL-based methods address multi-source SSL in adverse environments. In this survey, we consider as multi-source localization the scenario in which several sources overlap in time (\emph{i.e.}, they are simultaneously emitting), regardless of their type (e.g., there could be several speakers or several distinct sound events). The specific case of a multi-speaker conversation with or without speech overlap is strongly connected to the \textit{speaker diarization} problem (``who speaks when?'') \citep{tranter2006overview, anguera2012speaker, park_review_2021}. Speaker localization, diarization, and (speech) source separation are intrinsically connected problems, as the information retrieved from solving each of them can be useful for addressing the others \citep{vincent2018audio,kounades2017algorithm,jenrungrot_cone_2020}. An investigation of these connections is beyond the scope of this survey.

In the multi-source scenario, the source detection problem transposes to a source counting problem, but the same considerations as in the single-source scenario hold. In some works, the knowledge of the NoS is a working hypothesis, e.g., \citep{grumiaux_saladnet_2021, ma_exploiting_2015,he_adaptation_2019,perotin_crnn-based_2019,fahim_multi-source_2020,bohlender_exploiting_2021,grumiaux_improved_2021} and the sources' DoA can be directly estimated. If the NoS is unknown, one can apply a source counting system beforehand, e.g., with a dedicated DNN \citep{grumiaux_high-resolution_2020}. For example, \citet{tian_multiple_2020} trained a separate neural network to estimate the NoS in the recorded mixture signal, after which he used this information along with the output of the DoA estimation neural network. Alternatively, the NoS can be estimated alongside the DoAs, as in the single-source scenario, based on the SSL network output. When using a classification paradigm, the network output generally predicts the probability of the presence of a source within each discretized region of the space (see Section~\ref{sec:learningStrategies}). One can thus set a threshold on this estimated probability, which implicitly provides source counting.\footnote{Note that this problem is common to DL-based multi-source SSL methods and conventional methods for which a source activity profile is estimated and peak-picking algorithms are typically used to select the active sources.} Otherwise, the ground-truth or estimated NoS is typically used to select the corresponding number of classes having the highest probability. 

Finally, several DNN-based systems were purposefully designed to estimate the NoS alongside the DoAs. For example, the method proposed by \citet{nguyen_robust_2020} uses a neural architecture with two output branches: the first branch is used to estimate the NoS (up to four sources; the problem is formulated as a classification task), while the second branch is used to classify the azimuth into several regions. In the same spirit, we can mention the numerous systems presented at the DCASE Challenge, in which the SED task, jointly conducted with SSL, intrinsically provides an estimate of the NoS. Note that many DCASE Challenge candidate systems will be reviewed in the core of this survey.

\subsection{Moving sources}

%Source tracking is the problem of localizing moving sources, \emph{i.e.}, sources whose location evolves with time \SK{This is not entirely correct: one can track static sources as well, which can still be challenging due to clutter/false alarms.}. 

Source tracking is the problem of estimating the evolution of the sources' position(s) over time, especially when the sources are mobile. In this survey paper, we do not address the problem of tracking on its own, which is usually done in a separate algorithm using the sequence of DoA estimates obtained by applying SSL on successive time windows \citep{vo_multitarget_2015}. 
Still, several DL-based SSL systems have been shown to produce more accurate localization of moving sources when they were trained on a dataset that includes this type of source \citep{adavanne_localization_2019, diaz-guerra_robust_2021, guirguis_seld-tcn_2021, he_sounddet_2021}.
In other cases, as the number of real-world datasets with moving sources is limited and the simulation of signals with moving sources is cumbersome, a number of systems trained on static sources have been shown to retain fair to good performance for moving sources, e.g.,  \citep{opochinsky2021deep, grumiaux_improved_2021, sundar_raw_2020}.

\section{Conventional SSL methods}
\label{sec:traditional}

Before the advent of DL, a set of signal processing techniques were developed to address SSL. A detailed review of these techniques was made by \citet{dibiase_robust_2001}. A review in the specific robotics context was made by \citet{argentieri2015survey}. In this section, we briefly present the most common conventional SSL methods. As briefly stated in the introduction, the reason for this is twofold: first, conventional SSL methods are often used as baselines for DL-based methods; and second, many DL-based SSL methods use input features extracted with conventional methods (see Section~\ref{sec:inputFeatures}).

When the geometry of the microphone array is known, DoA estimation can be performed by estimating the time-difference of arrival (TDoA) of the sources between the microphones \citep{xu_high-accuracy_2013}. The generalized cross-correlation with phase transform (GCC-PHAT) is one of the most employed method when dealing with a 2-microphone array \citep{knapp_generalized_1976}. It is computed as the inverse Fourier transform of a weighted version of the cross-power spectrum (CPS) between the signals of the two microphones: 
\begin{equation}
r_{1,2}(\tau)=\sum_{f=0}^{F-1} \frac{X_1(f) X_2(f)^{*}}{|X_1(f) X_2(f)^{*}|} e^{j 2 \pi \frac{f \tau}{N}},
\end{equation}
where $X_i(f)$ are the $N$-point Fourier transform of the microphone signals $x_i(t)$, and $X_1(f) X_2(f)^{*}$ is the CPS ($^{*}$~denotes the complex conjugate). The TDoA estimate is then obtained by finding the time delay between the microphone signals that maximizes the GCC-PHAT function:
\begin{equation}
\hat{\tau} = \arg \max_{\tau} r_{1,2}(\tau).
\label{eq:TDoA-estimate-by-GCC-PHAT}
\end{equation}
The GCC approach has been extended to arrays with more than two microphones, showing in particular that the localization could be improved by taking advantage of the multiple microphone pairs  \citep{dibiase_robust_2001,benesty_microphone_2008}.

Building an acoustic power map $ P(\mathbf{x})$, with $\mathbf{x}$ the spatial coordinates, usually  a regular grid, is another way to retrieve the DoA of one or multiple sources, as local maxima of this map mainly correspond to the sources' DoA. The Steered-Response Power (SRP) map has been extensively used: it consists in pointing delay and sum beamformers towards each of the candidate grid positions and measuring the energy that arises from these directions. Its PHAT version, which reveals more robust to reverberation, is certainly the most popular. Practically, it can be derived from the average of the GCC-PHAT computed on all microphone pairs \citep{dibiase_robust_2001}:
\begin{equation}
P(\mathbf{x}) = \sum_{m_1=1}^{M}\sum_{m_2=m_1+1}^{M} r_{1,2}(\tau_{m_1,m_2}(\mathbf{x})),
\end{equation}
where $\tau_{m_1,m_2}(\mathbf{x})$ is the delay between the microphones $m_1$ and $m_2$ associated to the spatial position $\mathbf{x}$.

An alternative to building the SRP-based acoustic map -- which happens to be computationally expensive as it usually amounts to a grid search -- is localization by exploiting the sound intensity. The use of sound intensity for source localization has a long history, e.g., \citep{nehorai1994acoustic,hickling1993finding,jarrett20103d,tervo2009direction,raangs2002sound,basten2008acoustic}. In favorable acoustic conditions, sound intensity is parallel to the direction of the propagating sound wave (see Section~\ref{ss:intensity}), and hence the DoA can be efficiently estimated. Unfortunately, its accuracy quickly degrades in the presence of acoustic reflections \citep{daniel2020time}.

\textit{Subspace methods} are another classical family of localization algorithms. These methods rely on the computation of the (time-averaged) CPS matrix $\mathbf{R}(f)$ defined by:
\begin{equation}
\mathbf{R}(f) = \sum_{n=1}^N \mathbf{X}(f,n)\mathbf{X}(f,n)^{H},
\end{equation}
where $\mathbf{X}(f,n)$ is the STFT (or more generally a local discrete Fourier transform) of the multichannel signal vector defined in \eqref{eq:microphone-signal-sum-STFT} ($^H$ denotes the Hermitian operator), and its eigenvalue decomposition (EVD). Assuming that the target source signals and noise are uncorrelated, the multiple signal classification (MUSIC) method \citep{schmidt_multiple_1986} applies EVD to estimate the signal and noise subspaces. After Eq.~\eqref{eq:microphone-signal-sum-STFT}, the signal subspace bases are assumed to correspond to the columns of the mixing matrix $\mathbf{A}(f)$, which are the multichannel ATFs of the sources (often referred to as \emph{steering vectors} in this context). The signal or noise subspace bases are then used to probe a given direction for the presence of a source, i.e.~apply \emph{spatial filtering} or \emph{beamforming} \citep{van1988beamforming,benesty_microphone_2008}.  This time-demanding search can be relaxed using the Estimation of Signal Parameters via Rotational Invariance Technique (ESPRIT) algorithm \citep{roy_esprit-estimation_1989}, which exploits the structure of the source subspace to directly infer the source DoA. However, this often comes at the cost of producing less accurate predictions than MUSIC \citep{mabande2011comparison}. MUSIC and ESPRIT assume narrowband signals, although wideband extensions have been proposed, e.g., \citep{dmochowski2007broadband, hogg2021polynomial}. Subspace methods are robust to noise and can produce highly accurate estimates, but they are  sensitive to reverberation.

Methods based on probabilistic generative mixture models have been proposed by, e.g., \citet{roman2008binaural, mandel2009model, may2011probabilistic, woodruff2012binaural, schwartz2013speaker, dorfan2015tree, li2017multiple}. Typically, the models are variants of Gaussian mixture models (GMMs), with one Gaussian component per source to be localized or per candidate source position. In a very few papers (e.g., \citep{may2011probabilistic}), the model is trained offline with a dedicated training dataset. But most often, the model parameters are directly estimated ``at test time,'' that is using the multichannel signal containing the sources to localize. This is done by maximizing the data likelihood function with histogram-based or expectation-maximization (EM) algorithms exploiting the sparsity of sound sources in the time-frequency (TF) domain \citep{rickard_approximate_2002}, which can be computationally intensive. A GMM variant functioning directly in regression mode, \emph{i.e.}, a form of Gaussian mixture regression (GMR), was proposed for single-source localization by \citet{deleforge20122d} and later extended to multi-source localization (and possibly separation) \citep{deleforge2013variational,deleforge2015co}. The GMR is locally linear but globally non-linear and the estimation of the model parameters is done offline on training data. Hence the spirit is close to DNN-based SSL. White noise signals convolved with synthetic RIRs were used for training. The method was shown to generalize well to speech signals, which are sparser than noise in the TF domain, thanks to the use of a latent variable modeling the signal activity in each TF bin.

Mixture models are strongly connected to Bayesian inference, which considers the posterior distribution of model parameters given the observed data (hence involving both the likelihood function and a prior distribution of the model parameters). \citet{escolano2014bayesian} considered applying Bayesian inference on a Laplacian source mixture model, using GCC-PHAT features in a two-microphone array set-up. Interestingly, they used two levels of Bayesian inference: one for the estimation of the NoS (which is an hyper-parameter of the model), using Bayesian model selection, and one for the estimation of the model parameters (and thus the corresponding source DoAs), using posterior distribution evaluation. In this work, the evaluation of the involved distributions was done with sampling techniques, e.g., Markov Chain Monte Carlo (MCMC) methods. The same methodology was further applied by \citet{bush2018model} with a coprime array consisting of two superimposed, spatially undersampled, uniform linear arrays \citep{vaidyanathan2010sparse}, and by \citet{landschoot2019model} in the spherical harmonics domain using a spherical microphone array (see Section~\ref{sec:inputFeatures}). 

\textit{Compressive sensing} and sparse recovery methods are widely used in acoustics for different purposes \cite{gerstoft2018introduction, xenaki2014compressive}, including SSL \cite{yang2018sparse}. The main premise is that many high-dimensional signals admit a low-dimensional representation, which can be viewed through, e.g., \emph{sparse synthesis} \cite{candes2006stable} or \emph{sparse analysis} \cite{nam2013cosparse}  model. Concerning the SSL problem, the sparsity assumption is usually assumed in the spatial (or spatial beam) domain, e.g., \cite{chardon2012narrowband,noohi2013direction,fortunati2014single,kitic2014hearing,gerstoft2016multisnapshot}, and the resulting problem is addressed by convex optimization, greedy or Bayesian methods, e.g., \cite{foucart2013invitation,gerstoft2018introduction}. This concept has lead to prominent localization methods achieving remarkable performance. Nonetheless, despite their strong theoretical guarantees, compressive sensing methods suffer from two drawbacks. %however, the necessary assumptions to support these rigorous results are often violated in practice. Nonetheless, this concept has lead to prominent localization methods achieving remarkable performance, despite the two main drawbacks. 
For one, it is usually required that the sources coincide with points of some pre-defined grid, although grid-free methods have been proposed in some specific cases, e.g., \cite{xenaki2015grid,yang2015enhancing}. The second issue is shared with other conventional methods, \emph{i.e.}, the strong modeling assumptions reflected in, for example, the known structure of the (sub-Gaussian) dictionary matrix. Dictionary learning techniques have been proposed to alleviate the latter problem to some extent, e.g., \cite{wang2018acoustic,hahmann2021spatial,zea2021representation}. %} \addnote[LG-R2-Remark-2-SBL]{1}{
Sparse Bayesian learning (SBL) is a combination of the Bayesian framework with the principles of sparse representations and compressed sensing. %(in short, estimate high-dimensional sparse vectors using as few measurements as possible), a domain that has been applied to acoustics in the last decade \citep{xenaki2014compressive, gerstoft2018introduction}. 
It usually involves using sparse arrays such as the coprime array mentioned above and nested arrays \citep{pal2010nested}. SBL has been used for SSL by, e.g., \citet{nannuru2018sparse, zhang2014off,liu2012efficient,gerstoft2016multisnapshot, xenaki_sound_2018, ping2020three}.  

Finally, ICA is a class of algorithms aimed at retrieving the different source signals comprising a mixture by assuming and exploiting their mutual statistical independence. ICA has most often been used in audio processing for blind source separation, but it has also proven to be useful for multi-source SSL \citep{sawada_direction_2003}. As briefly stated before, in the multi-source scenario, SSL is closely related to the source separation problem, since localization can help separation, and separation can help localization \citep{gannot2017consolidated, vincent2018audio}. 
%A deep investigation of this topic is beyond the scope of this paper.

%%% ARCHITECTURES %%%
\section{Neural network architectures for SSL}
\label{sec:NN-archi}

In this section, we discuss the neural network architectures that have been proposed in the literature to address the SSL problem. However, we do not present the basics of these neural networks since they have been extensively described in the general DL literature, e.g., \citep{lecun2015deep,goodfellow_deep_2016, chollet2017deep}. The design of DNNs for a given application often requires investigating (and possibly combining) different architectures and tuning their hyperparameters. This was the case for SSL over the last decade, and the evolution of DL-based SSL techniques has followed the general evolution of DNNs toward more and more complex architectures or new efficient models adopted by the DL and SP communities at large, i.e., largely beyond the SSL problem (e.g., attention models). In other words, the DNN architectures used in SSL are often inherited from other works in other (connected or more distant) domains, simply because they were shown to work well on audio signals or other types or signals. In the same spirit, different models are often combined (in parallel and/or sequentially). 

We have thus organized the presentation according to the type of layers used in the networks, with a progressive and ``inclusive'' approach in terms of complexity: a network within a given category can contain layers from another previously presented category. We thus first present systems based on feedforward neural networks (FFNNs). We then focus on convolutional neural networks (CNNs) and recurrent neural networks (RNNs), which generally incorporate some feedforward layers. Next, we review architectures combining CNNs with RNNs, namely convolutional recurrent neural networks (CRNNs). Then, we focus on neural networks with residual connections and with attention mechanisms. Finally, we present SSL systems with an encoder-decoder architecture.

\subsection{Feedforward neural networks}

\begin{figure*}[!t]
\centering
\includegraphics[width=0.6\linewidth]{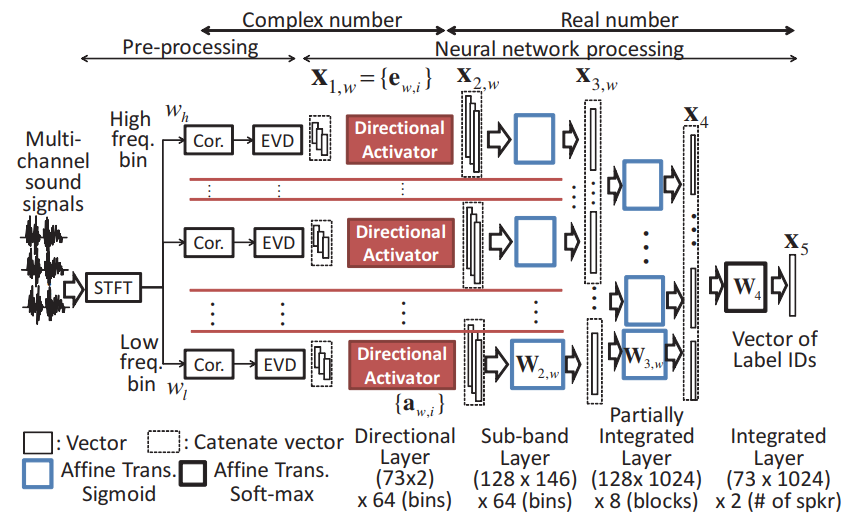}
\caption{The MLP architecture used by Takeda \emph{et al.}~in several papers \citep{takeda_sound_2016, takeda_discriminative_2016, takeda_unsupervised_2017, takeda_unsupervised_2018}. Multiple subband feedforward layers, indexed by $w$, are trained to extract features from the CPS matrix eigenvectors $\mathbf{e}_{w,i}$, which are used as directional activation functions. The obtained subband vectors $\mathbf{X}_{2,w}$ are integrated across subbands progressively via other feedforward layers, giving $\mathbf{X}_{3,w}$ and then $\mathbf{X}_4$. The output layer finally classifies its input in one of the candidate DoAs (the entries of the vector $\mathbf{X}_5$). Note: Reprinted from \citep{takeda_discriminative_2016}; copyright by IEEE; reprinted with permission.}
\label{fig:takedaNetwork}
\end{figure*}

The FFNN was the first and simplest type of artificial neural network to be designed. In such a network, data move in one direction from the input layer to the output layer, possibly via a series of hidden layers \citep{goodfellow_deep_2016,lecun2015deep}. Non-linear activation functions are usually used after each layer (possibly except for the output layer).
While this definition of FFNN is very general and may include architectures such as CNNs (discussed in the next subsection), here we mainly focus on architectures made of fully-connected layers known as Perceptron and Multi-Layer Perceptron (MLP) \citep{goodfellow_deep_2016,lecun2015deep}. A Perceptron has no hidden layer, while the notion of MLP is a bit ambiguous: some authors state that an MLP has one hidden layer, while others allow more hidden layers. In this paper, we call an MLP an FFNN with one or more hidden layers.

A few pioneering SSL methods using shallow neural networks (Perceptron or 1-hidden layer MLP) and applied in ``unrealistic'' setups (e.g., assuming direct-path sound propagation only) have been briefly mentioned in Section~\ref{subsec:acoustic-environments}.
One of the first uses of an MLP for SSL was proposed by \citet{kim_direction_2011}, who actually considered several MLPs. One network estimates the NoS, after which a distinct network is used for SSL for each considered NoS. The authors evaluated their method on reverberant data even though they assumed an anechoic setting. \citet{tsuzuki_approach_2013} proposed using a complex-valued MLP in order to process complex two-microphone-based features, which led to better results than using a real-valued MLP. \citet{youssef_learning-based_2013} also used an MLP to estimate the azimuth of a sound source from a binaural recording made with a robot head. The interaural time difference (ITD) and the interaural level difference (ILD) values (see Section~\ref{sec:inputFeatures}) were separately fed into the input layer and were each processed by a specific set of neurons. A single-hidden-layer MLP was used by \citet{xiao_learning-based_2015}, taking GCC-PHAT-based features as inputs and tackling SSL as a classification problem (see Section~\ref{sec:learningStrategies}), which showed an improvement over conventional methods on simulated and real data. A similar approach was proposed by \citet{vesperini_neural_2016}, but the localization was done by regression in the horizontal plane. 

Naturally, MLPs with deeper architecture (\emph{i.e.}, more hidden layers) have also been investigated for SSL.  \citet{roden_sound_2015} compared the performance of an MLP with two hidden layers and different input types, the number of hidden neurons being linked to the type of input features (see Section \ref{sec:inputFeatures} for more details). \citet{yiwere_distance_2017} used an MLP with three hidden layers (tested with different numbers of neurons) to output source azimuth and distance estimates. An MLP with four hidden layers was tested by \citet{he_deep_2018} for multi-source localization and speech/non-speech classification, showing similar results as a 4-layer CNN (see Section~\ref{sec:convolutionalNeuralNetworks}). 

\citet{ma_exploiting_2015} proposed using a different MLP for different frequency sub-bands, with each MLP having eight hidden layers. This idea is based on the assumption that, in the presence of multiple sources, each frequency band is mostly dominated by a single source, which enables the training to be done exclusively on single-source data. The output of each sub-band MLP corresponds to a probability distribution on azimuth regions, and the final azimuth estimations are obtained by integrating the probability values over the frequency bands.  Another system in the same vein was proposed by Takeda \emph{et al.}~in several papers  \citep{takeda_sound_2016, takeda_discriminative_2016, takeda_unsupervised_2017, takeda_unsupervised_2018}. In these works, the eigenvectors of the recorded signal interchannel correlation matrix were separately fed per frequency band into parallel branches of the network, particularly into specific fully-connected layers. Then, several additional fully-connected layers progressively integrated the frequency-dependent outputs (see Fig.~\ref{fig:takedaNetwork}). The authors showed that this specific architecture outperforms a more conventional 7-layer MLP and the classical MUSIC algorithm on anechoic and reverberant single- and multi-source signals. \citet{opochinsky_deep_2019} proposed a small 3-layer MLP to estimate the azimuth of a single source using the relative transfer function (RTF, see Section~\ref{sss:RTFs}) of the signal. Their approach is weakly supervised since one part of the loss function is computed without the ground truth DoA labels (see Section \ref{sec:learningStrategies}).

An indirect use of an MLP was explored by \citet{pak_sound_2019}, who used a 3-layer MLP to enhance the interaural phase difference (IPD) (see Section~\ref{sec:inputFeatures}) of the input signal, which was then used for DoA estimation.

\subsection{Convolutional neural networks}
\label{sec:convolutionalNeuralNetworks}

CNNs are a popular class of DNNs widely used for pattern recognition due to their property of being translation equivariant \citep{cohen2019gauge,goodfellow_deep_2016}. They have been successfully applied to various tasks, such as image classification, e.g.,  \citep{krizhevsky_imagenet_2017}, natural language processing (NLP), e.g.,  \citep{kim_convolutional_2014} or automatic speech recognition, e.g.,  \citep{waibel_phoneme_1989}. CNNs have also been used for SSL, as detailed below.

\begin{figure*}[!t]
\centering
\includegraphics[width=0.7\linewidth]{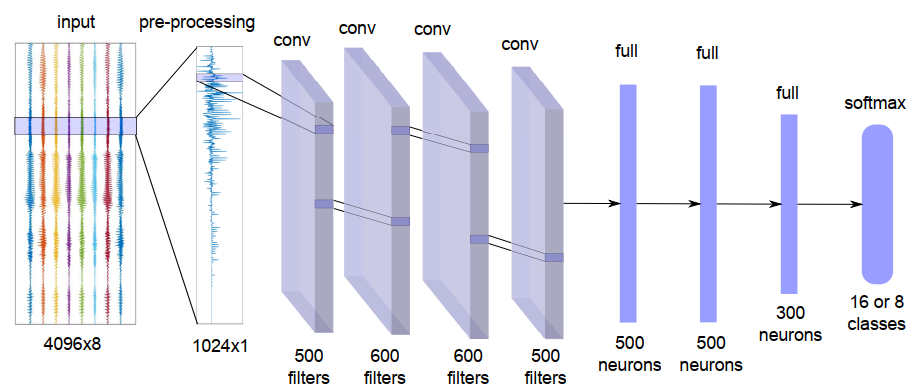}
\caption{The CNN architecture proposed by \citet{hirvonen_classication_2015} for SSL. The input is an 8-channel signal. For each short-term frame, the 8 magnitude spectra (of $128$ frequency bins) are concatenated to form a $1024 \times 1$ tensor, which is fed into a series of four convolutional layers with $500$ or $600$ learnable kernels. The extracted features then pass through several feedforward layers containing $500$ or $300$ neurons. The output layer contains $8$ neurons (or $16$ if the source type is also considered) and  estimates the probability of a source being present in eight candidate DoAs using a softmax activation function. Note: Reprinted from \citep{hirvonen_classication_2015}; copyright by the author; reprinted with permission.}
\label{fig:hirvonenNetwork}
\end{figure*}

% motivation for using a CNN?
To our knowledge, \citet{hirvonen_classication_2015} was the first to use a CNN for SSL. He employed this architecture to classify an audio signal containing one speech or musical source into one of eight spatial regions (see Fig.~\ref{fig:hirvonenNetwork}). This CNN is composed of four convolutional layers to extract feature maps from multichannel magnitude spectrograms (see Section~\ref{sec:inputFeatures}), followed by four fully-connected layers for classification. Classical pooling is not used because, according to the author, it does not seem relevant for audio representations. Instead, a 4-tap stride with a 2-tap overlap is used to reduce the number of parameters. This approach shows good performance on single-source signals and is capable of adapting to different configurations without hand-engineering. However, two topical issues of such a system were pointed out by the author: the robustness of the network with respect to a shift in source location, and the difficulty of interpreting the hidden features. 

Chakrabarty and Habets also designed a CNN to predict the azimuth of one \citep{chakrabarty_broadband_2017} or two  \citep{chakrabarty_multi-speaker_2019, chakrabarty_multi-speaker_2017} speakers in reverberant environments. The input features are the multichannel short-time Fourier transform (STFT) phase spectrograms (see Section~\ref{sec:inputFeatures}). In \citep{chakrabarty_broadband_2017}, they proposed using three successive convolutional layers with $64$ filters of size $2 \times 2$ to consider neighboring frequency bands and microphones. In \citep{chakrabarty_multi-speaker_2017}, they reduced the filter size to $2 \times 1$ ($1$ in the frequency axis) because of the W-disjoint orthogonality (WDO) assumption for speech signals, which assumes that several speakers are not simultaneously active in a same TF bin \citep{rickard_approximate_2002}. In \citep{chakrabarty_multi-speaker_2019}, they demonstrated that for an $M$-microphone array, the optimal number of convolutional layers for exploiting phase correlations between the neighboring microphones is $M-1$. 

\citet{he_deep_2018} compared a 4-layer MLP and a 4-layer CNN for the multi-speaker detection and localization task. The results showed similar accuracy for both architectures. A deeper architecture was proposed by \citet{yalta_sound_2017}, with 11 to 20 convolutional layers depending on the experiments. These deeper CNNs showed robustness against noise compared to MUSIC, as well as smaller training time, but this was partly due to the presence of residual blocks (see Section \ref{sec:residualNeuralNetworks}). A similar architecture was presented by \citet{he_joint_2018}, with many convolutional layers and some residual blocks, although with a specific multi-task configuration. The end of the network was split into two convolutional branches, one for azimuth estimation, and the other for  speech/non-speech signal classification. 

While most localization systems aim to estimate the azimuth or both the azimuth and elevation, \citet{thuillier_spatial_2018} investigated the estimation of only the elevation angle using a CNN with binaural input features: the ipsilateral and contralateral head-related transfer function (HRTF) magnitude responses (see Section \ref{sec:inputFeatures}). \citet{vera-diaz_towards_2018} chose to apply a CNN directly on raw multichannel waveforms, assembled side by side as an image, to predict the Cartesian coordinates $(x, y, z)$ of a single static or moving speaker. The successive convolutional layers contain around a hundred filters from size $7 \times 7$ for the first layers to $3 \times 3$ for the last layer. \citet{ma_phased_2018} also used a CNN to perform regression, but they used the CPS matrix as an input feature (see Section~\ref{sec:inputFeatures}). To estimate both the azimuth and elevation, \citet{nguyen_autonomous_2018} used a relatively small CNN (two convolutional layers) in regression mode, with binaural input features. A similar approach was considered by \citet{sivasankaran_keyword-based_2018} for speaker localization based on a CNN. They showed that injecting a speaker identifier, particularly a mask estimated for the speaker uttering a given keyword, alongside the binaural features at the input layer improved the DoA estimation. 

A joint VAD and DoA estimation CNN was developed by \citet{vecchiotti_deep_2018}. They showed that both problems can be handled jointly in a multi-room environment using the same architecture, although considering separate input features (GCC-PHAT and log-mel-spectrograms) in two separate input branches. These branches are then concatenated in a further layer. \citet{vecchiotti_detection_2019} extended this work by exploring several variant architectures and experimental configurations, and \citet{vecchiotti_end--end_2019} developed an end-to-end auditory-inspired system based on a CNN, with Gammatone filter layers included in the neural architecture. A method based on mask estimation was proposed by \citet{zhang_robust_2019}, in which a TF mask was estimated and used to either clean or be appended to the input features, facilitating the DoA estimation by a CNN.

\citet{nguyen_robust_2020} presented a multi-task CNN containing 10 convolutional layers with average pooling, inferring both the NoS and the sources' DoA. They evaluated their network on signals with up to four sources, showing very good performance on both simulated and real environments. A small 3-layer CNN was employed by \citet{varanasi_deep_2020} to infer both azimuth and elevation using signals decomposed with third-order spherical harmonics (see Section~\ref{sec:inputFeatures}). The authors tried several combinations of input features, including using only the magnitude and/or the phase of the spherical harmonic decomposition.

In the context of hearing aids, a CNN was applied to both VAD and DoA estimation by \citet{varzandeh_exploiting_2020}. This system is based on two input features, GCC-PHAT and periodicity degree, both fed separately into two convolutional branches. These two branches are then concatenated in a further layer, which is followed by feedforward layers. \citet{fahim_multi-source_2020} applied an 8-layer CNN to the so-called modal coherence of first-order Ambisonics input features (see Section~\ref{sec:inputFeatures}) for the localization of multiple sources in a reverberant environment. They proposed a new method to train a multi-source DoA estimation network with only single-source training data, showing an improvement over the system of \citet{chakrabarty_multi-speaker_2019}, especially for signals with three speakers. \citet{hao_spectral_2020} investigated a real-time implementation of SSL using a CNN with a relatively small architecture (three convolutional layers). 

\citet{krause_comparison_2021} investigated the use of several types of convolution. They reported that networks using 3D convolutions (on the time, frequency, and channel axes) achieved better localization accuracy compared to those based on 2D convolutions, complex convolutions, and depth-wise separable convolutions (all of them on the time and frequency axes), but with a high computational cost. They also showed that the use of depth-wise separable convolutions leads to a good trade-off between accuracy and model complexity (to our knowledge, they were the first to explore this type of convolutions).

\citet{bologni_acoustic_2021} proposed a neural network architecture including a set of 2D convolutional layers for frame-wise feature extraction, followed by several 1D convolutional layers in the time dimension for temporal aggregation. \citet{diaz-guerra_robust_2021} applied 3D convolutional layers on SRP-PHAT power maps computed for both azimuth and elevation estimation. They also used a couple of 1D causal convolutional layers at the end of the network to perform single-source tracking. Their whole architecture was designed to function in fully causal mode so that it can be adapted for real-time applications. \citet{wu2021sound} proposed using a supervised image mapping approach inspired from computer vision works and referred to as \textit{image translation}. The used a CNN (completed with residual layers, see Section~\ref{sec:residualNeuralNetworks}) to map an input 2D image (DoA features extracted by conventional beamforming and reshaped as a function of Cartesian coordinates $(x, y)$) into an output 2D image of the target source position (in which the pixel intensity is decreasing rapidly with the distance to the source), from which the source location is obtained.

As mentioned in the introduction, the DCASE Challenge includes a SELD task, and CNNs have also been used in some of the challenge candidate systems \citep{politis_overview_2020}. \citet{chytas_hierarchical_2019} used convolutional layers with hundreds of filters of size $4 \times 10$ for azimuth and elevation estimation in a regression mode. \citet{kong_cross-task_2019} compared different numbers of convolutional layers for SELD, while an 8-layer CNN was proposed by \citet{noh_three-stage_2019} to improve the results over the baseline.

An indirect use of a CNN was proposed by \citet{salvati_exploiting_2018}. They trained the neural network to estimate a weight for each of the narrow-band SRP components fed at the input layer in order to compute a weighted combination of these components. In their experiments, they showed on a few test examples that this allowed for a better fusion of the narrow-band components and reduced the effects of noise and reverberation, leading to better localization accuracy.

In the DoA estimation literature, a few works have explored the use of \textit{dilated convolutions} in DNNs. Dilated convolutions, also known as \textit{atrous} convolutions, are a type of convolutional layer in which the convolution kernel is wider than the classical one but zeros are inserted so that the number of parameters remains the same. Formally, a 1D dilated convolution with a dilation factor $l$ is defined by
\begin{equation}
    (x * k)(n) = \sum_{i} x(n-li)k(i),
\end{equation}
where $x$ is the input and $k$ the convolution kernel. The conventional linear convolution is obtained with $l=1$. This definition extends to multidimensional convolution. 

\citet{chakrabarty_multi-scale_2019} demonstrate that incorporating dilated convolutions with gradually increasing dilation factors reduces the optimal number of convolutional layers of their original CNN architecture \citep{chakrabarty_multi-speaker_2019} (discussed previously in this section). This leads to an architecture with similar SSL performance and lower computational cost.

\subsection{Recurrent neural networks}

RNNs are neural networks designed for modeling temporal sequences of data \citep{lecun2015deep,goodfellow_deep_2016}. Particular types of RNNs include long short-term memory (LSTM) cells \citep{hochreiter_long_1997} and gated recurrent units (GRUs) \citep{cho_learning_2014}. These two types of RNNs have become very popular thanks to their capability to circumvent the training difficulties that regular RNNs face, in particular the vanishing and exploding gradient problems \citep{lecun2015deep,goodfellow_deep_2016}. 

There are few published works on SSL using only RNNs, as recurrent layers are often combined with convolutional layers (see Section \ref{sec:ConvolutionalRecurrentNeuralNetworks}). \citet{nguyen_general_2021} used an RNN to align SED and DoA predictions, which were obtained separately for each possible sound event type. The RNN was ultimately used to determine which SED prediction matched which DoA estimation. A bidirectional LSTM network was used by \citet{wang_robust_2019} to estimate a TF mask to enhance the signal, further facilitating DoA estimation by conventional methods such as SRP or subspace methods.

\subsection{Convolutional recurrent neural networks}
\label{sec:ConvolutionalRecurrentNeuralNetworks}

% Here the motivation is stated.
CRNNs are neural networks containing one or more convolutional layers and one or more recurrent layers. CRNNs have been regularly exploited for SSL since 2018 because of the respective capabilities of these layers: The convolutional layers have proven to be suitable for extracting relevant features for SSL, and the recurrent layers are well designed for integrating the information over time.

\begin{figure}[!t]
\centering
\includegraphics[width=0.7\linewidth]{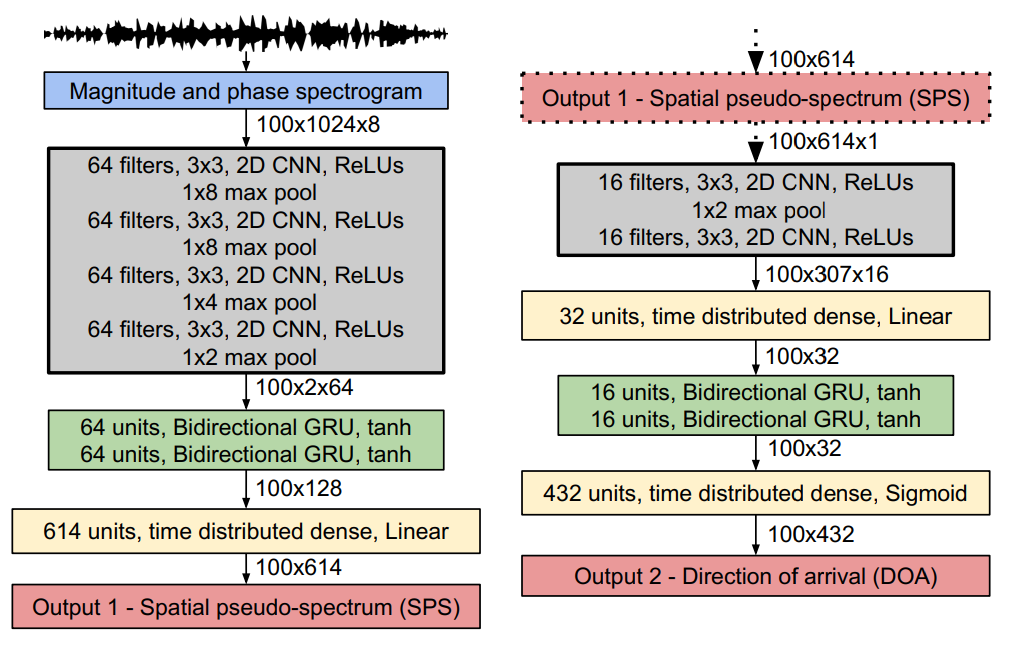}
\caption{The CRNN architecture of \citet{adavanne_direction_2018, adavanne_localization_2019, adavanne_sound_2019}, which has inspired numerous SELD systems. The input is the multichannel STFT-domain FOA magnitude and phase spectrogram. First, features are extracted by four successive convolutional layers with $64$ $3 \times 3$ kernels, each followed by a max-pooling layer. Then two BGRU layers with $64$ units each and tanh activations are used to capture the temporal evolution of the extracted features. An intermediate SPS output is then computed using a time distributed feedforward layer (\textit{i.e.}, this layer is computed separately on each vector of the temporal axis). Then, two $16$-kernel convolution layers followed by a $32$-unit time distributed feedfoward layer and two $16$-units BGRU layers process the estimated SPS. A final $432$-unit time distributed feedforward layer with sigmoid activation function is employed to infer the DoA. Note: Reprinted from \citep{adavanne_direction_2018}; copyright by IEEE; reprinted with permission.}
\label{fig:adavanneNetwork}
\end{figure}

In the series of papers \citep{adavanne_localization_2019,adavanne_direction_2018, adavanne_sound_2019}, Adavanne \emph{et al.}~used a CRNN for SELD, in a multi-task configuration, with first-order Ambisonics (FOA) input features (see Section~\ref{sec:inputFeatures}). In \citep{adavanne_direction_2018}, their architecture contained a series of successive convolutional layers, each followed by a max-pooling layer and two bidirectional GRU (BGRU) layers. Then, a feedforward layer provided an estimation of the spatial pseudo-spectrum (SPS) provided by the MUSIC algorithm \citep{schmidt_multiple_1986}, acting as an intermediary output (see Fig.~\ref{fig:adavanneNetwork}). This SPS was then fed into the second part of the neural network, which was composed of two convolutional layers, a dense layer, two BGRU layers, and a final feedforward layer for azimuth and elevation estimation by classification. The use of an intermediary SPS output has been proposed to help the neural network learn a representation that has proven to be useful for SSL using traditional methods. 

In \citep{adavanne_sound_2019} and \citep{adavanne_localization_2019}, this intermediary output was no longer used. Instead, the DoA was directly estimated using a block of convolutional layers, a block of BGRU layers, and a feedforward layer. This system is able to localize and detect several sound events even if they overlap in time, provided they are of different types (e.g., speech and car, see the discussion in Section~\ref{sec:sourcetypes}). This CRNN was the baseline system for Task 3 of the DCASE Challenge in 2019 and 2020. Therefore, it has inspired many other works, and many DCASE Challenge candidate systems were built on the system of \citet{adavanne_sound_2019} with various modifications and improvements. 

For example, \citet{lin_report_2019} added Gaussian noise to the input spectrograms to train the network to be more robust to noise. \citet{lu_sound_2019} integrated some additional convolutional layers and replaced the BGRU layers  with bidirectional LSTM layers. \citet{Leung2019} used the same architecture with all combinations of cross-channel power spectra, whereas the replacement of input features with group delays was tested by \citet{nustede_group_2019}. GCC-PHAT features were added as input features by \citet{maruri_gcc-phat_2019}. \citet{zhang_data_2019} used data augmentation during training and averaged the output of the network for a more stable DoA estimation. 
\citet{xue_multi-beam_2019} sent the input features separately into different branches of convolutional layers, log-mel, and constant Q-transform features on the one hand, and phase spectrograms and CPS features on the other hand (see Section~\ref{sec:inputFeatures}). \citet{Cao2019polyphonic} concatenated the log-mel spectrogram and GCC-PHAT features and fed them into two separate CRNNs for SED and DoA estimation (they also incorporated the intensity vector in \citep{cao_two-stage_2019}). In contrast to the baseline of \citet{adavanne_sound_2019}, more convolutional layers and one single BGRU layer were used. The convolutional part of the DoA network was transferred from the SED CRNN, which was followed by fine-tuning of the DoA branch, labelling this method as \textit{two-stage}. This led to a notable improvement in localization performance over the DCASE Challenge baseline of \citet{adavanne_sound_2019}. Small changes to this baseline were also tested by \citet{Pratik2019}, such as the use of Bark-scale spectrograms as input features, the modification of the activation function or pooling layers, and the use of data augmentation, resulting in noticeable improvements for some experiments. 

The same baseline neural architecture  of \citet{adavanne_sound_2019} was used by \citet{kapka_sound_2019}, with one separate (but identical, except for the output layer) CRNN instance for each subtask: source counting (up to two sources), DoA estimation of source 1 (if applicable), DoA estimation of source 2 (if applicable), and sound type classification. The authors showed that their method was more efficient than the baseline. 
\citet{krause_arborescent_2019} explored different manners of splitting the SED and DoA estimation tasks in a CRNN. While some configurations showed an improvement in SED, the localization accuracy was below the baseline for the reported experiments. \citet{park_reassembly_2019} investigated a combination of a gated linear unit (GLU, a convolutional block with a gated mechanism) and a trellis network (containing convolutional and recurrent layers, see the paper by \citet{bai_trellis_2019} for details), yielding better results than the baseline. The authors extended this work for the DCASE 2020 Challenge by improving the overall architecture and investigating other loss functions \citep{park_sound_2020}. A non-direct DoA estimation scheme was also derived by \citet{grondin_sound_2019}, who estimated the TDoA using a CRNN, from which they inferred the DoA.

We also found propositions of CRNN-based systems in the 2020 edition of the DCASE Challenge. \citet{singla_sequential_2020} used the same CRNN as in the baseline of \citet{adavanne_sound_2019}, except that they did not use two separated output branches for SED and DoA estimation. Instead, they concatenated the SED output with the output of the previous layer to estimate the DoA. \citet{song_localization_2020} used separated neural networks similar to the one of \citet{adavanne_sound_2019} to address NoS estimation and DoA estimation in a sequential way. Multiple CRNNs were trained by \citet{tian_multiple_2020}: one to estimate the NoS (up to two sources), another to estimate the DoA assuming one active source, and another (same as the baseline) to estimate the DoAs of two simultaneously active sources. \citet{cao_event-independent_2020} designed an end-to-end CRNN architecture to detect and estimate the DoA of possibly two instances of the same sound event. The addition of one-dimensional convolutional filters was investigated by \citet{ronchini_sound_2020} to exploit the information along the feature axes. \citet{sampathkumar_sound_2020} augmented the baseline system of \citet{adavanne_sound_2019} by providing the network with more input features (log-mel spectrograms, GCC-PHAT, and intensity vector, see Section~\ref{sec:inputFeatures}).

%Independently of the DCASE challenge, the CRNN of \citet{adavanne_sound_2019} was adapted by \citet{comminiello_quaternion_2019} to receive quaternion FOA input features (see Section \ref{sec:inputFeatures}), which slightly improved the CRNN performance. Perotin \emph{et al.} proposed using a CRNN with bidirectional LSTM layers on an FOA intensity vector to localize one \citep{perotin_crnn-based_2018} or two \citep{perotin_crnn-based_2019} speakers. They showed that this architecture achieves very good performance in simulated and real reverberant environments with static speakers.
Independently of the DCASE Challenge, the CRNN of \citet{adavanne_sound_2019} was adapted by \citet{comminiello_quaternion_2019} to receive quaternion FOA input features, which slightly improved the CRNN performance. Perotin \emph{et al.} proposed using a CRNN with bidirectional LSTM layers on the FOA pseudo-intensity vector to localize one \citep{perotin_crnn-based_2018} or two \citep{perotin_crnn-based_2019} speakers. They showed that this architecture achieves very good performance in simulated and real reverberant environments with static speakers (both types of input features are discussed in Section \ref{sec:inputFeatures}). This work was extended by \citet{grumiaux_improved_2021}, who obtained a substantial improvement in performance over the CRNN of \citet{perotin_crnn-based_2019} by adding more convolutional layers with less max-pooling, to localize up to three simultaneous speakers. 

Non-square convolutional filters and a unidirectional LSTM layer were used in the CRNN architecture of \citet{li_online_2018}. \citet{xue_sound_2020} presented a CRNN with two types of input features: the phase of the CPS and the signal waveforms. The former was first processed by a series of convolutional layers before being concatenated with the latter. Another improvement of the network of \citet{adavanne_sound_2019} was proposed by \citet{komatsu_sound_2021}, who replaced the classical convolutional blocks with GLUs, based on the hypothesis that GLUs are better suited for extracting relevant features from phase spectrograms. This has led to a notable improvement of localization performance compared to the baseline of \citet{adavanne_sound_2019}. \citet {bohlender_exploiting_2021} proposed an extension of the system of \citet{chakrabarty_multi-speaker_2019}, in which LSTMs and temporal convolutional networks (TCNs) replaced the last dense layer of the former architecture. A TCN was made of successive 1D dilated causal convolutional layers with increasing dilated factors \citep{lea_temporal_2017}. The authors showed that taking the temporal context into account with such temporal layers actually improves the localization accuracy.

Finally, we can mention the original approach of \citet{nguyen2020sequence} in which a two-step hybrid approach with two CRNNs is used: In the first step, a first CRNN is used for SED and a single-source histogram-based (conventional) method is used for DoA estimation. In the second step, a second CRNN-based network, referred to as sequence matching network (SMN), is used to match the estimated sequences from the SED and DoA branches. This approach is motivated by the fact that overlapping sounds often have different onsets and offsets, and by matching the outputs of the two branches, an estimated DoA can be associated with the corresponding sound class. This approach was extended to localize moving sources in the framework of the DCASE 2020 Challenge \citep{nguyen2020ensemble}, by adapting the resolution of the azimuth and elevation histograms and by using an ensemble of SMNs.

\subsection{Residual neural networks}
\label{sec:residualNeuralNetworks}

Residual neural networks were originally introduced by \citet{he_deep_2016}, who pointed out that designing very deep networks can lead the gradients to explode or vanish due to the non-linear activation functions, as well as the degradation of the overall performance. Residual connections are designed to enable a feature to bypass a layer block in parallel to the conventional process through this layer block. This allows the gradients to flow directly through the network, usually leading to a better training.

\begin{figure}[!t]
\centering
\includegraphics[]{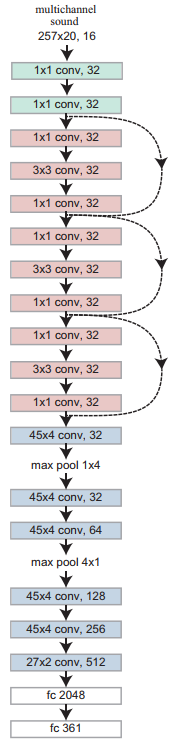}
\caption{The residual neural network architecture used by \citet{yalta_sound_2017}. Three residual blocks are employed in this network, which are each composed of two convolutional layers with $32$ $1 \times 1$ filters with another convolutional layer with $32$ $3 \times 3$ filters in-between. For all three residual blocks, the input is added to the output with a residual connection, showed with a dashed arrow in this diagram. The authors show that the use of residual connections not only reduces the learning cost, but also improves the model performance. Note: Reprinted from \citep{yalta_sound_2017}; under Creative Commons Attribution-NoDerivatives 4.0 International License.}
\label{fig:yaltaNetwork}
\end{figure}

To our knowledge, the first use of a network with residual connections for SSL was proposed by \citet{yalta_sound_2017}. As illustrated in Fig.~\ref{fig:yaltaNetwork}, this network includes three residual blocks, which are stacks of layers with one of the layers having residual connections with another layer deeper in the stack. Each of these blocks is made of three convolutional layers, the first and last of which are designed with $1 \times 1$ filters, with the middle layer designed with $3 \times 3$ filters. A residual connection is used between the input and output of each residual block. The same type of residual block was used for SSL by \citet{he_adaptation_2019, he_joint_2018} in parallel to sound classification as speech or non-speech. \citet{suvorov_deep_2018} used a series of 1D convolutional layers with several residual connections for single-source localization, directly from the multichannel waveform. 

\citet{pujol_source_2019,pujol_beamlearning_2021} integrated residual connections alongside 1D dilated convolutional layers with increasing dilation factors. They used the multichannel waveform as the network input. After the input layer, the architecture was divided into several subnetworks containing the dilated convolutional layers, which functioned as filter banks. 
\citet{ranjan_sound_2019} combined a modified version of the original ResNet architecture \citep{he_deep_2016} with recurrent layers for SELD. This was shown to reduce the DoA error by more than 20\textdegree ~compared to the baseline of \citet{adavanne_sound_2019}. Similarly, \citet{Bai_NWPU_task3_report} also used the ResNet model of \citep{he_deep_2016} followed by two GRU layers and two fully-connected layers for SELD. \citet{kujawski_deep_2019} also adopted the original ResNet architecture and applied it to the single-source localization problem. 

Another interesting architecture containing residual connections was proposed by \citet{naranjo-alcazar_sound_2020} for the DCASE 2020 Challenge. Before the recurrent layers (consisting of two BGRUs), three residual blocks successively processed the input features. These residual blocks contained two residual convolutional layers, followed by a squeeze-excitation module \citep{hu_squeeze-and-excitation_2020}. These modules aim to improve the modeling of interdependencies between input feature channels compared to classical convolutional layers. Similar squeeze-excitation mechanisms were used by \citet{sundar_raw_2020} for multi-source localization. Another combination of a residual network with squeeze-excitation blocks was reported by \citet{Huang_Aalto_task3_report}, who implemented it in the framework of a sample-level CNN (i.e., a CNN applied on the time-domain signal samples) \citep{lee2017sample}. The resulting blocks are further followed by two Conformer blocks (see the next subsection). The motivation for combining these different models was their observed effectiveness in other audio processing tasks such as SED.

\citet{shimada_sound_2020} and \citet{shimada_accdoa_2020} adapted the MMDenseLSTM architecture, originally proposed by \citet{takahashi_mmdenselstm_2018} for sound source separation, to the SELD problem. This architecture consists of a series of blocks made of convolutions and recurrent layers with residual connections. Their system showed very good performance among the other participants to the DCASE 2020 Challenge. \citet{wang_ustc-iflytek_2020} used an ensemble learning approach in which several variants of residual neural networks and recurrent layers were trained to estimate the DoA, achieving the best performance of the DCASE 2020 Challenge. 

\citet{guirguis_seld-tcn_2021} designed a neural network with a TCN in addition to classical 2D convolutions and residual connections. Instead of using recurrent layers as usually considered, the architecture was composed of TCN blocks that were made of several residual blocks, including a 1D dilated convolutional layer with an increasing dilated factor. The authors showed that replacing recurrent layers with TCNs made the hardware implementation of the network more efficient while slightly improving the SELD performance compared to the baseline of \citet{adavanne_sound_2019}.

\citet{yasuda_sound_2020} exploited a CRNN with residual connections in an indirect way for DoA estimation using an FOA pseudo-intensity vector input (see Section~\ref{ss:intensity}). A CRNN was first used to remove the reverberant part of the FOA pseudo-intensity vector, after which another CRNN was used to estimate a TF mask, which was applied to attenuate TF bins with a large amount of noise. The source DoA was finally estimated directly from the dereverberated and denoised pseudo-intensity vector.

\subsection{Attention-based neural networks}
\label{ssec:models-attention}

An \textit{attention mechanism} is a method that allows a neural network to put emphasis on vectors of a temporal sequence that are more relevant for a given task. Originally, attention was proposed by \citet{bahdanau_neural_2016} to improve sequence-to-sequence models such as RNNs for machine translation. The general principle is to allocate a different weight to the vectors of the input sequence when using a combination of these vectors for estimating a vector of the output sequence. The model is trained to compute the optimal weights that reflect both the link between vectors of the input sequence (self-attention) and the relevance of the input vectors to explain each output vector (attention at the decoder). This pioneering work has inspired the now popular \textit{Transformer} architecture proposed by \citet{vaswani_attention_2017}, which greatly improved the machine translation performance. In the Transformer, RNNs are removed, i.e.~they are totally replaced by attention models.

Attention models are now used in an increasing number of DL applications, including SSL. \citet{phan_audio_2020, phan_multitask_2020} submitted an attention-based neural system for the DCASE 2020 Challenge. Their architecture was made of several convolutional layers, followed by a BGRU, after which a self-attention layer was used to infer the activity and the DoA of several distinct sound events at each timestep. \citet{schymura_exploiting_2021} added an attention mechanism after the recurrent layers of a CRNN to output an estimation of the sound source activity and its azimuth/elevation. Compared to the baseline of \citet{adavanne_sound_2019}, the addition of attention demonstrated a better use of temporal information for SELD. An extension of the system of \citet{chakrabarty_multi-speaker_2019} based on attention mechanisms has been proposed by \citet{mack_signal-aware_2020}. Attention is employed to estimate binary masks to focus on frequency bins where the target source is dominant. The first attention stage appears right after the input layer (analogously to \citep{chakrabarty_multi-speaker_2019}, their network uses phase spectrograms as inputs), while the second attention stage takes place after new features have been extracted using convolutional layers. %features which consist of microphone phases, while the second attention stage takes place after new features have been extracted using convolutional layers.}
\citet{adavanne_differentiable_2021} used a self-attention layer after a GRU in order to estimate the association matrix which matches predictions and references. This solves the optimal assignment problem and resulted in large improvements in terms of localization error.

\begin{figure*}[!t]
\centering
\includegraphics[width=0.9\linewidth]{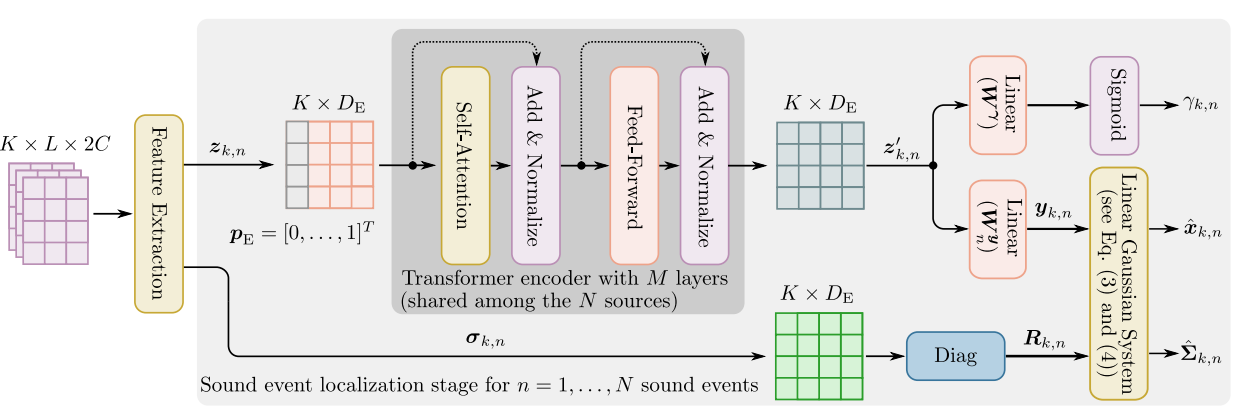}
\caption{The self-attention-based neural network architecture of \citet{schymura_pilot_2021}. The input is the multi-channel spectrogram shaped as a $K \times L \times 2C$ tensor, with $K$ the number of frequency bins, $L$ the number of frames, and $C$ the number of channels. A feature extraction is first done with convolutional layers (not detailed in the figure) to produce $\mathbf{z}_{k,n}$ to which is attached a positional encoding vector $\mathbf{p}_E$. Then, a Transformer encoder computes a new representation of shape $K \times D_E$, which is used to compute the source activity $\gamma_{k,n}$ and the mean $\mathbf{\hat{x}}_{k,n}$ of the multivariate Gaussian distributions representing the target sources' location (the corresponding covariance matrix $\mathbf{\hat{\Sigma}}_{k,n}$ is computed via a parallel (simpler) mechanism.) Note: Reprinted from \citep{schymura_pilot_2021}; copyright by the authors; reprinted with permission.}
\label{fig:schymuraNetwork}
\end{figure*}

Multi-head self-attention (MHSA), which is the parallel use of several Transformer-type attention models \citep{vaswani_attention_2017}, has also inspired SSL methods. In the DCASE 2021 Challenge, \citet{emmanuel_multiscale_2021} employed a MHSA layer right after several convolution modules tailored to learn varying spectral characteristics. \citet{yalta_dcase_2021} proposed using the whole encoder part of the Transformer architecture, in addition to several convolutional layers, to extract features from the input data. \citet{wang_four-stage_2021} adapted the Conformer architecture, originally designed by \citet{gulati_conformer_2020} for automatic speech recognition, to SSL. This architecture is composed of a feature extraction module based on ResNet and a MHSA module that learns local and global context representations. The authors demonstrated the benefit of using a specific data augmentation technique on this model. \citet{zhang_data_2021} also employed this architecture in the DCASE 2021 Challenge. As briefly mentioned in the previous subsection, Conformer blocks were also used in the architecture proposed by \citet{Huang_Aalto_task3_report}, where they followed a sample-level CNN with residual connections and squeeze-excitation. A Conformer block was also used in the architecture proposed for SELD by \citet{Ko_SKKU_task3_report}, after convolutional and fully-connected layers and before BGRU layers.
\citet{cao_improved_2021} positioned an 8-head attention layer after a series of convolutional layers to track the source location predictions over time for different sources (up to two sources in their experiments). \citet{schymura_pilot_2021} used three 4-head self-attention encoders along the time axis after a series of convolutional layers before estimating the activity and location of several sound events (see Fig.~\ref{fig:schymuraNetwork}). This neural architecture showed an improvement over the DCASE Challenge baseline of \citet{adavanne_sound_2019}. In the same line, \citet{Xinghao2021} replaced the conventional convolutional layers of the baseline with a combination of adaptive convolutional layers (using dilated convolutions with different dilation factors) and attention blocks. Another example of MHSA-based Transformer model for SSL can be found in the work of \citet{Park2021}. In this work, a pretrained model is fine-tuned with transfer learning. The output sequence corresponding to each 3s-sequence of input data is averaged to provide one DoA estimation. \citet{Sudarsanam2021} enriched the CRNN baseline of \citet{adavanne_sound_2019} with a set of several MHSA blocks followed by fully-connected layers. They provided an analysis of the influence of the number and dimension of the MHSA blocks (the optimal number was found to be 2) and the number of heads (optimal was 8), as well as the effect of positional embedding, normalization layers and residual connections.
\citet{grumiaux_saladnet_2021} showed that replacing the recurrent layers of a CRNN with self-attention encoders yielded a notable reduction in the computation time. Moreover, the use of MHSA slightly improved localization performance upon the baseline CRNN architecture of \citet{perotin_crnn-based_2019} for the considered multiple speaker localization task.

Finally, we can mention the use of cross-modal attention (CMA) models for SSL by \citet{Lee_SGU_task3_report}. A CMA model is the generalization of self-attention with two data streams in place of one, which is used in the Transformer decoder \citep{vaswani_attention_2017}.  \citet{Lee_SGU_task3_report} used two separate SED and DoA estimation CNN blocks to separately produce SED and DoA embeddings (this comes in contrast with most DCASE candidate systems where the first blocks are shared between SED and DoA estimation.) Then these embeddings are merged, first with a weighted linear combination and then with a second, more complex, alignment process using two mirrored CMA models. Finally, the SED and DoA outputs of the CMA modules are each sent to three parallel fully-connected networks for final estimation (this is because in the DCASE 2021 Challenge SELD Task, up to three sources can be simultaneously active).

In a general manner, it appears that attention modules, and MHSA in particular, have a tendency to replace the recurrent units in the recent SSL DNNs, following the ``Attention is all you need'' seminal line of \citet{vaswani_attention_2017}. This is because compared to RNNs, attention modules can model longer-term dependencies at a lower computational cost and can highly benefit from parallel computations, especially at training time. This tendency is also observed in other application domains, as we will discuss in Section~\ref{sec:conclusion}.

\subsection{Encoder-decoder neural networks}

An encoder-decoder network is an architecture made of two building blocks: an \textit{encoder}, which is fed by the input features and outputs a specific representation of the input data, and a \textit{decoder}, which transforms the new data representation from the encoder into the desired output data. Architectures following this principle have been largely explored in the DL literature due to their capacity to provide compact data representations in an unsupervised manner \citep{goodfellow_deep_2016}. 

\subsubsection{Autoencoder}

An autoencoder (AE) is an encoder-decoder neural network that is trained to output a copy of its input. Often, the dimension of the encoder's last layer output is small compared to the dimension of the data. This layer is then known as the \textit{bottleneck} layer and it provides a compressed encoding of the input data. Originally, AEs were made of feed-forward layers, but this term is also contemporaneously used to designate AE networks with other types of layers, such as convolutional or recurrent layers. 
To the best of our knowledge, the first use of an AE for DoA estimation was reported by \citet{zermini_deep_2016}. They used a simple AE to estimate TF masks for each possible DoA, which were then used for source separation. 
An interesting AE-based method was presented by \citet{huang_time-domain_2020}, in which an ensemble of AEs were trained to reproduce the multichannel input signal at the output, with one AE per candidate source position. Since the common latent information among the different channels is the dry signal, each encoder approximately deconvolves the signal from a given microphone. These dry signal estimates should be similar provided that the source is indeed at the assumed position; hence, the localization is performed by finding the AE with the most consistent latent representation. However, it is not clear whether this model can generalize well to unseen source positions and acoustic conditions.

\citet{moing_learning_2020} presented an AE with a large number of convolutional layers (and transposed convolutional layers, which are layers of the decoder that process the inverse operation of the corresponding convolutional layer at the encoder), which estimates the potential source activity of each subregion in the $(x,y)$ plane divided in a grid, making it possible to locate multiple sources. They evaluated several types of outputs  (binary, Gaussian-based, and binary followed by regression refinement), each of which showed promising results on the simulated and real data. An extension of this work was presented in \citep{moing_data-efficient_2021}, in which they proposed using adversarial training (see Section~\ref{sec:learningStrategies}) to improve network performance on real data, as well as on microphone arrays unseen in the training set, in an unsupervised training scheme. To do this, they introduced a novel \textit{explicit transformation} layer that helped the network to be invariant to the microphone array layout. Another encoder-decoder architecture was proposed by \citet{he_sounddet_2021}, in which a multichannel waveform was fed into a filter bank with learnable parameters, after which a 1D convolutional encoder-decoder network processed the filter bank output. The output of the last decoder was then fed separately into two branches, one for SED and the other for DoA estimation.

An encoder-decoder structure with one encoder followed by two separate decoders was proposed by \citet{wu_sslide_2021}. Signals recorded from several microphone arrays were first transformed in the short-term Fourier transform domain (see Section~\ref{sec:inputFeatures}) and then stacked in a 4D-tensor (whose dimensions were time, frequency, microphone array and microphone). This tensor was then sent to the encoder block, which was made of a series of convolutional layers followed by several residual blocks. The output of the encoder was then fed into two separate decoders, the first of which was trained to output a probability of source presence for each candidate $(x,y)$ region, while the second was trained in the same way but with a range compensation to make the network more robust. The same general encoder-decoder line was adopted in the 2D image mapping approach proposed by \citet{wu2021sound}. Note that here, the network is composed of convolutional layers at the encoder and transposed convolutional layers at the decoder, which is typical for image mapping applications in computer vision. 

An indirect use of an AE was proposed by \citet{vera-diaz_towards_2021}, who used convolutional and transposed convolutional layers to estimate the TDoA from GCC-based input features. The main idea was to rely on the encoder-decoder capacity to reduce the dimension of the input data so that the bottleneck representation forced the decoder to output a smoother version of the TDoA. This technique was shown to outperform the classical GCC-PHAT method in the reported experiments. This work was extended in the presence of two sources \citep{vera-diaz_acoustic_2021}.

\subsubsection{Variational autoencoder}
\label{sssec:VAE}

A variational autoencoder (VAE) is a generative model that was originally proposed by \citet{Kingma2014} and \citet{rezende2014stochastic} and is now very popular in the DL community. A VAE can be seen as a probabilistic version of an AE. Unlike a classical AE, a VAE learns a probability distribution of the data at the output of the decoder and also models the probability distribution of the so-called latent vector at the bottleneck layer, which makes the VAE strongly connected to the concept of unsupervised representation learning \citep{bengio2013representation}. New data can thus be obtained with the decoder by sampling these distributions.

To our knowledge, \citet{bianco_semi-supervised_2020} were the first to apply a VAE for SSL. Their VAE, made of convolutional layers, was trained to generate the phase of inter-microphone RTFs (see Section~\ref{sss:RTFs}), jointly with a classifier that estimates the speaker's DoA from the RTF phases. The interest of using a VAE is that this generative model, originally designed for unsupervised training, is here trained in a semi-supervised configuration using a large dataset of unlabeled RTF data together with a limited set of labeled data (RTF values + corresponding DoA labels). In such a limited labeled dataset configuration, this model was shown to outperform an SRP-PHAT-based method as well as a supervised CNN in reverberant scenarios. This semi-supervised (or weakly supervised) approach is further discussed in Section~\ref{ssec:perspectives-data}. An extension of this work has been further proposed in \citep{bianco2021semi}, with refined network architectures and more realistic acoustic scenarii.

\subsubsection{U-Net architecture}

\begin{figure}
\centering
\includegraphics[width=0.9\linewidth]{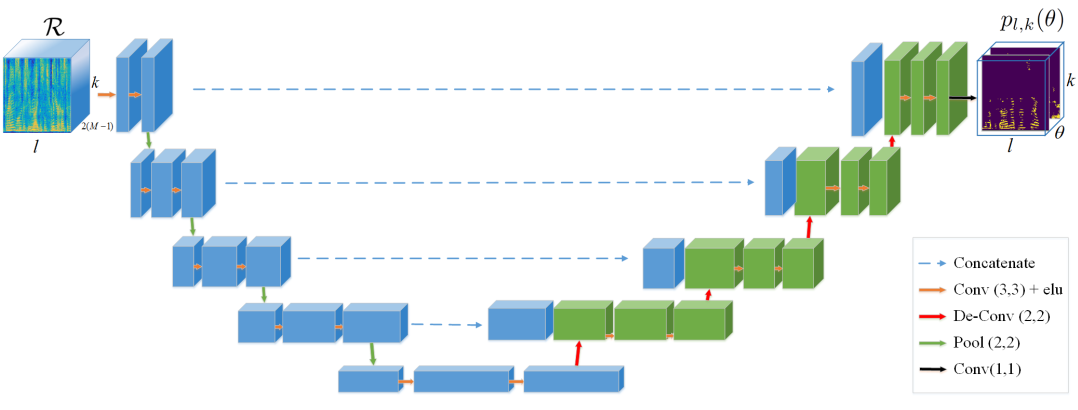}
\caption{The U-Net network architecture of \citet{chazan_multi-microphone_2019}. The input matrix $\mathcal{R}$ contains angular features extracted from the RTFs (see Section~\ref{sss:RTFs}) ($l$, $k$, and $M$ denote the time index, the frequency bin, and the number of microphones, respectively). Several stages of encoders (in blue) and decoders (in green) are used. At each encoder (or decoder) stage, two or three convolutional layers with $3 \times 3$ kernels are employed to compute a new representation which is used as the input of the next encoder (or decoder, respectively), except for the bottleneck stage from which the output is fed as input into the upper-stage decoder. Residual connections are used to concatenate one encoder output to the input of the same stage decoder, to alleviate the loss information problem. The output of this system consists of one TF mask $p_{l,k}(\theta)$ per considered DoA $\theta$. Note: Reprinted from \citep{chazan_multi-microphone_2019}; copyright by IEEE; reprinted with permission.}
\label{fig:chazanNetwork}
\end{figure}

A U-Net architecture is a particular fully-convolutional neural network originally proposed by \citet{ronneberger_u-net_2015} for biomedical image segmentation. In U-net, the input features are decomposed into successive feature maps throughout the encoder layers and then recomposed into ``symmetrical'' feature maps throughout the decoder layers, similarly to CNNs. Having the same dimension for feature maps at the same level in the encoder and decoder enables one to propagate information directly from an encoder level to the corresponding level of the decoder via residual connections. This leads to the typical U-shape schematization (see Fig.~\ref{fig:chazanNetwork}). 

Regarding SSL and DoA estimation, several works have been inspired by the original U-Net paper. \citet{chazan_multi-microphone_2019} employed such an architecture to estimate one TF mask per considered DoA (see Fig.~\ref{fig:chazanNetwork}), in which each TF bin was associated with a single particular DoA. This spectral mask was finally applied for source separation. This system was extended by \citet{hammer2021dynamically} to account for multiple moving speakers. Another joint localization and separation system based on a U-Net architecture was proposed by \citet{jenrungrot_cone_2020}. In this system, a U-Net was trained based on 1D convolutional layers and GLUs. The input is the multichannel raw waveform accompanied by an angular window that helps the network to perform separation on a particular zone. If the output of the network on the window is empty, no source is detected, otherwise, one or more sources are detected and the process is repeated with a smaller angular window, until the angular window reaches $2$\textdegree. This system shows interesting results on both synthetic and real reverberant data containing up to eight speakers.

For the DCASE 2020 Challenge, a U-Net with several BGRU layers in-between the convolutional blocks, was proposed for SELD by \citet{patel_single_2020}. The last transposed convolutional layer of this U-Net outputs a single-channel feature map per sound event, corresponding to its activity and DoA for all frames. This showed an improvement over the baseline of \citet{adavanne_sound_2019} in terms of DoA error. 
\citet{comanducci_source_2020} used a U-Net architecture in the second part of their proposed neural network to estimate the source coordinates $x$ and $y$. The first part, composed of convolutional layers, learns to map GCC-PHAT features to the so-called ray space (where source positions correspond to linear patterns, \emph{cf.} \citep{bianchi_ray_2016}), which is an intermediate representation used as the input of the U-Net architecture.

%%% INPUT FEATURES %%%
\section{Input features}
\label{sec:inputFeatures}

In this section, we provide an overview of the variety of input feature types found in the DL-based SSL literature. Generally, the considered features can be low-level signal representations such as waveforms or spectrograms, hand-crafted features such as binaural features, or they can be borrowed from traditional SP methods such as MUSIC or GCC-PHAT. 

Overwhelmingly, the input features for the SSL neural networks are based on some representation readily used in signal processing, often emphasizing spatial and/or TF information embedded in the signal. This seems to yield good results, despite the growing trend in other domains to learn the feature representation directly from raw data. One interpretation may be that the network architectures in SSL are usually of a relatively modest size, as compared to end-to-end models used in some other domains, e.g., NLP. A few publications have compared different types of input features for SSL, e.g., \citep{roden_sound_2015, krause_feature_2021}.

It is also quite common to provide the network with concatenated features of different nature (even if these carry redundant information), which usually has positive impact on performance. This can be attributed to the flexibility of the learning process, which seemingly adapts the network weights such that the pertinent information is efficiently ``routed'' from such an input to the upper layers of the network, where it is merged into an abstract, optimized feature representation.

% Ajouter rq sur communautÃ© DL/SP

We organized this section into the following feature categories: inter-channel, cross-correlation-based, spectrogram-based, Ambisonics,  intensity-based, and finally the direct use of the multichannel waveforms. Note that, as stated above, different kinds of features are often combined at the input layer of SSL neural networks.  

\subsection{Inter-channel features}
\label{ss:interchan}

\subsubsection{Relative transfer function (RTF)}
\label{sss:RTFs}

The RTF is a very general inter-channel feature that has been widely used for conventional (non-deep) SSL and other spatial audio processing such as source separation and beamforming \citep{gannot2017consolidated} and acoustic echo cancellation \citep{valero2017multi}, and is now considered for DL-based SSL as well. The RTF is defined for a given sound source position and for a microphone pair as the ratio $H(f)=A_2(f)/A_1(f)$ of the source-to-microphone ATFs of the two microphones, $A_2(f)$ and $A_1(f)$ (here we are working in the frequency or STFT domain and we recall that an ATF is the discrete Fourier transform of the corresponding RIR). It is thus strongly dependent on the source DoA (for a given recording set-up). In a multichannel set-up with more than two microphones, we can define an RTF for each microphone pair. Often, one microphone is used as a reference microphone, and the ATFs of all other microphones are divided by the ATF of this reference microphone.  

As an ATF ratio, an RTF is thus a vector with an entry defined for each frequency bin. If only one directional source is present in the recorded signals and if the (diffuse) background noise is negligible, Eq.~\eqref{eq:sensor-signal-STFT-domain} shows that an RTF estimate can be obtained for each STFT frame (indexed by $n$), each frequency bin, and each microphone pair (indexed by $i$ and $k$) by taking the ratio between the STFT transforms of the recorded waveforms of the two considered channels, $X_i(f,n)$ and $X_k(f,n)$:
\begin{equation}
 \hat{H}_{i,k}(f) = \frac{X_k(f,n)}{X_i(f,n)} \approx \frac{A_k(f)S(f,n)}{A_i(f)S(f,n)} = \frac{A_k(f)}{A_i(f)} = H_{i,k}(f),
 \label{eq:RTF-estimate}
\end{equation}
where $S(f,n)$ is the STFT of the source signal. 
%The above equation uses the fact that the convolution between the source signal and an RIR (see Eq.~\eqref{eq:sensor-signal-time-domain}) is approximated in the STFT domain by the product between $S(f,n)$ and the ATF. This approximation is referred to as the narrow-band assumption in the speech/audio source localization/separation literature \citep{gannot2017consolidated, vincent2018audio}, in which it is widely used even if the fact that the STFT analysis window is generally shorter than the RIRs length may question its accuracy. 
In the case where a background/sensor noise is present, more sophisticated RTF estimation procedures must be used, e.g., \citep{cohen2004relative, markovich2015performance, li2015estimation}. If multiple sources are present, things become more complicated, but using the natural sparsity of speech/audio signals in the TF domain, \emph{i.e.}, only at most one source is assumed to be active in each TF bin \citep{rickard_approximate_2002}, the same principle as for one active source can be applied separately in each TF bin. Therefore, a multiple set of estimated RTFs at different frequencies (and possibly at different time frames if the sources are static or not moving too fast) can be used for multi-source localization. The reader is referred to \citep{gannot2017consolidated} and references therein for more information on the RTF estimation problem.

 An RTF is a complex-valued vector. In practice, an equivalent real-valued pair of vectors is often used. We can use either the real and imaginary parts or the modulus and argument. Often, the log-squared value of the interchannel power ratio is used, \emph{i.e.}, the interchannel power ratio in dB, and the argument of the RTF estimate ideally corresponds to the difference of the ATF phases. Such RTF-based representations have been used in several DNN-based systems for SSL. For example, \citet{chazan_multi-microphone_2019}, \citet{hammer2021dynamically}, and \citet{bianco_semi-supervised_2020, bianco2021semi} used as input features the arguments of the measured RTFs obtained from all microphone pairs. 

\subsubsection{Binaural features}
\label{sss:binaural}

Binaural features have also been used extensively for SSL, in both conventional and deep systems \citep{argentieri2015survey}. These features correspond to a specific two-channel recording set-up, one which attempts to reproduce human hearing in the most realistic way possible. Toward this aim, a dummy head/body with in-ear microphones is used to mimic the source-to-human-ear propagation, and in particular the effects of the head and external ear (pinnae), which are important for source localization by the human perception system. In an anechoic binaural set-up environment, the (two-channel) source-to-microphone impulse response is referred to as the binaural impulse response (BIR). The frequency-domain representation of a BIR is the HRTF. Both BIR and HRTF are functions of the source DoA. 
To take into account the room acoustics in a real-world SSL application, BIRs are extended to binaural room impulse responses (BRIRs), which combine head/body effects and room effects (in particular reverberation, see further discussion on BRIR simulation in Section~\ref{subsec:syntheticData}). 

Several binaural features are derived from binaural recordings: The interaural level difference corresponds to the short-term log-power magnitude of the ratio between the two binaural channels in the STFT domain, $X_2(f,n)$ and $X_1(f,n)$:
\begin{equation}
 ILD(f,n) = 20 \log_{10} \displaystyle\left\lvert \frac{X_2(f,n)}{X_1(f,n)}\right\rvert.
 \label{eq:ILD}
\end{equation}
The interaural phase difference is the argument of this ratio:
\begin{equation}
 IPD(f,n) = \angle \frac{X_2(f,n)}{X_1(f,n)},
 \label{eq:IPD}
\end{equation}
and the interaural time difference is the delay that maximizes the cross-correlation between the two channels, similarly to the TDoA in \eqref{eq:TDoA-estimate-by-GCC-PHAT}. Just like the RTF, these features are actually vectors with frequency-dependent entries. 
In fact, the ILD and IPD are strongly related (not to say similar) to the log-power and argument of the RTF, as shown by comparing \eqref{eq:ILD} and \eqref{eq:IPD} with \eqref{eq:RTF-estimate}, the difference relying more on the set-up than on the features themselves. The RTF can be seen as a more general (multichannel) concept, whereas binaural features refer to the specific two-channel binaural setup.
As for the RTF, the ILD, IPD, and ITD implicitly encode the position of a source. Again, when several sources are present, the sparsity of speech/audio signals in the TF domain allows ILD/IPD/ITD values to provide information on the position of several simultaneously active sources.   

\citet{youssef_learning-based_2013} used ILD and ITD vectors fed separately into specific input branches of an MLP. \citet{ma_exploiting_2015} and \citet{yiwere_distance_2017} concatenated the cross-correlation of the two binaural channels with the ILD before feeding it into the input layer of their network. \citet{ma_exploiting_2015} justify this choice with two arguments. The first one is to avoid the noise-sensitivity of the peak-picking operation for the computation of the ITD, the second one is because of the systematic changes in the cross-correlation function according to the source azimuth. \citet{nguyen_autonomous_2018} used the IPD as the argument of a unitary complex number that was decomposed into real and imaginary parts. These parts were concatenated to the ILD for several frequency bins and several time frames, leading to a 2D tensor that was then fed into a CNN. \citet{pang2019multitask} also used a CNN to process ILD and IPD features in the TF domain, but the ILD and IPD 2D-tensors were directly concatenated at the input of the CNN. 
A system relying only on the IPD was proposed by \citet{pak_sound_2019}. An MLP was trained to output a clean version of the noisy input IPD in order to better retrieve the DoA using a conventional method. \citet{sivasankaran_keyword-based_2018} used as input features the concatenation of the cosine and sine of the IPDs for several frequency bins and time frames. This choice was based on a previous work that showed similar performance for this type of input feature compared to classical phase maps, but with a lower dimension. In an original way, \citet{thuillier_spatial_2018} employed unusual binaural features. They used the ipsilateral and contralateral spectra. These features were shown to be relevant for elevation estimation using a CNN. We finally found other DNN-based systems that used ILD, e.g., \citep{roden_sound_2015,zermini_deep_2016}, ITD, e.g., \citep{roden_sound_2015}, or IPD, e.g., \citep{shimada_sound_2020,shimada_accdoa_2020,zermini_deep_2016,subramanian_deep_2021} in addition to other types of features.

\subsection{Cross-correlation (CC)-based features}

Another manner for extracting and exploiting inter-channel information that depends on source location is to use features based on the  cross-correlation (CC) between the signals of different channels. In particular, as seen in Section~\ref{sec:traditional}, a variant of CC known as GCC-PHAT is a common feature used in classical localization methods \citep{knapp_generalized_1976}. It is less sensitive to speech signal variations than standard CC, but it may be adversely affected by noise and reverberation \citep{blandin2012multi}. Therefore, it has been used within the framework of neural networks, which was revealed to be robust to this type of disturbance/artefact. In several systems, GCC-PHAT has been computed for each microphone pair and several time delays, all concatenated to form a 1D vector used as the input of an MLP, e.g., \citep{xiao_learning-based_2015, vesperini_neural_2016, he_deep_2018}. Other architectures include convolutional layers to extract useful information from multi-frame GCC-PHAT features, e.g.,  \citep{he_deep_2018, vecchiotti_deep_2018,vecchiotti_detection_2019,noh_three-stage_2019,lu_sound_2019, maruri_gcc-phat_2019,Pratik2019,song_localization_2020,li_online_2018, comanducci_source_2020}.

Some SSL systems rely on the CPS, which we already mentioned in Section~\ref{sec:traditional} and which is linked to the CC by a Fourier transform operation (in practice, short-term estimates of the CPS are obtained by multiplying the STFT of one channel with the conjugate STFT of the other channel). \citet{Leung2019} and \citet{xue_sound_2020} sent the CPS into a CRNN architecture to improve localization performance over the baseline of \citet{adavanne_sound_2019} (see Section~\ref{sec:NN-archi}).  \citet{grondin_sound_2019} also used the cross-spectrum for each microphone pair in the convolutional block of their architecture, whereas GCC-PHAT features were concatenated in a deeper layer. The CPS was also used by \citet{ma_phased_2018} as an input feature. Acoustic imaging has traditionally shown some interest in the CPS feature to predict  localization and sound pressure level of competing sources; coupled with different architectures, from the simple MLP \citep{castellini_neural_2021} to the complex CNN DenseNet network \citep{xu_acoustic_2021}, authors have shown that the use of DNN could outperform traditional deconvolution methods, either in performance or computation time.

Traditional localization methods, such as MUSIC \citep{schmidt_multiple_1986} or ESPRIT \citep{roy_esprit-estimation_1989}, have been widely examined in the literature (see Section~\ref{sec:traditional}). These methods are based on the eigen-decomposition of the CC matrix of a multichannel recording. Several DNN-based SSL systems \citep{takeda_sound_2016,takeda_discriminative_2016,  takeda_unsupervised_2017, takeda_unsupervised_2018} have been inspired by these methods and reuse such features as input for their neural networks. 
\citet{nguyen_robust_2020} computed the spatial pseudo-spectrum based on the MUSIC algorithm and then used it as input features for a CNN.

Power map methods, which were discussed in Section~\ref{sec:traditional}, have also been used to derive input features for DNN-based SSL systems. \citet{salvati_exploiting_2018} proposed calculating the narrowband normalized steered response power for a set of candidate TDoAs corresponding to an angular grid and feeding it into a convolutional layer. This led to a localization performance improvement compared to the traditional SRP-PHAT method. Such power maps were also used by \citet{diaz-guerra_robust_2021} as inputs of 3D convolutional layers. In acoustic imaging, a SRP map is also a standard feature where finding the position and the acoustic level is the main goal. 
Some recent works used a CNN \citep{goncalves_deconvoluting_2021} or a U-Net \citep{lee_deep_2021} to produce clean deconvolved maps, hence going beyond the intrisic resolution of the array.
%Some recent works used such maps with different network architectures like CNN \citep{goncalves_deconvoluting_2021} or U-Net \citep{lee_deep_2021} to produce clean deconvolved maps, hence going beyond the intrisic resolution of the array. 

\subsection{Spectrogram-based features}
\label{subsec:spectrogram}

Alternatively to inter-channel features or CC-based features which already encode relative information between channels, another approach is to provide an SSL system directly with ``raw'' multichannel information, \emph{i.e.}, without any pre-processing in the channel dimension.  

This does not prevent some pre-processing in the other dimensions and, from an historical perspective, we notice that many models in this line use spectral or spectro-temporal features instead of raw waveforms (see next subsection) as inputs. In practice, (multichannel) STFT spectrograms are typically used  \citep{vincent2018audio}. These multichannel spectrograms are generally organized as 3D tensors, with one dimension for time (or frames), one for frequency (bins), and one for channel. The general spirit of DNN-based SSL methods is that the network should be able to ``see'' by itself and automatically extract and exploit the differences between TF spectrograms along the channel dimension while exploiting the  ``sparsity'' of TF signal representation.

In several works, the individual spectral vectors from the different STFT frames were provided independently to the neural model, meaning that the network did not take into account their temporal correlation (and a localization result is generally obtained independently for each frame). Thus, in that case, the network input is a matrix of size $M \times K$, with $M$ being the number of microphones, and $K$ being the number of considered STFT frequency bins. \citet{hirvonen_classication_2015} concatenated the log-spectra of eight channels for each individual analysis frame and sent it into a CNN as a 2D matrix. \citet{chakrabarty_multi-speaker_2019,chakrabarty_broadband_2017,chakrabarty_multi-speaker_2017, chakrabarty_multi-scale_2019} and \citet{mack_signal-aware_2020} used the multichannel phase spectrogram as input features, disregarding the magnitude information. This choice is motivated by the fact that it allows to easily generate a training dataset from white noise signals.
As an extension of this work, phase maps were also exploited by \citet{bohlender_exploiting_2021}. 

When several consecutive frames are considered, the STFT coefficients for multiple timesteps and multiple frequency bins form a 2D matrix for each recording channel. Usually, these spectrograms are stacked together in a third dimension to form the 3D input tensor. Several systems considered only the magnitude spectrograms, e.g.,  \citep{yalta_sound_2017,wang_robust_2019,patel_single_2020,pertila_robust_2017}, while others considered only the phase spectrogram, e.g.,  \citep{zhang_robust_2019,subramanian_deep_2021}. When considering both magnitude and phase, they can also be stacked in a third dimension (as well as channels). This representation has been employed in many DNN-based SSL systems, e.g.,  \citep{he_neural_2021,guirguis_seld-tcn_2021,krause_comparison_2021,lin_report_2019,maruri_gcc-phat_2019,zhang_data_2019,kapka_sound_2019,krause_arborescent_2019,schymura_pilot_2021}.

\citet{yang2021supervised} dedicated different input branches of their CRNN to magnitude and phase features. Other authors have proposed to decompose the complex-valued spectrograms into real and imaginary parts, e.g.,  \citep{hao_spectral_2020,he_joint_2018, moing_learning_2020,kucuk_real-time_2019}. Finally, \citet{Leung2019} tried several combinations of features computed from the complex multi-channel spectrogram, including the magnitude and phase, the real and imaginary parts and the CPS. They claim that providing this redundant information could help the neural network for better localization.

While basic (STFT) spectrograms consider equally-spaced frequency bins, mel-scale spectrograms and Bark-scale spectrograms are represented with a non-linear sub-bands division, corresponding to a perceptual scale (low-frequency sub-bands have a higher resolution than high-frequency sub-bands) \citep{peeters_large_2004}. Mel-spectrograms were preferred to STFT spectrograms in several SSL neural networks, e.g., \citep{vecchiotti_deep_2018, kong_cross-task_2019, cao_two-stage_2019, Cao2019polyphonic, ranjan_sound_2019}. The Bark scale was also  explored for spectrograms in the SSL system of \citet{Pratik2019}. 

\subsection{Ambisonic signal representation}

In the SSL literature, numerous systems utilize the Ambisonics format, \emph{i.e.}, the spherical harmonics (SH) decomposition coefficients \citep{jarrett2017theory}, to represent the input signal. Ambisonics is a multichannel format that is increasingly used due to its capability to represent the spatial properties of a sound field, while being agnostic to the microphone array configuration \citep{zotter_ambisonics_2019}.

The SH decomposition is done for the acoustic pressure measured on the surface of a sphere $\mathbb{S}^2$, concentric with the microphone array. For a fixed sound source in far field, the decomposition coefficient of order $\ell$ and degree $m\in[-\ell,\ell]$, in the STFT domain, is given as follows \cite{jarrett2017theory}:
\begin{equation}\label{eqAmbiTF}
    B_{\ell,m}(f,n) = \int_{\Omega \in \mathbb{S}^2} X(f,n,\Omega) Y^*_{\ell,m}(\Omega) d\Omega,
\end{equation}
where $X(f,n,\Omega)$ and $Y_{\ell,m}(\Omega)$ are the acoustic pressure and the SH function, at the direction $\Omega$, respectively. In practice, this integral is approximated by a quadrature rule, since the number of microphones consisting an array is finite. Such approximation implies that the pressure $X(f,n,\Omega)$ is assumed to be an (almost) ``order-limited'' function on the sphere \cite{rafaely_fundamentals_2019}, meaning that $B_{\ell>L,m}(f,n)=0$, for some maximal order $L$ (that depends on the number of microphones in the array). Hence, for FOA ($L=1$), the Ambisonics representation \eqref{eqAmbiTF} counts only $4$ coefficients (channels) per TF bin. Alternatively, the Higher-Order Ambisonics (HOA), $L>1$, signals have more than four channels.

The plane wave, bearing an amplitude $S(f,n)$, and coming from a direction $\Omega$, admits a simple SH representation $B_{\ell,m}(f,n) = S(f,n) Y_{\ell,m}(\Omega)$ \cite{rafaely_fundamentals_2019}. Therefore, as opposed to other types of microphone arrays, the Ambisonic channels are in phase, since the spatial response of each channel $Y_{\ell,m}(\Omega)$ is TF-independent.\footnote{In practice, the spatial response of Ambisonic microphones is approximately frequency-independent only within certain bandwidth (dictated by the HOA order), due to spatial aliasing in the high frequency range, and noise amplification at lower frequencies \cite{zotter_ambisonics_2019}.} Analogous to \eqref{eq:microphone-signal-sum}, the multichannel Ambisonic spectrogram $\mathbf{B}(f,n)$, due to $J$ sources and reverberation, is given by the multivariate expression: 
\begin{equation}\label{eq:AmbiTimeDomain}
    \mathbf{B}(f,n) = \sum\limits_{j=1}^{J}\sum\limits_{r=0}^{\infty} A_{jr}(f,n)  S_j(f,n) \mathbf{Y}(\Omega_{jr}) + \mathbf{N}(f,n),
\end{equation}
where $A_{jr}$ is the amplitude of the $r$\textsuperscript{th} reflection of the source $S_j$ (with $r=0$ corresponding to the direct path), $\mathbf{Y}$ is the vector whose entries are appropriate spherical harmonics $Y_{\ell,m}$ for all considered Ambisonic orders, and $\mathbf{N}$ is the additive noise vector. Note that the complex-valued amplitudes $A_{jr}$ account for the attenuation and phase shift of a corresponding plane wave component.
%Depending on the \addnote[PA-R2-Other-b]{1}{considered order}, one can deal with FOA (consisting of the first four Ambisonics channels) or higher-order Ambisonics (HOA, with more than four Ambisonics channels). %In the former case, the real-valued encoding vector $\mathbf{a}_j$, in the so-called FuMa/N3D format \cite{zotter_ambisonics_2019}, is given by
%\begin{equation}\label{eq:FOAencoding}
%    \mathbf{a}_j = \begin{bmatrix} 1 \\ 
%\sqrt{3} \cos\theta \cos\phi \\ \sqrt{3} \sin\theta \cos\phi \\ \sqrt{3} \sin\phi \end{bmatrix}.
%\end{equation}

%As with other microphone arrays, one can use a TF representation of the Ambisonics signal obtained by applying the STFT on each Ambisonics channel:
%\begin{equation}
%    \mathbf{X}(f,n) = \mathbf{A} \mathbf{B}(f) \mathbf{S}(f,n) + \mathbf{N}(f,n),
%\end{equation}
%where $\mathbf{A} = \left[ \begin{smallmatrix} \mathbf{a}_1 b_1 \; \mathbf{a}_2 b_2 \; \hdots \; \mathbf{a}_J b_J \end{smallmatrix} \right]^T$. This way, 

The FOA spectrograms, decomposed into magnitude and phase components, have been used by, e.g., \citet{adavanne_localization_2019, guirguis_seld-tcn_2021,adavanne_direction_2018,  adavanne_sound_2019, kapka_sound_2019} and \citet{krause_arborescent_2019}. \citet{varanasi_deep_2020} and  \citet{poschadel_direction_2021,poschadel_multisource_2021} used third-order Ambisonics spectrograms. \citet{poschadel_direction_2021, poschadel_multisource_2021} compared the performance of a CRNN with HOA spectrograms from order $1$ to $4$, showing that the higher the order, the better the localization accuracy of the network (but still below the performance of the so-called FOA \textit{pseudo-intensity} features, which we will discuss in Section~\ref{ss:intensity}). They used the phase and magnitude for both elevation and azimuth estimation. Another way of representing the Ambisonics format was proposed by \citet{comminiello_quaternion_2019}. Based on the FOA spectrograms, they proposed considering them as quaternion-based input features, which proved to be a suitable representation in previous works \citep{parcollet_quaternion_2018}. To cope with this type of input feature, a neural network was adapted from the one of \citet{adavanne_sound_2019}, showing an improvement over the baseline.

\subsection{Intensity-based features}
\label{ss:intensity}

Sound intensity is an acoustic quantity defined as the product of sound pressure and particle velocity \citep{jacobsen2013fundamentals,rossing2007springer}. In the frequency or TF domain, sound intensity is a complex vector whose real part (known as ``active'' intensity) is proportional to the gradient of the phase of sound pressure, \emph{i.e.}, it is orthogonal to the wavefront. This is a useful property that has been extensively used for SSL, e.g.,  \citep{nehorai1994acoustic,hickling1993finding,jarrett20103d,tervo2009direction,evers2014multiple,kitic2018tramp,pavlidi20153d}. The imaginary part (``reactive'' intensity) is related to oscillatory local energy transfers, and its physical interpretation is less obvious \citep{maysenholder1993reactive}. Hence, it has been largely ignored by the SSL community, even though it is relevant in room acoustics \cite{nolan2019experimental}. While the pressure is directly measurable by regular microphones, particle velocity requires specific sensors, such as acoustic vector-sensors \citep{nehorai1994acoustic,jacobsen2013fundamentals}, e.g., the ``Microflown'' transducer \cite{de2003overview}. Otherwise, it has to be approximated using the acoustic pressure measurements. Under certain conditions, particle velocity can be assumed to be proportional to the spatial gradient of sound pressure \citep{rossing2007springer,merimaa2006analysis}, which allows for the estimation by, e.g., the finite difference method \citep{tervo2009direction} or using the FOA channels discussed in the previous section \citep{zotter_ambisonics_2019}. The latter approximation is often called (FOA) complex \emph{pseudo-intensity} vector \cite{jarrett20103d}:
\begin{equation}
    \mathbf{I}(f,n) =  B_{0,0}(f,n) \mathbf{B}_{\ell=1,m}(f,n)^*,
\end{equation}
where $\mathbf{B}_{\ell=1,m}^*(f,n)$ is the vector of first-order SH coefficients, excluding the zero-order $B_{0,0}(f,n)$. In free field conditions, assuming the presence of a single source at the TF bin $(f,n)$, the entries of $\Re \left( \mathbf{I}(f,n) \right)$ are the Cartesian coordinates of a vector colinear with the DoA of the source (with $\Re$ denoting the real part of a complex value).

%The intensity vector has been used as an input feature in a number of recent neural models (especially those based on Ambisonic representations) and has led to very good SSL performance. 
The first use of an Ambisonics pseudo-intensity vector for DL-based SSL was reported by \citet{perotin_crnn-based_2018}, showing superiority in performance compared to the use of the raw Ambisonics waveforms and traditional Ambisonics-based methods. Interestingly, the authors demonstrated that using both active and reactive intensity improves SSL performance. Moreover, they normalized the intensity vector of each frequency band by its energy, which can be shown to yield features similar to RTFs in the spherical harmonics domain \citep{daniel2020time,jarrett2017theory}. \citet{yasuda_sound_2020} proposed using two CRNNs to refine the input FOA pseudo-intensity vector. The first CRNN is trained to estimate denoising and separation masks under the assumption that there are two active sources and that the WDO hypothesis holds. The second CRNN estimates another mask to remove the remaining unwanted components (e.g., reverberation). The two networks, hence, produce an estimate of the ``clean'' intensity vector for each active source (the NoS is estimated by their system as well). The pseudo-intensity vector has consequently been used in several other recent works, e.g.,  \citep{grumiaux_saladnet_2021, perotin_crnn-based_2019,  grumiaux_improved_2021, nguyen_general_2021, cao_two-stage_2019, park_sound_2020, song_localization_2020, cao_improved_2021, tang_regression_2019, perotin_regression_2019}. 

Sound intensity was also explored by \citet{liu_deep_2021} without the Ambisonics representation. The authors computed the instantaneous complex sound intensity using an average of the sound pressure across the four considered channels and two orthogonal particle velocity components using the differences in sound pressure for both microphone pairs. They kept only the real part of the estimated sound intensity (active intensity) and applied a PHAT weighting to improve the robustness against reverberation.

\subsection{Waveforms}

Since 2018, several authors have proposed directly providing their neural network models with the raw multichannel recorded signal waveforms. This idea relies on the DNN's capability to find the best representation for SSL without the need of hand-crafted features or pre-processing of any kind. This is in line with the general trend of DL to go toward an \textit{end-to-end} approach that is observed in many other applications, including in speech/audio processing. Of course, this goes together with the always increasing size of networks, datasets and computational power. 

To our knowledge, \citet{suvorov_deep_2018} were the first to apply this idea. They trained their neural network directly with the recorded eight-channel waveforms, stacking many 1D convolutional layers to extract high-level features for the final DoA classification. \citet{vera-diaz_towards_2018}, \citet{vecchiotti_end--end_2019}, \citet{chytas_hierarchical_2019}, \citet{cao_event-independent_2020}, and \citet{pujol_source_2019,pujol_beamlearning_2021} sent the raw multichannel waveforms into 2D convolutional layers. \citet{Huang_Aalto_task3_report} sent the raw multichannel waveforms (in microphone format and FOA format) into a 1D CNN with residual connections and squeeze-excitation blocks. Note that this model is used for SELD and the authors motivate the use of raw waveform inputs by the fact that ``SED and DOA may have some common features that are better preserved in the raw audio [wave]form.''  \citet{huang_time-domain_2020} sent the multichannel waveforms into an AE. \citet{jenrungrot_cone_2020} shifted the waveforms of each channel to make them temporally aligned according to the TDoA before being injected into the input layer of their network. In the same vein, \citet{huang_dnn-based_2018, huang_time-domain_2019} proposed time-shifting the multichannel signal by calculating the time delay between the microphone position and the candidate source location, which requires scanning for all candidate locations.

A potential disadvantage of waveform-based features is that the architectures exploiting such data are often more complex, as one part of the network needs to be dedicated to feature extraction. Moreover, some papers have reported that learning the ``optimal'' feature representations from raw data becomes more difficult when noise is present in the input signals \citep{wichern2019wham} or may even harm generalization, in some cases \citep{sato2021does}.
However, it is interesting to mention that the visual inspection of the learned weights of the input layers of some end-to-end (waveform-based) neural networks has revealed that they resemble the filterbanks that are usually applied in the pre-processing stage of SSL (see Section~\ref{subsec:spectrogram}) and other various classical speech/audio processing tasks \citep{sainath2017multichannel, luo2019conv}.

\subsection{Other types of features}

%In the DNN-based SSL literature, several systems have proposed unusual types of features that do not belong to one of the categories described above. \SK{``Several'', but only one reference?}

\citet{varzandeh_exploiting_2020} have proposed unusual types of features that do not belong to one of the categories described above. Particularly, they have used a periodicity degree feature together with GCC-PHAT features in a CNN. The periodicity degree is computed for a given frame and period. It is equal to the ratio between the harmonic power signal for the given period and the total power signal. This conveys information about the harmonic content of the source signal to the CNN.

%%% OUTPUT STRATEGIES %%%
\section{Output strategies}
\label{sec:output}

In this section, we discuss the different strategies proposed in the literature to obtain a final DoA estimate. We generally divide the strategies into two categories: classification and regression. When the SSL network is designed for the classification task, the source location search space is generally divided into several zones, corresponding to different classes, and the neural network outputs a probability value for each class. As for regression, the goal is to directly estimate (continuous) source position/direction values, which are usually either Cartesian coordinates $(x, y, z)$, or spherical coordinates $(\theta, \phi, r)$ (although the source-microphone distance $r$ is very rarely considered). However, the latter is an important factor as it can affect the estimation accuracy (for instance, due to the influence of the direct-to-reverberant ratio (DRR) \cite{vincent2018audio}). Therefore, in order to obtain a robust model, the training dataset needs to be sufficiently diverse such that the network is exposed to sources at different directions, but also at different source-microphone distances. In the last subsection, we report a few non-direct methods in which the neural network does not estimate the location of a source in its output layer. Instead, it either helps another (conventional) algorithm to finally retrieve the desired DoA, or the location estimate is a byproduct of some intermediate network layer. %does not directly provide the location of the source(s) but instead helps another algorithm to finally retrieve the desired DoA.
A reader particularly interested in the comparison between the classification and regression approaches may consult the papers of \citet{tang_regression_2019} and  \citet{perotin_regression_2019}.

\subsection{DoA estimation via classification}

Many systems treat DoA estimation as a classification problem, \emph{i.e.}, each class represents a certain zone in the considered search space. In other words, space is divided into several subregions, usually of similar size, and the neural network is trained to produce a probability of active source presence for each subregion. Such a classification problem is often addressed by using a feedforward layer as the last layer in the network, with as many neurons as the number of considered subregions. Two activation functions are generally associated with the final layer neurons: the softmax and sigmoid functions. Softmax ensures that the sum of all neuron outputs is $1$, so it is suitable for a single-source localization scenario. With a sigmoid, all neuron outputs are within $[0,1]$ independently from each other, which is suitable for multi-source localization. The last layer output is often referred to as the \textit{spatial (pseudo)-spectrum}, whose peaks correspond to a high probability of source activity in the corresponding zone. 

As already mentioned in Section~\ref{sec:numberOfSources}, the final DoA estimate(s) is/are generally extracted using a peak picking algorithm: If the number of sources $J$ is known, the selection of the $J$ highest peaks gives the multi-source DoA estimation; if the NoS is unknown, usually the peaks above a certain user-defined threshold are selected, leading to a joint NoS and localization estimations. Some preprocessing, such as spatial spectrum smoothing or angular distance constraints, can be used for better DoA estimation. Hence, such a classification strategy can be readily used for single-source and/or multi-source localization, as the neural network is trained to estimate a probability of source activity in each zone, regardless of the NoS.

\subsubsection{Spherical coordinates}

Regarding the quantization of the source location space, namely the localization \textit{grid}, different approaches have been proposed. Most early works focused on estimating only the source's \textit{azimuth} $\theta$ relative to the microphone array position, dividing the $360$\textdegree~azimuth space into $N_{\theta}$ regions of equal size, leading to a grid quantization step of $\frac{360}{N_{\theta}}$. Without being exhaustive, we found in the literature many different values for $N_{\theta}$, e.g., $N_{\theta}=7$ \citep{roden_sound_2015}, $N_{\theta}=8$ \citep{hirvonen_classication_2015}, $N_{\theta}=20$ \citep{suvorov_deep_2018}, $N_{\theta}=37$ \citep{vecchiotti_end--end_2019}, $N_{\theta}=72$ \citep{ma_exploiting_2015}, and $N_{\theta}=360$ \citep{xiao_learning-based_2015}. Some other works did not consider the whole $360$\textdegree~azimuth space. For example, \citet{chazan_multi-microphone_2019} focused on the region $[0,180]$ with $N_{\theta}=13$.

Estimating the elevation $\phi$ alone has not been frequently investigated in the literature, probably because of the lack of interesting applications in indoor scenarios. To the best of our knowledge, only one paper focused on estimating the elevation alone \citep{thuillier_spatial_2018}. The authors divided the whole elevation range into nine regions of equal size. The majority of recent SSL neural networks are trained to estimate both source azimuth and elevation, whenever the microphone array geometry makes it possible. To do this, several options have been proposed in the literature. One can use two separate output layers, each with the same number of neurons as the number of subregions in the corresponding dimension. For example, the output layer of the neural architecture proposed by \citet{fahim_multi-source_2020} is divided into two branches with fully connected layers, one for azimuth estimation ($N_{\theta}$ neurons), and the other for elevation estimation ($N_{\phi}$ neurons). One can also have a single output layer where each neuron corresponds to a zone in the unit sphere, \emph{i.e.}, a unique pair $(\theta,\phi)$, e.g.,  \citep{grumiaux_saladnet_2021, perotin_crnn-based_2019}. Finally, one can directly design two separate neural networks, with each estimating the azimuth or the elevation angle, e.g.,  \citep{varanasi_deep_2020}. 

However, most of the neural networks following the classification strategy for joint azimuth and elevation estimation are designed so that the output corresponds to a 2D grid on the unit sphere. For example, \citet{perotin_crnn-based_2018, perotin_crnn-based_2019} and \citet{grumiaux_improved_2021} used a quasi-uniform spherical grid with $429$ classes, each represented by a unique neuron in the output layer of their network.  \citet{adavanne_direction_2018} sampled the unit sphere in the whole azimuth axis but in the limited elevation range of $[-60\degree,60\degree]$, yielding an output vector corresponding to $432$ classes.

Distance estimation has barely been investigated in the SSL literature, highlighting the fact that it is a difficult problem. \citet{roden_sound_2015} addressed the distance estimation along with azimuth or elevation prediction by dividing the distance range into five candidate classes. \citet{yiwere_distance_2017} quantized the distance range into four classes and estimated it along with three possible azimuth values. In the paper by \citet{takeda_sound_2016}, the azimuth axis was classified with $I=72$ classes along with the distance and height of the source, but these last two quantities were classified into a very small set of possible pairs: $(30,30)$, $(90,30)$ and $(90,90)$ (in centimeters). \citet{bologni_acoustic_2021} trained a CNN to classify a single-source signal into a 2D map representing the azimuth and distance dimensions.

\subsubsection{Cartesian coordinates}

A few works applied the classification paradigm to estimate the Cartesian coordinates. \citet{moing_data-efficient_2021,moing_learning_2020} and \citet{ma_phased_2018} divided the horizontal $(x,y)$ plane into small regions of the same size, with each being a class in the output layer. However, this representation suffers from a decreasing angular difference between the regions that are far from the microphone array, which is probably why regression is usually preferred for estimating Cartesian coordinates.

\subsection{DoA estimation via regression}

In regression SSL networks, the source location estimate is directly given by the continuous value provided by one or several output neurons (whether we consider Cartesian or spherical coordinates, and how many source coordinates are of interest). This technique offers the advantage of a potentially more accurate DoA estimation since there is no quantization. Its drawback is twofold. First, the NoS needs to be known or assumed, as there is no way to estimate if a source is active or not based on a localization regression. Second, regression-based SSL usually faces the well-known source permutation problem \citep{subramanian_deep_2021}, which occurs in the multi-source localization configuration and is common with DL-based source separation methods. Indeed, during the computation of the loss function at the training time, there is an ambiguity in the association between target and actual output -- in other words, which estimate should be associated with which target? This issue also arises during the evaluation. One possible solution is to force the SSL network training to be permutation invariant \citep{subramanian_deep_2021}, in line with what was proposed for audio source separation \citep{yu_permutation_2017}.

As for classification, when using regression, there is a variety of possibilities for the type of coordinates to be estimated. The choice among these possibilities is driven more by the context or the application than by design limitations, since regression generally requires only a few output neurons.

\subsubsection{Spherical coordinates}

\citet{tsuzuki_approach_2013} proposed a complex-valued neural approach for SSL. The output of the network is a complex number of unit amplitude whose argument is an estimate of the azimuth of the source. A direct regression scheme was employed by \citet{nguyen_autonomous_2018} with a two-neuron output layer that predicts the azimuth and elevation values in a single-source environment.
The system of \citet{opochinsky_deep_2019} performed only azimuth estimation. Regarding the DCASE 2019 Challenge  \citep{politis_overview_2020}, a certain number of candidate systems have used two neurons per event type to estimate the azimuth and elevation of the considered event, e.g., \citep{chytas_hierarchical_2019, cao_two-stage_2019, park_reassembly_2019}, while the event activity was jointly estimated in order to extract (or not) the corresponding coordinates. \citet{Sudo2019} proposed representing the output as a quaternion including the cosinus and sinus of the azimuth and elevation angles, from which they retrieve the DoA angle values. This enables to tackle the problem of discontinuity at angle interval boundaries (for instance, at $-180\degree$ and $180\degree$).

In the system of \citet{maruri_gcc-phat_2019}, azimuth and elevation estimations were done separately in two network branches, each containing a specific dense layer.  \citet{sundar_raw_2020} proposed a regression method relying on a preceding classification step: dividing the azimuth space into $I$ equal subregions, with the output of the neural network being made of $3I$ neurons. Assuming there is at most one active source per subregion, three neurons are associated with each of them: one neuron is trained to detect the presence of a source, while the other two neurons estimate the distance and azimuth of that source. The loss function for training is a weighted sum of categorical cross-entropy (for the classification task) and mean square error (for the regression task).

\subsubsection{Cartesian coordinates}
\label{sssec:output-reg-Cartesian}

Another way to predict the DoA with regression is to estimate the Cartesian coordinates of the source(s). \citet{vesperini_neural_2016} designed their network output layer with only two neurons to estimate the coordinates $x$ and $y$ in the horizontal plane, with an output range normalized within $[0,1]$, which represents the scaled version of the room size in each dimension. Following the same idea, \citet{vecchiotti_deep_2018, vecchiotti_detection_2019} also used two neurons to estimate $(x,y)$ but added a third one to estimate the source activity.

The estimation of the three Cartesian coordinates $(x,y,z)$ has been investigated in several systems. \citet{vera-diaz_towards_2018} and \citet{krause_comparison_2021} designed the output layer with three neurons to estimate the coordinates of a single source with regression. \citet{adavanne_localization_2019,adavanne_sound_2019} chose the same strategy. However, they performed SELD for several types of event, and thus there are three output neurons to provide $(x,y,z)$ estimates for each event type, plus another output neuron to estimate whether or not this event is active. The hyperbolic tangent activation function is used for the localization neurons to keep the output values in the $[-1,1]$ range, leading to a DoA estimate on the unit sphere. The same strategy was followed in an extension of this work by \citet{comminiello_quaternion_2019}.

In \citet{shimada_accdoa_2020}, the authors proposed the activity-coupled cartesian DoA (ACCDOA) representation which encodes the DoA with the source activity in a single vector, separately for each sound class to be localized. More specifically, the ACCDOA vector encodes the Cartesian coordinates $(x, y, z)$, is then normalized and then multiplied by the source activity ($\in [0,1]$). Using a threshold, the active sources can be detected using this vector norm, and their respective DoAs can be retrieved from the normalized Cartesian coordinates. This ACCDOA output representation has then been used in other works, e.g.,  \citep{shimada_sound_2020, Sudarsanam2021, shimada_ensemble_2021, nguyen_spectrotemporally_2021, emmanuel_multiscale_2021, Naranjo-Alcazar_UV_task3_report}. 

\subsection{Non-direct DoA estimation}

Neural networks have also been used in the regression mode to estimate intermediate quantities, which are then used by a non-neural algorithm to predict the final DoA.

\citet{pertila_robust_2017} proposed using a CNN in the regression mode to estimate a TF mask. This mask was then applied to the noisy multichannel spectrogram to obtain an estimate of the clean multichannel spectrogram, and a classical SRP-PHAT method was next applied to retrieve the final DoA. Another TF mask estimation was done by \citet{wang_robust_2019} using a bidirectional LSTM network to improve traditional DoA estimation methods, such as GCC-PHAT or MUSIC. \citet{pak_sound_2019} trained an MLP to remove unwanted artefacts of the IPD input features. The cleaned feature was then used to estimate the DoA with a non-neural method. \citet{yasuda_sound_2020} proposed a method to filter out reverberation and other non-desired effects from the intensity vector by TF mask estimation. The filtered intensity vector led to a better DoA estimation than an intensity-based conventional method. \citet{yang2021supervised} used a two-stage neural network system to estimate the direct-path RTF (DP-RTF), that is, the part of the RTF that corresponds to the direct source-to-microphone propagation \citep{li2016estimation}. In \citep{yang2021supervised}, the source DoA is the direction parameter of a DP-RTF taken from a dictionary of pre-computed DP-RTFs, corresponding to the closest match with the network estimate.

\citet{huang_dnn-based_2018,huang_time-domain_2019} employed neural networks on multichannel waveforms, shifted in time with a delay corresponding to a certain candidate source location, to estimate the original dry signal. Doing this for a set of candidate locations, they then calculated the sum of CC coefficients between the estimated dry source signals for all candidate source locations. The final estimated location was obtained as the one leading to the maximum sum.

A joint localization and separation scheme was proposed by \citet{jenrungrot_cone_2020}. The neural network was trained to estimate the signal coming from a certain direction within a certain angular window, whose parameters were injected as an input to each layer. Thus, the network acted like a radar and scanned through all directions, then progressively reduced the angular window up to a desired angular resolution.

Several works proposed employing neural networks for a better prediction of the TDoA, which is then used to determine the DoA as often done in traditional methods. \citet{grondin_sound_2019} estimated the TDoA in the regression mode using a hyperbolic tangent activation function at the output layer. \citet{vera-diaz_towards_2021} used an AE to estimate a function from GCC-based features (similar to TDoA) that exhibited a clear peak corresponding to the estimated DoA. Their work was extended in the presence of two sources \citep{vera-diaz_acoustic_2021}. In \citet{comanducci_time_2020}, the authors employed a U-Net in a regression manner to clean GCC-based features from noise and reverberation.

\citet{subramanian2021directional} proposed a neural system based on a stacked localization network, parametric beamformers and a speech recognition network. Since each of these modules is differentiable, the system is trained in the end-to-end mode, using an ASR-specific cost function. Despite being optimized for the ASR, the trained system also exhibits a very good performance in terms of source separation and localization, whose predictions are the intermediate results, retrievable at the output of the corresponding processing modules.

%%% DATASETS %%%
\section{Data}
\label{sec:data}

In this section, we detail the different approaches taken to deal with data during model training or testing. Because we are dealing with indoor domestic/office environments, noise and reverberation are common in real-world signals. We successively inspect the use of synthetic and recorded datasets in DNN-based SSL.

\subsection{Synthetic data}
\label{subsec:syntheticData}
A well-known limitation of supervised learning (see Section~\ref{sec:learningStrategies}) for SSL is the lack of labeled training data. In a general manner, it is difficult to produce datasets of recorded signals with corresponding source position metadata in diverse spatial configurations (and possibly with diverse spectral content) that would be sufficiently large for efficient SSL neural model training. Therefore, one often has to \textit{simulate} a large amount of data to obtain an efficient SSL system. 

To generate realistic data, taking into account reverberation, one needs to simulate the room acoustics. This is usually done by synthesizing the RIR that models the sound propagation for a ``virtual'' source-microphone pair. This is done for all microphones of the array (and for a large number of source positions and microphone array positions, see below). Then, a ``dry'' (\emph{i.e.}, clean reverberation-free monophonic) source signal is convolved with this RIR to obtain the simulated microphone signal (this is done for every channel of the microphone array). 
As already stated in Section~\ref{subsec:intro-general-principle}, the foundation of SSL relies on the fact that the relative location of a source with respect to the microphone array position is implicitly encoded in the (multichannel) RIR, and an SSL DNN learns to extract and exploit this information from examples. Therefore, such data generation has to be done with many different dry signals and for a large number of simulated RIRs with different source and microphone array positions. The latter must be representative of the configurations in which the SSL system will be used in practice. Moreover, other parameters, such as room dimensions and reverberation time, may have to be varied to take into account other factors of variations in SSL. 

One advantage of this approach is that many dry signal datasets exist, in particular for speech signals, e.g., \citep{garofolo_timit_1993,lamel_bref_1991, WSJ0}. Therefore, many SSL methods are trained with dry speech signals convolved with simulated RIRs. \citet{chakrabarty_broadband_2017,chakrabarty_multi-speaker_2017} used white noise as the dry signal for training and speech signals for testing. This approach is reminiscent of the work of \citet{deleforge2013variational,deleforge2015co} based on a GMR and as already mentioned in Section~\ref{sec:traditional}. Using white noise as the dry signal enables the acquisition of training data that are ``dense'' in the TF domain. However, \citet{vargas_improved_2021} showed that training on speech or music signals leads to better results than noise-based training, even when the signals are simulated with a generative adversarial network (GAN). Furthermore, the results of \citet{krause2021data} indicate that using speech, noise and sound events data altogether leads to better localization performance, even compared to ``matched'' training and test signals.

As for RIR simulation, there exist several methods (and variants thereof) and many acoustic simulation softwares. Detailing these methods and software implementations is out of the scope of this article, but an interested reader may consult appropriate references, e.g., \citep{rindel2000use,svensson2002computational,siltanen2010rays}. Let us only mention that the simulators based on the image source method (ISM) \citep{allen_image_1979} have been widely used in the SSL community, probably due to the fact that they offer a relatively good trade-off between the simulation fidelity, in particular regarding the ``head'' of an RIR, \emph{i.e.}, the direct propagation and early reflections \citep{rindel2000use}, and computational complexity.
Among publicly available libraries, the RIR generator of \citet{habets_room_2006}, the related signal generator \citep{habets_signal_generator}, the Roomsim toolbox of \citet{campbell2005matlab} and its extension to mobile sources called Roomsimove \cite{vincent2008roomsimove}, the Spherical Microphone Impulse Response (SMIR) generator of \citet{jarrett2012rigid}, the Pyroomacoustics toolbox of \citet{scheibler_pyroomacoustics_2018}, and the Multichannel Room Acoustics
Simulator (MCRoomSim) of \citet{wabnitz2010room}, are very popular. Such libraries have been used by, e.g.,  \citet{chakrabarty_multi-speaker_2019,perotin_crnn-based_2019,grumiaux_improved_2021,nguyen_robust_2020,varanasi_deep_2020,salvati_exploiting_2018,li_online_2018,bianco_semi-supervised_2020}. An efficient open-source implementation of the ISM method, relying on Graphic Processing Unit (GPU) acceleration, has been recently presented by \citet{diaz2021gpurir} and used in \citet{diaz-guerra_robust_2021} to simulate moving sources.

Other improved models based on the ISM have also been used to simulate impulse responses, such as the one presented by \citet{hirvonen_classication_2015}. This model relies on that of \citet{lehmann_diffuse_2010}, which adds a diffuse reverberation model to the original ISM method. \citet{hubner_efficient_2021} proposed a low-complexity model-based training data generation method that includes a deterministic model for the direct path and a statistical model for late reverberation. %This method shows similar performance compared to the usual ISM, while being computationally more efficient. 
It has been demonstrated that the SSL neural network, trained using the data generated by this method, achieves comparable localization performance as the same architecture trained on a dataset generated by the usual ISM. However, the proposed simulation method is computationally more efficient. An investigation of several simulation methods was done by \citet{gelderblom_synthetic_2021}, with extensions of ISM, namely ISM with directional sources, and ISM with a diffuse field due to scattering. \citet{gelderblom_synthetic_2021} compared the simulation algorithms via the training of an MLP (in both regression and classification modes) and showed that ISM with scattering effects and directional sources leads to the best SSL performance. More sophisticated software, such as ICARE\textsuperscript{\textregistered} \cite{bouatouch2006real}, often combine ISM with efficient ray-tracing and statistical methods, permitting simulation of more complicated room geometries and acoustic effects. Note, however, that none of the methods based on approximating the sound propagation by geometrical acoustics is capable of precisely simulating certain wave phenomena, such as diffraction  \cite{kuttruff2016room}.

Training and testing binaural SSL systems requires either directly using signals recorded in a binaural setup (see next subsection) or using a dataset of two-channel BIRs and convolving these BIRs with (speech/audio) dry signals, just like for simulations in conventional set-up.
Most of the time, the BIRs are recorded ones (see next subsection; there exist a few BIR simulators, but we will not detail this quite specific aspect here). 
To take into account the room acoustics in a real-world SSL application, BIR effects are often combined with RIR effects. This is not obtained by trivially cascading the BIR and RIR filters, since the BIR depends on the source DoA, %and a RIR contains many reverberation components from many directions
meaning that one would have to integrate it with RIR components from many incoming directions \cite{bernschutz2016microphone}. However, such a process is included in several RIR simulators, which are able to produce the corresponding combined response, called the binaural room impulse response (BRIR), e.g., \citep{campbell2005matlab}. Recall that BIRs are often manipulated in the frequency domain (referred as HRTFs), where they are a function of both frequency and source DoA. 

\subsection{Real data}

Collecting real labeled data is crucial to assessing the robustness of an SSL neural network in a real-world environment. However, it is a cumbersome task. As of today, only a few datasets of such recordings exist. Among them, several impulse response datasets are publicly available and have been used to generate training and/or testing data. 

The distant-speech interaction for robust home applications (DIRHA) simulated corpus presented by \citet{cristoforetti_dirha_2014} has been used to simulate microphone speech signals based on real RIRs, recorded in a multi-room environment \citep{vesperini_neural_2016,vecchiotti_deep_2018}. Another database consisting of recorded RIRs from three rooms with different acoustic characteristics is publicly available \citep{hadad_multichannel_2014}, using three microphone array configurations to capture signals from several source azimuth positions in the range $[-90\degree,90\degree]$. The RIR dataset published by \citet{c5cn-jv76-21} is intended to be used for DoA estimation, and contains measurements from a three-channel array. Other RIR datasets have been published by, e.g., \citet{szoke_building_2019},  \citet{eaton_ace_2015}, \citet{hahmann2021acoustic}, \citet{koyama2021meshrir}, \citet{kristoffersen2021deep}, and \citet{VerburgRiezu2021}. The last four ones were initially designed for sound field analysis and synthesis, and they contain measurements from single-channel microphones (\emph{i.e.}, not microphone arrays). However, the acquired RIRs correspond to multiple positions within a room, and could be potentially used to emulate microphone arrays.

As for BIR dataset recordings, a physical head-and-torso simulator (HATS) (aka ``dummy head'') is used, with ear microphones plugged into the dummy head ears. To isolate head and torso effects from other environment effects such as reverberation, binaural recordings are generally made in an anechoic room. For example, the dataset published by \citet{thiemann_multiple_2015} was collected using four different dummy heads and used for SSL by \citet{roden_sound_2015}.

The Surrey Binaural Room Impulse Responses database was published by \citet{francombe_iosr_2017} and has been used for SSL by, e.g., \citet{ma_exploiting_2015} to synthesize signals for evaluating the proposed method. This database has been recorded using a HATS in four room configurations, with sound coming from loudspeakers. It thus combines binaural effects with room effects. 

Several challenges have also been organized for some years, and evaluation datasets with real recordings have been constituted to assess the candidate systems. 
Datasets were created for the SELD task of the DCASE Challenge, in 2019 \citep{Adavanne2019multi}, 2020 \citep{Politis2020dataset}, and 2021 \citep{politis_dataset_2021}.
These datasets contains sound events in reverberant and noisy environments, synthesized from recordings of real RIRs. These data come in two four-microphone spatial audio formats: tetrahedral microphone array and FOA. The dataset comprises 12 sound event types, including, e.g., barking dog, female/male speech or ringing, with up to three simultaneous events overlapping. In the 2019 dataset, the sources are static, whereas they are both static and moving in the 2020 and 2021 datasets, with more diverse acoustic conditions. Finally, in the 2021 edition of the DCASE dataset, additional sound events have been added to the recordings to play the role of (directional) interferers (that are not bound to be classified). These datasets have been used in many SSL systems, e.g., \citep{cao_two-stage_2019, park_reassembly_2019, grondin_sound_2019,  cao_event-independent_2020, naranjo-alcazar_sound_2020, shimada_sound_2020, wang_ustc-iflytek_2020, mazzon_first_2019}. Very recently, another SELD challenge focused on 3D sound has been announced \citep{guizzo2021l3das21}, where a \emph{pair} of FOA microphones was used to capture a large number of RIRs in an office room, from which the audio data were generated.

The acoustic source LOCAlization and TrAcking (LOCATA) challenge \citep{evers2020locata} has been one of the most comprehensive challenges targeting the localization of speech sources. The challenge tasks include single and multiple SSL, each of which in a setting where the sources and/or microphones are static or mobile. The recordings have been made using several types of microphone arrays, namely the planar array from \citet{brutti2010woz}, the \textit{em32 Eigenmike}$^{\text{\textregistered}}$  spherical array, a hearing aid, and a set of microphones mounted on a robot head. The ground truth data include position information obtained through an optical tracking system, hand-labeled VAD metadata, and dry (or close-talking) source signals. This dataset has been used in a number of works to validate the effectiveness of a proposed method on ``real-life'' recordings, e.g., \citep{grumiaux_improved_2021,diaz-guerra_robust_2021,sundar_raw_2020,pak_sound_2019,varanasi_deep_2020,tang_regression_2019,yang2021learning}. 

A few audio-visual datasets have also been developed and are publicly available, in which the audio data are enriched with video information. This type of dataset is dedicated to the development and testing of audio-visual localization and tracking techniques, which are out of the scope of this survey paper. Among these corpora, the AV16.3 corpus \citep{lathoud_av163_2004} and the CHIL database \citep{stiefelhagen_clear_2007} have provided an evaluative basis for several (purely audio) SSL systems \citep{vera-diaz_towards_2018,vera-diaz_towards_2021, vera-diaz_acoustic_2021} by considering only the audio part of the audiovisual dataset. 

Finally, we also found a series of papers in which neural networks were tested using real data specifically recorded for the presented work in the researchers' own laboratories, e.g., \citep{chazan_multi-microphone_2019, grumiaux_saladnet_2021, he_neural_2021, perotin_crnn-based_2018, grumiaux_improved_2021, nguyen_robust_2020, he_deep_2018,  varanasi_deep_2020,  moing_learning_2020,   perotin_regression_2019}.

\subsection{Data augmentation techniques}
\label{sec:dataAugmentationTechniques}

To limit the massive use of simulated data, which can limit the robustness of the network on real-world data, and to overcome the limitation in the amount of real data, several authors have proposed resorting to data augmentation techniques. Without producing more recordings, data augmentation allows for the creation of additional training examples, often leading to improved network performance. 

For the DCASE Challenge, many submitted systems were trained using data augmentation techniques on the train dataset. \citet{mazzon_first_2019} proposed and evaluated three techniques to augment the training data, taking advantage of the FOA representation used by their SSL neural network: swap or inversion of FOA channels, label-oriented rotation (the rotation is applied to result in the desired label), or channel-oriented rotation (the rotation is directly applied with a desired matrix). Interestingly, the channel-oriented rotation method gave the worst results in their experiments, while the other two methods showed an improvement in neural network performance. \citet{zhang_data_2019} applied the SpecAugment method of \citet{park_specaugment_2019}, which led to new data examples by masking certain time frames or frequencies of a spectrogram, or both at the same time. This method was also employed by, e.g., \citet{yalta_dcase_2021, shimada_ensemble_2021, Bai_NWPU_task3_report, krause2021data}. In the work of \citet{Pratik2019}, new training material was created with the \textit{Mixup} method of \citet{zhang_mixup_2018}, which relies on convex combinations of an existing training data pair. \citet{noh_three-stage_2019} used pitch shifting and block mixing data augmentation \citep{salamon_deep_2017}. The techniques of \citet{mazzon_first_2019} and \citet{zhang_data_2019} were employed by \citet{shimada_sound_2020, shimada_ensemble_2021} to create new mixtures, along with another data augmentation method proposed by \citet{takahashi_deep_2016}, which is based on random mixing of two training signals.

\citet{wang_four-stage_2021} applied four new data augmentation techniques to the DCASE dataset \citep{politis_dataset_2021}. The first one applies the benefit of the FOA format to changing the location of the sources by swapping audio channels. The second method is based on the extraction of spatial and spectral information on the sources, which are then modified and recombined to create new training examples. The third one relies on mixing multiple examples, resulting in new multi-source labelled mixtures. The fourth technique is based on random TF masking. The authors evaluated the benefits of these data augmentation methods both when used separately and when applied sequentially.

%%% LEARNING STRATEGIES %%%
\section{Learning strategies}
\label{sec:learningStrategies}

In a general manner, when training a neural network to accomplish a certain task, one needs to choose a training paradigm that often depends on the type and amount of available data. In the DNN-based SSL literature, most of the systems rely on supervised learning, although several examples of semi-supervised and weakly supervised learning can also be found.

\subsection{Supervised learning}

When training a neural network with supervised learning, the training dataset must contain the output target (also known as the label, especially in the classification mode) for each corresponding input data. A cost function (or loss function) is used to quantify the error between the output target and the actual output of the neural network for a given input data, and training consists of minimizing the average loss function over the training dataset. We have seen in Section~\ref{sec:output} that in a single-source SSL scenario with the classification paradigm, a softmax output function is generally used. In that case, the cost function is generally the categorical cross-entropy, e.g.,  \citep{perotin_crnn-based_2018, yalta_sound_2017, chakrabarty_broadband_2017}. When dealing with multiple sources, still with the classification paradigm, sigmoid activation functions and a binary cross-entropy loss function are used, e.g., \citep{perotin_crnn-based_2019, grumiaux_improved_2021, chakrabarty_multi-speaker_2017}. With a regression scheme, the choice for the cost function is the mean square error in most systems, e.g., \citep{he_neural_2021, nguyen_autonomous_2018, krause_comparison_2021, salvati_exploiting_2018, adavanne_sound_2019,  shimada_accdoa_2020,  pertila_robust_2017}. We also sometimes witness the use of other cost functions, such as the angular error \citep{perotin_regression_2019} and the $\ell_1$-norm \citep{jenrungrot_cone_2020}.

The limitation of supervised training is that the training relies on a great amount of labeled training data, whereas only a few real-world datasets with limited size have been collected for SSL. These datasets are not sufficient for robust training with DL models. To cope with these issues, one can opt for a data simulation method, as seen in Section~\ref{subsec:syntheticData}, or data augmentation techniques, as seen in Section~\ref{sec:dataAugmentationTechniques}. Otherwise, alternative training strategies can be employed, such as semi-supervised and weakly supervised learning, as presented hereafter.

\subsection{Semi-supervised and weakly supervised learning}

Unsupervised learning refers to model training with a dataset that does not contain labels. In the present SSL framework, this means that we would have a dataset of recorded acoustic signals without the knowledge of sources position/direction, and hence unsupervised learning alone is not applicable to SSL in practice. Semi-supervised learning refers to when part of the learning is done in a supervised manner, and another part is done in an unsupervised manner. Usually the network is pre-trained with labeled data training and refined (or fine-tuned) using unsupervised learning, \emph{i.e.}, without resorting to labels. In the SSL literature, semi-supervised learning has been proposed to improve the performance of the neural network on conditions unseen during supervised training or on real data, compared to its performance when trained only in the supervised manner. It can be seen as an alternative manner to enrich a labeled training dataset of too limited size or conditions (see Section~\ref{sec:data}).

For example, \citet{takeda_unsupervised_2017} and \citet{takeda_unsupervised_2018} adapted a pre-trained neural network to unseen conditions in a unsupervised way. For the cost function, the cross-entropy was modified to be computed only with the estimated output, so that the overall entropy was minimized. They also applied a parameter selection method dedicated to avoid overfitting, as well as early stopping. \citet{bianco_semi-supervised_2020} combined supervised and unsupervised learning using a VAE-based system. A generative network was trained to infer the phase of RTFs, which were used as input features in a classifier network. The cost function directly encompasses a supervised term and an unsupervised term and, during the training, the examples can come with or without labels. 

\citet{moing_data-efficient_2021} proposed a semi-supervised approach to adapt the network to real-world data after it was trained with a simulated dataset. This strategy was implemented with \textit{adversarial training} \citep{goodfellow_generative_2014}. In the present SSL context, a discriminator network was trained to label incoming data as synthetic or real, and the generator network learned to fool the discriminator. This enabled the adaptation of the DoA estimation network to infer from real data.

A different kind of training, named weakly supervised, was used by \citet{he_adaptation_2019,he_neural_2021}. The authors fine-tuned a pre-trained neural network by adapting the cost function to account for weak labels, which is the NoS, presumably known. This helped to improve the network performance by reducing the amount of incoherent predictions. Weak supervision was also used by \citet{opochinsky_deep_2019}. Under the assumption that only a few training data come with labels, a triplet loss function is computed. For each training step, three examples are drawn: a \textit{query} sample, acting as a usual example, a \textit{positive} sample close to the query sample, and a \textit{negative} sample from a more remote source position. The triplet loss (named so because of these three components) is then derived so that the network learns to infer the position of the positive sample closer to the query sample than the negative sample.

% LG: attention: nouveau fichier, j'ai inclus la conclusion dedans !!!!
%%% CONCLUSION %%%
\section{Conclusions and perspectives}
\label{sec:conclusion}

In this paper, we have presented a comprehensive overview of the literature on SSL techniques based on DL methods from 2011 to 2021. We attempted to categorize the many publications in this domain according to different characteristics of the methods in terms of source (mixture) configuration, neural network architecture, input data type, output strategy, training and test datasets, and learning strategy. 
%As we have seen through this survey, most research is oriented toward finding an appropriate neural architecture encompassing several constraints (NoS, moving sources, high reverberation, real-time implementation, etc.) More recently, part of the scientific effort has been geared towards the adaptation of systems trained on synthetic data to perform better on real-world data.
Tables~II--V summarize our survey: They gather the references of the reviewed DL-based SSL papers with the main  characteristics of the proposed methods (the ones that were used in our taxonomy of the different methods) being reported into different columns. We believe these tables can be very useful for a quick search of methods with a given set of characteristics.

% New: perspectives

To conclude this survey paper, we can comment on some current trends and draw a series of perspectives on the future directions that would be interesting to investigate to improve the performance of SSL systems and gain a better understanding of their behavior. Note that some of these perspectives appeal to general methodological issues in deep learning that are common to many applications, and some others are more specific to SSL. Note also that this list of research directions is not meant to be exhaustive.

\subsection{Adaptation to (limited sets of) real-world data}
\label{ssec:perspectives-data}

In a general manner, we observe a drop in performance when DNNs trained on simulated data are tested on real-world signals. This effect is well-known in the DL research in general, it is a particular case of the poor generalization capability of DNNs in the case of significant train-test data mismatch \citep{lecun2015deep, goodfellow_deep_2016}. We recall that this problem remains particularly crucial in SSL due to the difficulty of developing massive labeled datasets (i.e., with reliable annotations of ground-truth sources location) and the use of simulated training data. This is valid for training datasets generated using the usual ``shoebox'' acoustic simulations. Such geometry is rarely encountered is real-world environments. Moreover, the placement of the simulated microphone array is often unrealistic (e.g., it is floating in the air, whereas a practical recording device is often positioned on a table, leading to strong reflections). 

A first approach to tackle this problem is to consider more sophisticated room acoustics simulators, capable of taking into account more complex room geometries and acoustic phenomena, such as scattering or diffraction, see the related discussion in Section~\ref{subsec:syntheticData}.
However, this presents the limitation of a heavier computation cost, which should be balanced with the amount of data to be generated. Another line of research is to progressively train the network with more and more realistic signals, e.g., first with signals generated with simulated SRIRs, then fine-tuning the network with signals generated with real SRIRs, then further fine-tuning it with recorded data. This is in line with the general methodology of domain adaptation (DA) \citep{kouw2019review} and transfer learning \citep{bengio2012deep,zhuang2020comprehensive} used in many applications of DL, which aims at improving the performance of a network on a particular domain (in our case, real-world data) after it has been trained on another domain (here, simulated data). For SSL, the idea is to ``optimize'' the model to the target acoustic environment and/or sound sources. DA is a promising research field on its own, and it has only recently attracted the attention of the SSL community. To our best knowledge, the adversarial approach of \citet{moing_data-efficient_2021} and the entropy-based adaptation of \citet{takeda_unsupervised_2017} are the only representatives of DA for SSL. %Therein, an adversarial configuration is used to ``steer'' a parametric deep generative model to produce the features that resemble the target, unlabeled, domain data.}

Another line of research would be to inspire from weakly-supervised SSL methods based on \textit{manifold learning} \citep{laufer2020data}. The general principle is that the high-dimensional multichannel observed data live in a low-dimensional acoustic space, controlled by a limited number of latent variables (mainly, room dimensions, source and microphone positions, and reflection coefficients). This low-dimensional space, or manifold, can be identified using a large set of unlabeled data and unsupervised data dimension reduction techniques. Then a limited set of labeled data can be used to identify the relationship between observed data and source positions ``in the manifold,'' and thus estimate the source positions from new observed data (using, e.g., interpolation techniques). This principle was largely developed by \citet{laufer2020data}, who proposed several non-deep manifold identification techniques and corresponding SSL algorithms. The same principle can be applied with a DL approach, in particular with deep latent-variable generative models such as the VAE, in the line of the semi-supervised VAE-SSL model of \citet{bianco_semi-supervised_2020,bianco2021semi} already mentioned in Section~\ref{sssec:VAE} (see also an example of weakly supervised VAE-based source-filter decomposition of speech signals by \citet{sadok2022}). To our knowledge, SSL based on ``deep manifold learning'' is still a largely under-considered and open topic in the literature, yet it offers a promising direction to deal with limited annotated datasets.

\subsection{Flexibility of the trained models}

As opposed to conventional SP techniques, which can be parameterized to adapt to the changes in the system setup, DL methods for SSL generally assume identical setups for the training and the inference phase. 
Particularly, the number, geometrical arrangement and the directivity of the microphones composing an array, are usually assumed to be fixed. This is a serious disadvantage, since the network needs to be retrained for different microphone arrays, despite the fact that the task (SSL) remains the same.  A partial remedy is to use array-agnostic inputs, such as Ambisonics, e.g., \citep{adavanne_direction_2018,perotin_crnn-based_2019,grumiaux_improved_2021}, CPS eigenvectors, e.g., \citep{takeda_sound_2016} or spatial pseudo-spectra, e.g., \citep{nguyen_robust_2020,wu2021sound}. Another possibility is to adopt array-invariant techniques from end-to-end multichannel speech enhancement, e.g., \cite{luo2020end}.

Moreover, one could apply transfer learning and DA techniques, discussed previously, that could enable the models trained for a particular microphone array to adapt to another. Such techniques could not only be beneficial for the changes in microphone array setups, but also for the changes in the input signal parameters, such as the sampling rate, frame and overlap length, as well as the type of the STFT window function. A radical approach would be to make the method inherently independent to parameterization, by treating the input signal as a point cloud, as recently suggested by \citet{subramani2021point}.

%Quantification of uncertainty (perhaps already there with VAE etc.?) 

\subsection{Multi-task learning}

Multi-task training is a general methodology to improve the performance of a DNN-based system on a given task by training the model to jointly and simultaneously tackle several other tasks \citep{zhang2021survey,ruder2017overview}. It has been observed in practice that this often leads to better performance on the first target task. This principle is most often implemented in the following manner: An early part of the model (e.g., a common feature extraction module composed of several layers or several layer blocks) is common for the different tasks, then the model splits into different branches, each one specialized in one of the different tasks. The common part is assumed to allow the discovery of an efficient signal representation, and the fact that this representation is used for several downstream tasks somehow reinforce the efficiency of the representation extraction. 

This principle can be applied to SSL. In fact, it has already been extensively illustrated in this survey with the SELD Task of the DCASE Challenge \citep{politis_overview_2020} and the many candidates that have been proposed to this challenge (and that we have reported in this survey). The vast majority of the candidate DNNs follow the above architecture, with a common feature extraction module followed by two SED and SSL branches. In 2021, the ACCDOA representation was adopted by many researchers, see Section~\ref{sssec:output-reg-Cartesian}, and allowed for a joint SED and SSL process up to the very last model layer. We believe that combining the SSL task with other tasks (alternately to SED or in addition to it) such as source separation or ASR could lead to further advances. For example jointly proceeding to source counting in addition to SSL in the work of \citet{grumiaux_improved_2021} was shown to improve the SSL performance (note that here source counting is explicit and high-resolution, i.e., it consists in estimating the number of active sources at the short-term frame level, whereas it is most often implicit and generally made on a much larger time scale in the SELD task of the DCASE Challenge). Other examples of multi-task learning for SSL can be found in the works of \citet{wu_sslide_2021, wu2021sound}. Combining SSL with source separation in a DL framework is further discussed below. 

Somewhat different from multi-task approaches, the end-to-end \emph{task-oriented} learning of the entire processing chain (stacked localization, DoA-parameterized beamforming and ASR blocks) of \citet{subramanian2021directional} represents a refreshing idea to address the lack of DoA-annotated data. For instance, by using pre-trained ASR blocks, and by ``freezing'' all but the localization part of the system during training, one could use the abundant labeled speech corpora as a proxy information for the localization task. Such systems could incorporate both neural network modules and processing blocks based on conventional SP, as discussed in the next subsection.

\subsection{Combination of DL and conventional SP techniques}

In this survey paper, we have seen how the DL-based data-driven approach to the SSL problem has somehow replaced the conventional SP approach over the last decade. Yet, conventional methods are able to ``explicitly'' exploit strong prior knowledge on the physical underlying processes via signal and propagation models, whereas the exploitation of the spatial information contained in the mixture signal is done mostly ``implicitly'' by DNNs. Therefore, a major perspective for SSL is to get the best of both worlds, i.e., the combination of DL with conventional multichannel SP techniques. 

This can be inspired by what has been done in, e.g., speech enhancement and speech/audio source separation. In the single-channel configuration, DL-based speech enhancement and separation are mostly based on the masking approach in the TF domain. Binary masks or soft masks (reminiscent of the well-known single-channel Wiener filter) are estimated with DNNs from the noisy signal and applied to it to obtain a cleaned version, see the review by \citet{wang2018supervised}.
For multichannel speech enhancement and separation, a straightforward approach is to input the multichannel signal in the mask estimation network. However, more clever strategies can be elaborated. For example,  \citet{erdogan2016improved}, \citet{heymann2016neural} and \citet{higuchi2017deep} proposed combining the DNN-based single-channel masking with beamforming techniques \citep{van1988beamforming}. In these works, the TF-domain masks estimated by a DNN are used to select speech-dominant against noise-dominant TF points, which are then used to estimate speech and noise spatial covariance matrices, respectively, which are finally used to build beamforming filters. These papers report better ASR scores than with direct TF masking or basic beamforming applied separately. This approach was extended by \citet{perotin2018multichannel} with an additional first stage of beamforming in the HOA domain to improve the mask estimation. A joint end-to-end optimization of the mask estimator, the beamformer, and possibly an ASR acoustic model, was considered in the TF domain by \citet{meng2017deep} and  \citet{heymann2017beamnet}, and in the time domain by \citet{li2016neural}. Closer to source separation than to beamforming, \citet{nugraha2016multichannel} combined a DNN trained to estimate a clean speech spectrogram from a noisy speech spectrogram with the source separation technique based on the spatial covariance matrix (SCM) model and Wiener filtering of \citet{duong2010un}. \citet{leglaive2019semi} proposed an unsupervised multichannel speech enhancement system combining a VAE for modeling the (single-channel) clean speech signal and the SCM model for modeling the spatial characteristics of the multi-channel signal. 

Although we can find many examples of combination of DL-based and SP-based approaches for beamforming and source separation, to our knowledge and as shown by our survey, this principle has been poorly applied to SSL so far. Yet, powerful deep models, and in particular deep generative models such as GANs \citep{goodfellow_generative_2014}, VAEs \citep{Kingma2014}, and dynamical VAEs \citep{girin2021dynamical} are now avaible to model the temporal and/or spectral characteristics of sounds, and can be combined with SP-based models.  
Morevover, as already mentioned earlier in this survey, the connection between audio source separation, diarization, and SSL is strong, reciprocal (each task can help to solve the other ones), and is already exploited in many conventional systems \citep{vincent2018audio, gannot2017consolidated}. Future works may thus consider jointly sound source localization, diarization and separation/enhancement in an hybrid approach combining powerful DL models and conventional SP techniques. General frameworks for the joint optimization of DNN parameters and ``conventional'' parameters are now established and can be exploited \citep{engel2020ddsp, shlezinger2020model}.

\subsection{Moving sources and deep tracking}

In this survey, we poorly considered the case of moving sound sources and the necessity to rely in this case on tracking algorithms. These algorithms take as input the results of SSL obtained individually on each time frame and connect them through time. This is generally based on the use of a model of the source dynamics. In the multi-source case, dynamical models are often combined with source appearance models (which would model the sound texture or the different speakers' voice in the case of audio signals), resulting in the formation of source tracks with a consistent source ``identity'' for each of these tracks. Tracking algorithms also estimate the tracks ``birth'' and ``death,'' i.e., the time at which the corresponding sources are activated or inactivated. Such multi-object tracking (MOT) algorithms have a long history and their detailed description is beyond the scope of this paper; for a good overview of this domain, see the review papers of \citet{vo_multitarget_2015} and \citet{luo2021multiple}. 
%In the particular case of speech sources, it could be interesting to incorporate speaker spectral signatures into a tracking system for even better performance, which would bring us at the intersection with speaker recognition and diarization.

More recently, deep approaches to the MOT problem have emerged, an evolution mostly driven by the computer vision community \citep{CIAPARRONE202061}. For example, RNNs have been used in place of the traditional Kalman filter to model object dynamics for MOT in videos, e.g., \citep{babaee2018occlusion, sadeghian2017tracking, liang2018lstm, saleh2021probabilistic, xiang2019online}. The current trend is to replace RNNs with Transformer-like models, as discussed at the end of Section~\ref{ssec:models-attention}, e.g., \citep{xu2021transcenter, sun2020transtrack, meinhardt2021trackformer}.
The combination of a deep appearance model (automatic speaker recognition, SED) with a deep dynamical model in a sound source tracking system is a largely open problem and certainly a key ingredient for future developments in robust multi-source acoustic scene analysis in adverse acoustic environments and complex scenarios. Given the problem of annotated data scarsity in SSL, DL-based sound source localization and tracking may inspire from the unsupervised deep approaches to the MOT problem recently proposed by, e.g., \citep{Luiten_2020_WACV, karthik2020simple, he2019tracking, crawford2020exploiting, lin2022unsupervised}.

%%% 2011 to 2018 %%%
\begin{table*}[ht!]
    \caption{Summary of DL-based SSL systems published from 2011 to 2018, organized in chronological then alphabetical order. \textbf{Type}: R = regression, C = classification. \textbf{Learning}: S = supervised, SS = semi-supervised, WS = weakly supervised. \textbf{Sources}: \textbf{NoS} = considered number of sources, \textbf{Kno.} indicates if the NoS is known or not before estimating the DoA (\cmark = yes, \xmark = no), \textbf{Mov.} specifies if moving sources are considered. \textbf{Data}: SA = synthetic anechoic, RA = real anechoic, SR = synthetic reverberant, RR = real reverberant.}
    %\resizebox{1.\columnwidth}{!}{%
    \begin{adjustbox}{width=1.\textwidth}
    \begin{tabular}{|c|c|c|c|c|c|c|ccc|cccc|cccc|}
    \hline \hline
        \multirow{3}{*}{\textbf{Author}} & \multirow{3}{*}{\textbf{Year}} & \multirow{3}{*}{\textbf{Architecture}} & \multirow{3}{*}{\textbf{Type}} &
        \multirow{3}{*}{\textbf{Learn-}} & \multirow{3}{*}{\textbf{Input features}} & \multirow{3}{*}{\textbf{Output}} &  \multicolumn{3}{c|}{\textbf{Sources}}           & \multicolumn{8}{c|}{\textbf{Data}}   \\
         &  &  &  &  &  &  & \multirow{2}{*}{\textbf{NoS}} & \multirow{2}{*}{\textbf{Kno.}} & \multirow{2}{*}{\textbf{Mov.}} & \multicolumn{4}{c}{\textbf{Train}}  & \multicolumn{4}{c|}{\textbf{Test}}  \\
         &  &  &  & \textbf{ing}  &  &  &  &  &  & SA & RA & SR & RR & SA & RA & SR & RR \\ \hline
        \citet{kim_direction_2011} & 2011 & MLP & R & S & Power of multiple beams & $\theta$ & 1-5 & \xmark & \xmark & \cmark & \xmark & \xmark & \xmark & \xmark & \xmark & \xmark & \cmark \\ \hline
        \citet{tsuzuki_approach_2013} & 2013 & MLP & R & S & Time delay, phase delay, sound pressure diff. & $\theta$ & 1 & \xmark & \xmark & \cmark & \xmark & \xmark & \xmark & \xmark & \cmark & \xmark & \xmark \\ \hline
        \citet{youssef_learning-based_2013} & 2013 & MLP & R & S & ILD, ITD & $\theta$ & 1 & \xmark & \xmark & \cmark & \xmark & \cmark & \xmark & \cmark & \xmark & \cmark & \cmark \\ \hline
         \citet{hirvonen_classication_2015} & 2015 & CNN & C & S & Magnitude spectrograms & $\theta$ & 1 & \cmark & \xmark & \xmark & \xmark & \cmark & \xmark & \xmark & \xmark & \cmark & \xmark \\ \hline
        \citet{ma_exploiting_2015} & 2015 & MLP & C & S & Binaural cross-correlation + ILD & $\theta$ &  1-3 & \cmark & \xmark & \xmark & \cmark & \xmark & \xmark & \xmark & \xmark & \xmark & \cmark \\ \hline
        \multirow{2}{*}{\citet{roden_sound_2015}} & \multirow{2}{*}{2015} & \multirow{2}{*}{MLP} & \multirow{2}{*}{C} & \multirow{2}{*}{S} & ILD, ITD, binaural magnitude + phase spectrogr., & \multirow{2}{*}{$\theta$ / $\phi$ / $r$} & \multirow{2}{*}{1} & \multirow{2}{*}{\cmark} & \multirow{2}{*}{\xmark} & \multirow{2}{*}{\xmark} & \multirow{2}{*}{\xmark} & \multirow{2}{*}{\xmark} & \multirow{2}{*}{\cmark} & \multirow{2}{*}{\xmark} & \multirow{2}{*}{\xmark} & \multirow{2}{*}{\xmark} & \multirow{2}{*}{\cmark} 
        \\ & & & & & binaural real + imaginary spectrograms & & & & & & & & & & & & \\ \hline
        \citet{xiao_learning-based_2015} & 2015 & MLP & C & S & GCC-PHAT & $\theta$ & 1 & \cmark & \xmark & \xmark & \xmark & \cmark & \xmark & \xmark & \xmark & \cmark & \cmark \\ \hline
        \citet{takeda_sound_2016}& 2016 & MLP & C & S & Complex eigenvectors from correlation matrix & $\theta$, $z$, $r$ & 0-1 & \cmark & \xmark & \xmark & \cmark & \xmark & \cmark & \xmark & \cmark & \xmark & \cmark \\ \hline
         \citet{takeda_discriminative_2016}& 2016 & MLP & C & S & Complex eigenvectors from correlation matrix & $\theta$ & 0-2 & \xmark & \xmark & \xmark & \cmark & \xmark & \cmark & \xmark & \cmark & \xmark & \cmark \\ \hline
        \citet{vesperini_neural_2016} & 2016 & MLP & R & S & GCC-PHAT & $x$, $y$ & 1 & \cmark & \xmark & \xmark & \xmark & \cmark & \cmark & \xmark & \xmark & \cmark & \cmark \\ \hline
        \citet{zermini_deep_2016}& 2016 & AE & C & S & Mixing vector + ILD + IPD & $\theta$ &  & \cmark & \xmark & \xmark & \xmark & \xmark & \cmark & \xmark & \xmark & \xmark & \cmark \\ \hline
        \citet{chakrabarty_broadband_2017}& 2017 & CNN & C & S & Phase map & $\theta$ & 1 & \cmark & \xmark & \xmark & \xmark & \cmark & \xmark & \xmark & \xmark & \cmark & \cmark \\ \hline
        \citet{chakrabarty_multi-speaker_2017} & 2017 & CNN & C & S & Phase map & $\theta$ & 2 & \cmark & \xmark & \xmark & \xmark & \cmark & \xmark & \xmark & \xmark & \cmark & \xmark \\ \hline
      \citet{pertila_robust_2017}& 2017 & CNN & R & S & Magnitude spectrograms & TF Mask & 1 & \cmark & \cmark & \xmark & \xmark & \xmark & \cmark & \xmark & \xmark & \xmark & \cmark \\ \hline
        \citet{takeda_unsupervised_2017} & 2017 & MLP & C & SS & Complex eigenvectors from correlation matrix & $\theta$, $\phi$ & 1 & \cmark & \xmark & \xmark & \cmark & \xmark & \cmark & \xmark & \cmark & \xmark & \cmark \\ \hline
        \citet{yalta_sound_2017}& 2017 & Res. CNN & C & S & Magnitude spectrograms & $\theta$ & 1 & \cmark & \xmark & \xmark & \xmark & \xmark & \cmark & \xmark & \xmark & \xmark & \cmark \\ \hline
        \citet{yiwere_distance_2017} & 2017 & MLP & C & S & Binaural cross-correlation + ILD & $\theta$, $d$ & 1 & \cmark & \xmark & \xmark & \xmark & \xmark & \cmark & \xmark & \xmark & \xmark & \cmark \\ \hline
         \citet{adavanne_direction_2018}& 2018 & CRNN & C & S & Magnitude + phase spectrograms & SPS, $\theta$, $\phi$ & $\infty$ & \cmark & \xmark & \xmark & \xmark & \cmark & \xmark & \cmark & \xmark & \cmark & \xmark \\ \hline
        \citet{he_deep_2018}& 2018 & MLP, CNN & C & S & GCC-PHAT & $\theta$ & 0-2 & \xmark/\cmark & \xmark & \xmark & \xmark & \xmark & \cmark & \xmark & \xmark & \xmark & \cmark \\ \hline
        \citet{he_joint_2018}& 2018 & Res. CNN & C & S & Real + imaginary spectrograms & $\theta$ & $\infty$ & \xmark & \xmark & \xmark & \xmark & \xmark & \cmark & \xmark & \xmark & \xmark & \cmark \\ \hline
        \citet{huang_dnn-based_2018} & 2018 & DNN & R & S & Waveforms & dry signal & 1 & \cmark & \xmark & \xmark & \xmark & \cmark & \xmark & \xmark & \xmark & \cmark & \xmark \\ \hline
        \citet{li_online_2018} & 2018 & CRNN & C & S & GCC-PHAT & $\theta$ & 1 & \xmark & \xmark & \xmark & \xmark & \cmark & \xmark & \xmark & \xmark & \cmark & \xmark \\ \hline
        \citet{ma_phased_2018}& 2018 & CNN & C & S & CPS & $x$, $y$ & 3 & \cmark & \xmark & \cmark & \xmark & \xmark & \xmark & \cmark & \xmark & \xmark & \xmark \\ \hline
        \citet{nguyen_autonomous_2018}& 2018 & CNN & R & S & ILD + IPD & $\theta$, $\phi$ & 1 & \cmark & \xmark & \xmark & \xmark & \xmark & \cmark & \xmark & \xmark & \xmark & \cmark \\ \hline
        \citet{perotin_crnn-based_2018}& 2018 & CRNN & C & S & Intensity  & $\theta$, $\phi$ & 1 & \cmark & \xmark & \xmark & \xmark & \cmark & \xmark & \xmark & \xmark & \cmark & \cmark \\ \hline
        \citet{salvati_exploiting_2018}& 2018 & CNN & C/R & S & Narrowband SRP components & SRP weights & 1 & \cmark & \xmark & \multicolumn{4}{c|}{?}  & \xmark & \xmark & \cmark & \cmark \\ \hline
         \citet{sivasankaran_keyword-based_2018}& 2018 & CNN & C & S & IPD & $\theta$ & 1 & \cmark & \xmark & \xmark & \xmark & \cmark & \xmark & \xmark & \xmark & \cmark & \xmark \\ \hline
        \citet{suvorov_deep_2018} & 2018 & Res. CNN & C & S & Waveforms & $\theta$ & 1 & \cmark & \xmark & \xmark & \xmark & \xmark & \cmark & \xmark & \xmark & \xmark & \cmark \\ \hline
       \citet{takeda_unsupervised_2018}& 2018 & MLP & C & SS & Complex eigenvectors from correlation matrix & $\theta$ & 1 & \cmark & \xmark & \xmark & \cmark & \xmark & \cmark & \xmark & \cmark & \xmark & \cmark \\ \hline
        \citet{thuillier_spatial_2018}& 2018 & CNN & C & S & Ipsilateral + contralateral ear input signal & $\phi$ & 1 & \cmark & \xmark & \xmark & \xmark & \xmark & \cmark & \xmark & \xmark & \xmark & \cmark \\ \hline
        \citet{vecchiotti_deep_2018} & 2018 & CNN & R & S & GCC-PHAT + mel spectrograms & $x$, $y$ & 1 & \cmark & \xmark & \xmark & \xmark & \xmark & \cmark & \xmark & \xmark & \xmark & \cmark \\ \hline
        \citet{vera-diaz_towards_2018}& 2018 & CNN & R & S & Waveforms & $x$, $y$, $z$ & 1 & \cmark & \cmark & \xmark & \xmark & \cmark & \cmark & \xmark & \xmark & \xmark & \cmark \\ \hline
\end{tabular}
%}
\end{adjustbox}
\end{table*}

%%% 2019 %%%
\begin{table*}[ht!]
    \caption{Summary of DL-based SSL systems published in 2019, organized in alphabetical order. See Table~I's caption for acronyms specification.}
    %\resizebox{1.\columnwidth}{!}{%
    \begin{adjustbox}{width=\textwidth}
    \begin{tabular}{|c|c|c|c|c|c|c|ccc|cccc|cccc|}
    \hline \hline
        \multirow{3}{*}{\textbf{Author}} & \multirow{3}{*}{\textbf{Year}} & \multirow{3}{*}{\textbf{Architecture}} & \multirow{3}{*}{\textbf{Type}} &
        \multirow{3}{*}{\textbf{Learn-}} & \multirow{3}{*}{\textbf{Input features}} & \multirow{3}{*}{\textbf{Output}} &  \multicolumn{3}{c|}{\textbf{Sources}}           & \multicolumn{8}{c|}{\textbf{Data}}   \\
         &  &  &  &  &  &  & \multirow{2}{*}{\textbf{NoS}} & \multirow{2}{*}{\textbf{Kno.}} & \multirow{2}{*}{\textbf{Mov.}} & \multicolumn{4}{c}{\textbf{Train}}  & \multicolumn{4}{c|}{\textbf{Test}}  \\
         &  &  &  & \textbf{ing}  &  &  &  &  &  & SA & RA & SR & RR & SA & RA & SR & RR \\ \hline
        \citet{adavanne_sound_2019}& 2019 & CRNN & R & S & FOA magnitude + phase spectrograms & $x$, $y$, $z$ & 1 & \cmark & \xmark & \cmark & \cmark & \cmark & \cmark & \cmark & \cmark & \cmark & \cmark \\ \hline
       \citet{adavanne_localization_2019}& 2019 & CRNN & R & S & FOA magnitude + phase spectrograms & $x$, $y$, $z$ & 1 & \cmark & \cmark & \cmark & \cmark & \cmark & \cmark & \cmark & \cmark & \cmark & \cmark \\ \hline
        \citet{cao_two-stage_2019}& 2019 & CRNN & R & S & Log-Mel spectrogr. + GCC-PHAT + intensity  & $\theta$, $\phi$ & 1 & \cmark & \xmark & \xmark & \xmark & \xmark & \cmark & \xmark & \xmark & \xmark & \cmark \\ \hline
        \citet{Cao2019polyphonic}& 2019 & CRNN & R & S & Log-Mel spectrogr. + GCC-PHAT & $\theta$, $\phi$ & 1 & \cmark & \xmark & \xmark & \xmark & \xmark & \cmark & \xmark & \xmark & \xmark & \cmark \\ \hline
         \citet{chakrabarty_multi-scale_2019} & 2019 & CNN & C & S & Phase map & $\theta$ & 2 & \cmark & \xmark & \xmark & \xmark & \cmark & \xmark & \xmark & \xmark & \cmark & \xmark \\ \hline
         \citet{chakrabarty_multi-speaker_2019} & 2019 & CNN & C & S & Phase map & $\theta$ & 2 & \cmark & \xmark & \xmark & \xmark & \cmark & \xmark & \xmark & \xmark & \cmark & \cmark \\ \hline
         \citet{chazan_multi-microphone_2019}& 2019 & U-net & C & S & Phase map of the RTF between each mic pair & $\theta$ & $\infty$ & \xmark & \xmark & \xmark & \xmark & \cmark & \xmark & \xmark & \xmark & \xmark & \cmark \\ \hline
        \citet{chytas_hierarchical_2019}& 2019 & CNN & R & S & Waveforms & $\theta$, $\phi$ & 1 & \cmark & \xmark & \xmark & \xmark & \xmark & \cmark & \xmark & \xmark & \xmark & \cmark \\ \hline
        \citet{comminiello_quaternion_2019}& 2019 & CRNN & R & S & Quaternion FOA & $x$, $y$, $z$ & 1 & \cmark & \xmark & \cmark & \xmark & \xmark & \xmark & \cmark & \xmark & \xmark & \xmark \\ \hline
        \citet{grondin_sound_2019} & 2019 & CRNN & R & S & CPS + GCC-PHAT & $\theta$, $\phi$ & 1 & \cmark & \xmark & \xmark & \xmark & \xmark & \cmark & \xmark & \xmark & \xmark & \cmark \\ \hline
        \citet{he_adaptation_2019}& 2019 & Res. CNN & C & WS & Real + imaginary spectrograms & $\theta$ &  1-2 & \cmark & \xmark & \xmark & \xmark & \cmark & \xmark & \xmark & \xmark & \xmark & \cmark \\ \hline
        \citet{huang_time-domain_2019}& 2019 & CNN & R & S & Waveforms & dry signal & 1 & \cmark & \xmark & \xmark & \xmark & \cmark & \xmark & \xmark & \xmark & \cmark & \xmark \\ \hline
        \citet{kapka_sound_2019}& 2019 & CRNN & R & S & Magnitude + phase spectrograms & $x$, $y$, $z$ &  1-2 & \xmark & \xmark & \xmark & \xmark & \xmark & \cmark & \xmark & \xmark & \xmark & \cmark \\ \hline
        \citet{kong_cross-task_2019}& 2019 & CNN & R & S & Log-Mel magnitude FOA spectrograms & $\theta$, $\phi$ & 1 & \cmark & \xmark & \xmark & \xmark & \xmark & \cmark & \xmark & \xmark & \xmark & \cmark \\ \hline
        \citet{krause_arborescent_2019}& 2019 & CRNN & R & S & Magnitude / phase spectrograms & $\theta$, $\phi$ & 1 & \cmark & \xmark & \xmark & \xmark & \xmark & \cmark & \xmark & \xmark & \xmark & \cmark \\ \hline
        \citet{kucuk_real-time_2019}& 2019 & CNN & C & S & Real + imaginary spectrograms & $\theta$ & 1 & \cmark & \xmark & \xmark & \xmark & \cmark & \cmark & \xmark & \xmark & \cmark & \cmark \\ \hline
        \citet{kujawski_deep_2019}& 2019 & Res. CNN & R & S & Beamforming map & $x$, $y$ & 1 & \cmark & \xmark & \cmark & \xmark & \xmark & \xmark & \cmark & \xmark & \xmark & \xmark \\ \hline
        \citet{Leung2019} & 2019 & CRNN & R & S & CPS + real/imag. spectro + mag./phase spectro & $\theta$, $\phi$ & 1 & \cmark & \xmark & \xmark & \xmark & \xmark & \cmark & \xmark & \xmark & \xmark & \cmark \\ \hline
        \citet{lin_report_2019}& 2019 & CRNN & C & S & Magnitude and phase spectrograms & $\theta$, $\phi$ & 1 & \cmark & \xmark & \xmark & \xmark & \xmark & \cmark & \xmark & \xmark & \xmark & \cmark \\ \hline
        \citet{lu_sound_2019} & 2019 & CRNN & R & S & GCC-PHAT & $\theta$, $\phi$ & 1 & \cmark & \xmark & \xmark & \xmark & \xmark & \cmark & \xmark & \xmark & \xmark & \cmark \\ \hline
        \citet{maruri_gcc-phat_2019}& 2019 & CRNN & R & S & GCC-PHAT + magnitude + phase spectrograms & $\theta$, $\phi$ & 1 & \cmark & \xmark & \xmark & \xmark & \xmark & \cmark & \xmark & \xmark & \xmark & \cmark \\ \hline
        \citet{mazzon_first_2019}& 2019 & CRNN & R & S & Mel-spectrograms + GCC-PHAT/intensity  & $\theta$, $\phi$ & 1 & \cmark & \xmark & \xmark & \xmark & \xmark & \cmark & \xmark & \xmark & \xmark & \cmark  \\ \hline
        \citet{noh_three-stage_2019} & 2019 & CNN & C & S & GCC-PHAT & $\theta$, $\phi$ & 1 & \cmark & \xmark & \xmark & \xmark & \xmark & \cmark & \xmark & \xmark & \xmark & \cmark \\ \hline
        \citet{nustede_group_2019}& 2019 & CRNN & R & S & Group delays & $\theta$, $\phi$ & 1 & \cmark & \xmark & \xmark & \xmark & \xmark & \cmark & \xmark & \xmark & \xmark & \cmark \\ \hline
        \citet{opochinsky_deep_2019}& 2019 & MLP & R & WS & RTFs & $\theta$ & 1 & \cmark & \xmark & \xmark & \xmark & \cmark & \xmark & \xmark & \xmark & \cmark & \xmark \\ \hline
        \citet{pak_sound_2019}& 2019 & MLP & R & S & IPD & (clean) IPD &  & \cmark & \xmark & \xmark & \xmark & \cmark & \xmark & \xmark & \xmark & \cmark & \xmark \\ \hline
        \citet{pang2019multitask} & 2019 & CNN & R & S & ILD + IPD & $\theta$, $\phi$ & 1 & \cmark & \xmark & \xmark & \xmark & \xmark & \cmark & \xmark & \xmark & \xmark & \cmark \\ \hline \citet{park_reassembly_2019}& 2019 & CRNN & R & S & Log-Mel spectrograms + intensity  & $\theta$, $\phi$ & 1 & \cmark & \xmark & \xmark & \xmark & \xmark & \cmark & \xmark & \xmark & \xmark & \cmark \\ \hline
        \citet{perotin_crnn-based_2019}& 2019 & CRNN & C & S & FOA pseudo-intensity   & $\theta$, $\phi$ & 2 & \cmark & \xmark & \xmark & \xmark & \cmark & \xmark & \xmark & \xmark & \cmark & \cmark \\ \hline
        \citet{perotin_regression_2019}& 2019 & CRNN & C/R & S & FOA pseudo-intensity   & $\theta$, $\phi$ / $x$, $y$, $z$ & 1 & \cmark & \xmark & \xmark & \xmark & \cmark & \xmark & \xmark & \xmark & \cmark & \cmark \\ \hline
        \citet{Pratik2019}& 2019 & CRNN & R & S & GCC-PHAT + Mel/Bark spectrograms & $\theta$, $\phi$ & 1 & \cmark & \xmark & \xmark & \xmark & \xmark & \cmark & \xmark & \xmark & \xmark & \cmark \\ \hline
        \citet{pujol_source_2019}& 2019 & Res. CNN & R & S & Waveforms & $x$, $y$ & 1 & \cmark & \xmark & \xmark & \xmark & \cmark & \xmark & \cmark & \xmark & \cmark & \xmark \\ \hline
        \citet{ranjan_sound_2019}& 2019 & Res. CRNN & C & S & Log-Mel spectrograms & $\theta$, $\phi$ & 1 & \cmark & \xmark & \xmark & \xmark & \xmark & \cmark & \xmark & \xmark & \xmark & \cmark \\ \hline
        \multirow{2}{*}{\citet{Sudo2019}} & \multirow{2}{*}{2019} & \multirow{2}{*}{CRNN} & \multirow{2}{*}{R} & \multirow{2}{*}{S} & \multirow{2}{*}{cos(IPD), sin(IPD)} & $cos(\theta)$, $sin(\theta)$, & \multirow{2}{*}{1} & \multirow{2}{*}{\cmark} & \multirow{2}{*}{\xmark} & \multirow{2}{*}{\xmark} & \multirow{2}{*}{\xmark} & \multirow{2}{*}{\xmark} & \multirow{2}{*}{\cmark} & \multirow{2}{*}{\xmark} & \multirow{2}{*}{\xmark} & \multirow{2}{*}{\xmark} & \multirow{2}{*}{\cmark} \\ 
        & & & & & & $cos(\phi)$, $sin(\phi)$ & & & & & & & & & & & \\ \hline
        \citet{tang_regression_2019}& 2019 & CRNN & C/R & S & FOA pseudo-intensity   & $\theta$, $\phi$ / $x$, $y$, $z$ & 1 & \cmark & \xmark & \xmark & \xmark & \cmark & \xmark & \xmark & \xmark & \xmark & \cmark \\ \hline
         \citet{vecchiotti_detection_2019}& 2019 & CNN & R & S & GCC-PHAT + Mel-spectrograms & $x$, $y$ & 1 & \cmark & \xmark & \xmark & \xmark & \cmark & \cmark & \xmark & \xmark & \cmark & \cmark \\ \hline
         \citet{vecchiotti_end--end_2019} & 2019 & CNN & C & S & Waveforms & $\theta$ & 1 & \cmark & \xmark & \xmark & \cmark & \xmark & \cmark & \cmark & \cmark & \xmark & \cmark \\ \hline
        \citet{wang_robust_2019}& 2019 & RNN & R & S & Magnitude spectrograms & TF Mask & 1 & \cmark & \xmark & \xmark & \xmark & \cmark & \xmark & \xmark & \xmark & \xmark & \cmark \\ \hline
        \citet{xue_multi-beam_2019}& 2019 & CRNN & R & S & Log-Mel spectr. + CQT + phase spectrogr. + CPS & $\theta$, $\phi$ & 1 & \cmark & \xmark & \xmark & \xmark & \xmark & \cmark & \xmark & \xmark & \xmark & \cmark \\ \hline
       \citet{zhang_data_2019}& 2019 & CRNN & R & S & Magnitude and phase spectrograms & $\theta$, $\phi$ & 1 & \cmark & \xmark & \xmark & \xmark & \xmark & \cmark & \xmark & \xmark & \xmark & \cmark \\ \hline
        \citet{zhang_robust_2019}& 2019 & CNN & C & S & Phase spectrograms & $\theta$ & 1 & \cmark & \xmark & \xmark & \xmark & \cmark & \xmark & \xmark & \xmark & \cmark & \xmark \\ \hline \hline
\end{tabular}
%}
\end{adjustbox}
\end{table*}

\begin{table*}[ht!]
    \caption{Summary of DL-based SSL systems published in 2020, organized in alphabetical order. See Table~I's caption for acronyms specification.}
    %\resizebox{1.\columnwidth}{!}{%
    \begin{adjustbox}{width=\textwidth}
    \begin{tabular}{|c|c|c|c|c|c|c|ccc|cccc|cccc|}
    \hline \hline
        \multirow{3}{*}{\textbf{Author}} & \multirow{3}{*}{\textbf{Year}} & \multirow{3}{*}{\textbf{Architecture}} & \multirow{3}{*}{\textbf{Type}} &
        \multirow{3}{*}{\textbf{Learn.}} & \multirow{3}{*}{\textbf{Input features}} & \multirow{3}{*}{\textbf{Output}} &  \multicolumn{3}{c|}{\textbf{Sources}}           & \multicolumn{8}{c|}{\textbf{Data}}   \\
         &  &  &  &  &  &  & \multirow{2}{*}{\textbf{NoS}} & \multirow{2}{*}{\textbf{Kno.}} & \multirow{2}{*}{\textbf{Mov.}} & \multicolumn{4}{c|}{\textbf{Train}}  & \multicolumn{4}{c|}{\textbf{Test}}  \\
         &  &  &  &  &  &  &  &  &  & SA & RA & SR & RR & SA & RA & SR & RR \\ \hline
        \citet{bianco_semi-supervised_2020} & 2020 & VAE & C & SS & RTFs & $\theta$ & 1 & \cmark & \xmark & \xmark & \xmark & \cmark & \xmark & \xmark & \xmark & \cmark & \xmark \\ \hline
        \citet{cao_event-independent_2020}& 2020 & CRNN & R & S & FOA waveforms & $\theta$, $\phi$ & 0-2 & \xmark & \cmark & \xmark & \xmark & \xmark & \cmark & \xmark & \xmark & \xmark & \cmark \\ \hline
        \citet{comanducci_source_2020} & 2020 & CNN/U-Net & C & S & GCC-PHAT & $x$, $y$ & 1 & \cmark & \xmark & \cmark & \xmark & \cmark & \xmark & \xmark & \xmark & \cmark & \cmark \\ \hline
        \citet{comanducci_time_2020} & 2020 & U-Net & R & S & GCC & Clean GCC & 1 & \cmark & \xmark & \cmark & \xmark & \cmark & \xmark & \xmark & \xmark & \cmark & \cmark \\ \hline
        \citet{fahim_multi-source_2020} & 2020 & CNN & C & S & FOA modal coherence & $\theta$, $\phi$ &  1-7 & \cmark & \xmark & \xmark & \xmark & \cmark & \xmark & \xmark & \xmark & \cmark & \cmark \\ \hline
        \citet{hao_spectral_2020}& 2020 & CNN & C & S & Real + imaginary spectrograms + spectral flux & $\theta$ & 1 & \cmark & \xmark & \xmark & \xmark & \xmark & \cmark & \xmark & \xmark & \xmark & \cmark \\ \hline
        \citet{huang_time-domain_2020}& 2020 & AE & R & S & Waveforms & $\theta$ & 1 & \cmark & \xmark & \cmark & \xmark & \xmark & \xmark & \cmark & \xmark & \xmark & \xmark \\ \hline
        \citet{hubner_efficient_2021}& 2020 & CNN & C & S & Phase map & $\theta$ & 1 & \cmark & \xmark & \xmark & \xmark & \cmark & \xmark & \xmark & \xmark & \xmark & \cmark \\ \hline
        \citet{jenrungrot_cone_2020}& 2020 & U-Net & R & S & Waveforms & $\theta$ &  0-8 & \xmark & \cmark & \xmark & \xmark & \cmark & \xmark & \xmark & \xmark & \cmark & \cmark \\ \hline
        \citet{mack_signal-aware_2020}& 2020 & CNN + attention & C & S & Phase map & $\theta$ & 2 & \cmark & \xmark & \xmark & \xmark & \cmark & \xmark & \xmark & \xmark & \cmark & \xmark \\ \hline
        \citet{moing_learning_2020}& 2020 & AE & C,R & S & Real + imaginary spectrograms & $x$, $y$ & 1-3 & \xmark & \xmark & \xmark & \xmark & \cmark & \cmark & \xmark & \xmark & \cmark & \cmark \\ \hline
         \citet{moing_data-efficient_2021}& 2020 & AE & C & SS & Real + imaginary spectrograms & $x$, $y$ & 1-3 & \xmark & \xmark & \cmark & \xmark & \xmark & \cmark & \xmark & \xmark & \xmark & \cmark \\ \hline
         \citet{naranjo-alcazar_sound_2020}& 2020 & Res. CRNN & R & S & Log-Mel magnitude spectrograms + GCC-PHAT & $x$, $y$, $z$ & 1 & \cmark & \cmark & \xmark & \xmark & \xmark & \cmark & \xmark & \xmark & \xmark & \cmark \\ \hline
        \citet{nguyen_robust_2020} & 2020 & CNN & C & S & Spatial pseudo-spectrum & $\theta$ & 0-4 & \xmark & \xmark & \xmark & \xmark & \cmark & \xmark & \xmark & \xmark & \cmark & \cmark \\ \hline
        \citet{nguyen2020ensemble} & 2020 & CRNN & R & S & DoAs from histogram-based method & $\theta$, $\phi$ & 1 & \cmark & \cmark & \xmark & \xmark & \xmark & \cmark & \xmark & \xmark & \xmark & \cmark \\ \hline
        \citet{nguyen2020sequence} & 2020 & CRNN & R & S & DoAs from histogram-based method & $\theta$, $\phi$ & 1 & \cmark & \xmark & \xmark & \xmark & \xmark & \cmark & \xmark & \xmark & \xmark & \cmark \\ \hline
        \citet{park_sound_2020}& 2020 & CRNN & R & S & Log-Mel energy + intensity   & $\theta$, $\phi$ & 1 & \cmark & \cmark & \xmark & \xmark & \xmark & \cmark & \xmark & \xmark & \xmark & \cmark \\ \hline
       \citet{patel_single_2020}& 2020 & U-Net & R & S & Mel-spectrograms & $x$, $y$, $z$ & 1 & \cmark & \cmark & \xmark & \xmark & \xmark & \cmark & \xmark & \xmark & \xmark & \cmark \\ \hline
        \multirow{2}{*}{\citet{phan_audio_2020}} & \multirow{2}{*}{2020} & \multirow{2}{*}{CRNN + SA} & \multirow{2}{*}{R} & \multirow{2}{*}{S} & FOA log-Mel spectrograms + active/reactive & \multirow{2}{*}{$x$, $y$, $z$} & \multirow{2}{*}{1} & \multirow{2}{*}{\cmark} & \multirow{2}{*}{\cmark} & \multirow{2}{*}{\xmark} & \multirow{2}{*}{\xmark} & \multirow{2}{*}{\xmark} & \multirow{2}{*}{\cmark} & \multirow{2}{*}{\xmark} & \multirow{2}{*}{\xmark} & \multirow{2}{*}{\xmark} & \multirow{2}{*}{\cmark} \\
        &  &  &  &  & intensity, or GCC-PHAT & &  & & & & &  &  &  &  &  &  \\ \hline
        \multirow{2}{*}{\citet{phan_multitask_2020}} & \multirow{2}{*}{2020} & \multirow{2}{*}{CRNN + SA} & \multirow{2}{*}{R} & \multirow{2}{*}{S} & FOA log-Mel spectrograms + active/reactive  & \multirow{2}{*}{$x$, $y$, $z$} & \multirow{2}{*}{1} & \multirow{2}{*}{\cmark} & \multirow{2}{*}{\cmark} & \multirow{2}{*}{\xmark} & \multirow{2}{*}{\xmark} & \multirow{2}{*}{\xmark} & \multirow{2}{*}{\cmark} & \multirow{2}{*}{\xmark} & \multirow{2}{*}{\xmark} & \multirow{2}{*}{\xmark} & \multirow{2}{*}{\cmark} \\
        &  &  &  &  & intensity, or GCC-PHAT & &  & & & & &  &  &  &  &  &  \\ \hline
       \citet{ronchini_sound_2020} & 2020 & CRNN & R & S & FOA log-Mel spectrograms + log-Mel intensity & $x$, $y$, $z$ & 1 & \cmark & \cmark & \xmark & \xmark & \xmark & \cmark & \xmark & \xmark & \xmark & \cmark \\
        \hline
        \multirow{2}{*}{\citet{sampathkumar_sound_2020}} & \multirow{2}{*}{2020} & \multirow{2}{*}{CRNN} & \multirow{2}{*}{R} & \multirow{2}{*}{S} & MIC + FOA Mel spectrograms + active intensity   & \multirow{2}{*}{$\theta$, $\phi$} & \multirow{2}{*}{1} & \multirow{2}{*}{\cmark} & \multirow{2}{*}{\cmark} & \multirow{2}{*}{\xmark} & \multirow{2}{*}{\xmark} & \multirow{2}{*}{\xmark} & \multirow{2}{*}{\cmark} & \multirow{2}{*}{\xmark} & \multirow{2}{*}{\xmark} & \multirow{2}{*}{\xmark} & \multirow{2}{*}{\cmark} \\
        & & & & & + GCC-PHAT & & & & & & & & & & & & \\ \hline
        \citet{shimada_accdoa_2020} & 2020 & Res. CRNN & R & S & FOA magnitude spectrograms + IPD & ACCDOA & 1 & \cmark & \xmark & \xmark & \xmark & \xmark & \cmark & \xmark & \xmark & \xmark & \cmark \\ \hline
        \citet{shimada_sound_2020}& 2020 & Res. CRNN & R & S & FOA magnitude spectrograms + IPD & ACCDOA & 1 & \cmark & \cmark & \xmark & \xmark & \xmark & \cmark & \xmark & \xmark & \xmark & \cmark \\ \hline
        \citet{singla_sequential_2020} & 2020 & CRNN & R & S & FOA log-Mel spectrograms + log-Mel intensity & $x$, $y$, $z$ & 1 & \cmark & \cmark & \xmark & \xmark & \xmark & \cmark & \xmark & \xmark & \xmark & \cmark \\
        \hline
        \citet{song_localization_2020}& 2020 & CRNN & R & S & GCC-PHAT + FOA active intensity & $x$, $y$, $z$ & 1 & \xmark & \cmark & \xmark & \xmark & \xmark & \cmark & \xmark & \xmark & \xmark & \cmark \\ \hline
        \citet{sundar_raw_2020}& 2020 & Res. CNN & C/R & S & Waveforms & $d$, $\theta$ & 1-3 & \xmark & \cmark & \cmark & \xmark & \cmark & \xmark & \cmark & \xmark & \cmark & \cmark \\ \hline
        \citet{tian_multiple_2020}& 2020 & CRNN & ? & S & Ambisonics & ? & ? & \xmark & \cmark & \xmark & \xmark & \xmark & \cmark & \xmark & \xmark & \xmark & \cmark \\ \hline
        \citet{varanasi_deep_2020}& 2020 & CNN & C & S & 3rd spherical harmonics (phase or phase+magnitude) & $\theta$, $\phi$ & 1 & \cmark & \cmark & \xmark & \xmark & \cmark & \xmark & \xmark & \xmark & \cmark & \cmark \\ \hline
         \citet{varzandeh_exploiting_2020}& 2020 & CNN & C & S & GCC-PHAT + periodicity degree & $\theta$ &  0-1 & \xmark & \xmark & \xmark & \cmark & \xmark & \xmark & \xmark & \xmark & \xmark & \cmark \\ \hline
          \citet{vera-diaz_towards_2021}& 2020 & AE & R & S & GCC-PHAT & time-delay & 1 & \cmark & \cmark & \xmark & \xmark & \xmark & \cmark & \xmark & \xmark & \xmark & \cmark \\ \hline
        \multirow{2}{*}{\citet{wang_ustc-iflytek_2020}} & \multirow{2}{*}{2020} & \multirow{2}{*}{Res. CRNN} & \multirow{2}{*}{R} & \multirow{2}{*}{S} & FOA pseudo-intensity + FOA log-Mel spectrograms & \multirow{2}{*}{$x$, $y$, $z$} & \multirow{2}{*}{1} & \multirow{2}{*}{\cmark} & \multirow{2}{*}{\cmark} & \multirow{2}{*}{\xmark} & \multirow{2}{*}{\xmark} & \multirow{2}{*}{\xmark} & \multirow{2}{*}{\cmark} & \multirow{2}{*}{\xmark} & \multirow{2}{*}{\xmark} & \multirow{2}{*}{\xmark} & \multirow{2}{*}{\cmark} \\
        & & & & & + GCC-PHAT  & & & & & & & & & & & & \\ \hline
        \citet{xue_sound_2020}& 2020 & CRNN & C & S & CPS + waveforms + beamforming output & $\theta$, $\phi$ & 1 & \cmark & \xmark & \xmark & \xmark & \xmark & \cmark & \xmark & \xmark & \xmark & \cmark \\ \hline
        \citet{yasuda_sound_2020}& 2020 & Res. CRNN & R & S & FOA log-Mel spectrograms + intensity & denoised IV & 2 & \cmark & \xmark & \xmark & \xmark & \xmark & \cmark & \xmark & \xmark & \xmark & \cmark \\ \hline
    \end{tabular}
%}
\end{adjustbox}
\end{table*}

\begin{table*}[ht!]
    \caption{Summary of DL-based SSL systems published in 2021, organized in alphabetical order. See Table~I's caption for acronyms specification.}
    %\resizebox{1.\columnwidth}{!}{%
    \begin{adjustbox}{width=\textwidth}
    \begin{tabular}{|c|c|c|c|c|c|c|ccc|cccc|cccc|}
    \hline \hline
        \multirow{3}{*}{\textbf{Author}} & \multirow{3}{*}{\textbf{Year}} & \multirow{3}{*}{\textbf{Architecture}} & \multirow{3}{*}{\textbf{Type}} &
        \multirow{3}{*}{\textbf{Learn.}} & \multirow{3}{*}{\textbf{Input features}} & \multirow{3}{*}{\textbf{Output}} &  \multicolumn{3}{c|}{\textbf{Sources}}           & \multicolumn{8}{c|}{\textbf{Data}}   \\
         &  &  &  &  &  &  & \multirow{2}{*}{\textbf{NoS}} & \multirow{2}{*}{\textbf{Kno.}} & \multirow{2}{*}{\textbf{Mov.}} & \multicolumn{4}{c|}{\textbf{Train}}  & \multicolumn{4}{c|}{\textbf{Test}}  \\
         &  &  &  &  &  &  &  &  &  & SA & RA & SR & RR & SA & RA & SR & RR \\ \hline
        \citet{adavanne_differentiable_2021}& 2021 & CRNN + SA & R & S & FOA Mel spectrograms + intensity + GCC-PHAT & x,y,z & 2 & \cmark & \cmark & \xmark & \xmark & \xmark & \cmark & \xmark & \xmark & \xmark & \cmark \\ \hline
        \citet{Bai_NWPU_task3_report}& 2021 & Res. CRNN & R & S & Log-Mel spectrograms + intensity & $x$, $y$, $z$ & 1 & \cmark & \cmark & \xmark & \xmark & \xmark & \cmark & \xmark & \xmark & \xmark & \cmark \\ \hline
        \citet{bianco2021semi} & 2021 & VAE & C & SS & RTF & $\theta$ & 1 & \cmark & \xmark & \xmark & \xmark & \cmark & \xmark & \xmark & \xmark & \cmark & \cmark \\ \hline
        \citet{bohlender_exploiting_2021}& 2021 & CNN/CRNN & C & S & Phase map & $\theta$ &  1-3 & \cmark & \xmark & \xmark & \xmark & \cmark & \xmark & \xmark & \xmark & \xmark & \cmark \\ \hline
        \citet{bologni_acoustic_2021}& 2021 & CNN & C & S & Waveforms & $\theta$, $d$ & 1 & \cmark & \xmark & \xmark & \xmark & \cmark & \xmark & \xmark & \xmark & \cmark & \xmark \\ \hline
        \citet{cao_improved_2021}& 2021 & SA & R & S & Log-Mel spectrograms + intensity & $x$, $y$, $z$ & 0-2 & \xmark & \cmark & \xmark & \xmark & \xmark & \cmark & \xmark & \xmark & \xmark & \cmark \\ \hline
        \citet{castellini_neural_2021} & 2021 & MLP & R & S & real + imaginary CPS & $x$, $y$ & 1-3 & \cmark & \xmark & \cmark & \xmark & \xmark & \xmark & \xmark & \xmark & \xmark & \cmark \\ \hline
        \citet{diaz-guerra_robust_2021}& 2021 & CNN & R & S & SRP-PHAT power map & $x$, $y$, $z$ & 1 & \cmark & \cmark & \xmark & \xmark & \cmark & \xmark & \xmark & \xmark & \cmark & \cmark \\ \hline
        \citet{emmanuel_multiscale_2021}& 2021 & CNN + SA & R & S & Log-spectrograms + intensity & ACCDOA & 1 & \cmark & \cmark & \xmark & \xmark & \xmark & \cmark & \xmark & \xmark & \xmark & \cmark \\ \hline
        \citet{gelderblom_synthetic_2021}& 2021 & MLP & C/R & S & GCC-PHAT & $\theta$ & 2 & \cmark & \xmark & \xmark & \xmark & \cmark & \xmark & \xmark & \xmark & \xmark & \cmark \\ \hline
        \citet{goncalves_deconvoluting_2021} & 2021 & CNN & R & S & Magnitude CPS & $x$, $y$ & 1-10 & \xmark & \xmark & \cmark & \xmark & \xmark & \xmark & \cmark & \xmark & \xmark & \xmark \\ \hline
        \citet{grumiaux_improved_2021}& 2021 & CRNN & C & S & Intensity & $\theta$, $\phi$ &  1-3 & \cmark & \cmark & \xmark & \xmark & \cmark & \xmark & \xmark & \xmark & \cmark & \cmark \\ \hline
        \citet{grumiaux_saladnet_2021}& 2021 & CNN + SA & C & S & Intensity & $\theta$, $\phi$ &  1-3 & \cmark & \xmark & \xmark & \xmark & \cmark & \xmark & \xmark & \xmark & \cmark & \cmark \\ \hline
        \citet{guirguis_seld-tcn_2021}& 2021 & TCN & R & S & Magnitude + phase spectrograms & $x$, $y$, $z$ & 1 & \cmark & \cmark & \cmark & \cmark & \cmark & \cmark & \cmark & \cmark & \cmark & \cmark \\ \hline
        \citet{hammer2021dynamically} & 2021 & U-net & C & S & Phase map of the RTF between each mic pair & $\theta$ & $\infty$ & \xmark & \cmark & \xmark & \xmark & \cmark & \xmark & \xmark & \xmark & \xmark & \cmark \\ \hline
        \citet{he_neural_2021}& 2021 & Res. CNN & C & WS & Magnitude + phase spectrograms & $\theta$ &  1-4 & \cmark/\xmark & \xmark & \xmark & \xmark & \cmark & \cmark & \xmark & \xmark & \xmark & \cmark \\ \hline
        \citet{he_sounddet_2021}& 2021 & CNN & R & S & Waveforms & $x$, $y$, $z$ & 1 & \cmark & \cmark & \cmark & \cmark & \cmark & \cmark & \cmark & \cmark & \cmark & \cmark \\ \hline
        \citet{Huang_Aalto_task3_report}& 2021& Res. CNN + SA & R & S & Waveforms & ACCDOA & 1 & \cmark & \cmark & \xmark & \xmark & \xmark & \cmark & \xmark & \xmark & \xmark & \cmark \\ \hline
        \citet{komatsu_sound_2021}& 2021 & CRNN & R & S & FOA magnitude + phase spectrograms & $\theta$, $\phi$ & 1 & \cmark & \cmark & \cmark & \cmark & \cmark & \cmark & \cmark & \cmark & \cmark & \cmark \\ \hline
        \citet{krause_comparison_2021}& 2021 & CNN & R & S & Magnitude + phase spectrograms & $x$, $y$, $z$ & 1 & \cmark & \xmark & \xmark & \xmark & \cmark & \xmark & \xmark & \xmark & \cmark & \xmark \\ \hline
        \citet{krause_feature_2021}& 2021 & CRNN & R & S & Misc. & $\theta$, $\phi$ & 1 & \cmark & \xmark & \xmark & \xmark & \xmark & \cmark & \xmark & \xmark & \xmark & \cmark \\ \hline
        %\citet{lee_zero-shot_2021}& 2021 & MLP & R & SS & Mel-spectrograms & $x$, $y$ & 1 & \cmark & \xmark & \xmark & \xmark & \cmark & \cmark & \xmark & \xmark & \cmark & \cmark \\ \hline
        \citet{lee_deep_2021} & 2021 & U-Net & R & S & SRP power map & $x$,$y$ & 1-3 & \xmark & \xmark & \cmark & \xmark & \xmark & \xmark & \xmark & \cmark & \xmark & \xmark \\ \hline
        \citet{Lee_SGU_task3_report}& 2021 & CNN + attention & C & S & Log-Mel spectrograms + intensity & $\theta$ & 1 & \cmark & \cmark & \xmark & \xmark & \cmark & \xmark & \xmark & \xmark & \cmark & \cmark \\ \hline
        \citet{liu_deep_2021}& 2021 & CNN & C & S & Intensity & $\theta$ & 1 & \cmark & \xmark & \xmark & \xmark & \cmark & \xmark & \xmark & \xmark & \cmark & \cmark \\ \hline
        \citet{Naranjo-Alcazar_UV_task3_report}& 2021 & Res. CRNN & R & S & Log-Mel spectrograms + GCC-PHAT & ACCDOA & 1 & \cmark & \cmark & \xmark & \xmark & \xmark & \cmark & \xmark & \xmark & \xmark & \cmark \\ \hline
        \citet{nguyen_general_2021}& 2021 & CRNN & C & S & Intensity/GCC-PHAT & $\theta$, $\phi$ & 1 & \cmark & \cmark & \cmark & \cmark & \cmark & \cmark & \cmark & \cmark & \cmark & \cmark \\ \hline
        \citet{nguyen_spectrotemporally_2021}& 2021 & CNN + RNN/SA & R & S & Log-spectrograms + DRR + SCM eigenvectors & ACCDOA & 1 & \cmark & \cmark & \xmark & \xmark & \xmark & \cmark & \xmark & \xmark & \xmark & \cmark \\ \hline
        \citet{Park2021}& 2021 & SA & R & S & log-Mel spectrograms + intensity & $x$, $y$, $z$ & 1 & \cmark & \cmark & \xmark & \xmark & \xmark & \cmark & \xmark & \xmark & \xmark & \cmark \\ \hline
        \citet{poschadel_direction_2021}& 2021 & CRNN & C & S & HOA magnitude + phase spectrograms & $\theta$, $\phi$ & 1 & \cmark & \xmark & \xmark & \xmark & \cmark & \xmark & \xmark & \xmark & \cmark & \cmark \\ \hline
        \citet{poschadel_multisource_2021}& 2021 & CRNN & C & S & HOA magnitude + phase spectrograms & $\theta$, $\phi$ & 2-3 & \cmark & \xmark & \xmark & \xmark & \cmark & \xmark & \xmark & \xmark & \cmark & \cmark \\ \hline
        \citet{pujol_beamlearning_2021}& 2021 & Res. CNN & R & S & Waveforms & $\theta$, $\phi$ & 1 & \cmark & \xmark & \xmark & \xmark & \cmark & \xmark & \xmark & \cmark & \cmark & \cmark \\ \hline
        \citet{Ko_SKKU_task3_report} & 2021 & CRNN + SA & R & S & Log-Mel spectrograms + intensity & $\theta$, $\phi$ & 1 & \cmark & \cmark & \cmark & \cmark & \cmark & \cmark & \cmark & \cmark & \cmark & \cmark \\ \hline
        \citet{schymura_pilot_2021}& 2021 & CNN + SA & R & S & Magnitude + phase spectrograms & $\theta$, $\phi$ & 1 & \cmark & \xmark & \cmark & \xmark & \cmark & \cmark & \cmark & \xmark & \cmark & \cmark \\ \hline
        \citet{schymura_exploiting_2021}& 2021 & CNN + AE + attent. & R & S & FOA magnitude + phase spectrograms & $\theta$, $\phi$ & 1 & \cmark & \xmark & \cmark & \cmark & \cmark & \cmark & \cmark & \cmark & \cmark & \cmark \\ \hline
        \citet{shimada_ensemble_2021}& 2021 & Res. CRNN + SA & R & S & IPD & ACCDOA & 1 & \cmark & \cmark & \xmark & \xmark & \xmark & \cmark & \xmark & \xmark & \xmark & \cmark \\ \hline
        \citet{subramanian2021directional} & 2021 & CRNN & C/R & S & Phase spectrogram & $\theta$ & 2 & \cmark & \xmark & \xmark & \xmark & \cmark & \xmark & \xmark & \xmark & \cmark & \xmark \\ \hline
        \citet{subramanian_deep_2021} & 2021 & CRNN & C &  & Phase spectrograms, IPD & $\theta$ & 2 & \cmark & \xmark & \xmark & \xmark & \cmark & \xmark & \xmark & \xmark & \cmark & \xmark \\ \hline
        \citet{Sudarsanam2021} & 2021 & SA & R & S & Log-Mel spectrograms + intensity & ACCDOA & 1 & \cmark & \cmark & \xmark & \xmark & \xmark & \cmark & \xmark & \xmark & \xmark & \cmark \\ \hline
        \citet{vargas_improved_2021}& 2021 & CNN & C & S & Phase map & $\theta$ & 1 & \cmark & \xmark & \xmark & \xmark & \cmark & \cmark & \xmark & \xmark & \cmark & \cmark \\ \hline
        \citet{vera-diaz_acoustic_2021}& 2021 & AE & R & S & GCC-PHAT & time-delay & 2 & \cmark & \cmark & \xmark & \xmark & \xmark & \cmark & \xmark & \xmark & \xmark & \cmark \\ \hline
        \citet{wang_four-stage_2021}& 2021 & SA & R & S & Mel-spectr. + intensity/Mel-spectr. + GCC-PHAT & $x$, $y$, $z$ & 1 & \cmark & \cmark & \xmark & \xmark & \xmark & \cmark & \xmark & \xmark & \xmark & \cmark \\ \hline
        \citet{wu_sslide_2021}& 2021 & AE & R & S & Likelihood surface & $x$, $y$ & 1 & \cmark & \xmark & \xmark & \xmark & \cmark & \cmark & \xmark & \xmark & \cmark & \cmark \\ \hline
        \citet{wu2021sound}& 2021 & CNN AE & R & S & Beamforming heatmap image & $x$, $y$ & 1 & \cmark & \xmark & \xmark & \xmark & \cmark & \cmark & \xmark & \xmark & \cmark & \cmark \\ \hline
        \citet{Xinghao2021}& 2021 & CNN + SA & R & S & Log-Mel spectrograms + intensity & ACCDOA & 1 & \cmark & \cmark & \xmark & \xmark & \xmark & \cmark & \xmark & \xmark & \xmark & \cmark \\ \hline
        \citet{xu_acoustic_2021} & 2021 & DenseNet & R & S & Real CPS & $x$, $y$ & 6-25 & \xmark/\cmark & \xmark & \cmark & \xmark & \xmark & \xmark & \xmark & \xmark & \cmark & \xmark \\ \hline
        \citet{yalta_dcase_2021}& 2021 & SA & R & S & Log-Mel spectrograms + intensity & $x$, $y$, $z$ & 1 & \cmark & \cmark & \xmark & \xmark & \xmark & \cmark & \xmark & \xmark & \xmark & \cmark \\ \hline
        \citet{yang2021supervised} & 2021 & CRNN & C & S & Log-magnitude and phase spectrograms & $\theta$ & 1 & \cmark & \xmark & \xmark & \xmark & \cmark & \xmark & \xmark & \xmark & \cmark & \xmark \\ \hline
        \citet{yang2021learning} & 2021 & CRNN & C & S & Log-magnitude and phase spectrograms & $\theta$ & 1 & \cmark & \cmark & \xmark & \xmark & \cmark & \xmark & \xmark & \xmark & \cmark & \cmark \\ \hline
        \citet{zhang_data_2021}& 2021 & CNN + SA & R & S & Log-spectrograms + intensity + GCC-PHAT & $x$, $y$, $z$ & 1 & \cmark & \cmark & \xmark & \xmark & \xmark & \cmark & \xmark & \xmark & \xmark & \cmark \\ \hline
\end{tabular}
%}
\end{adjustbox}
\end{table*}

%% before appendix (optional) and bibliography:
\section*{Acknowledgements}
This work was funded by the French Association for Technological Research (ANRT CIFRE contract 2019/0533) and partially funded by the Multidisciplinary Institute in Artificial Intelligence MIAI@Grenoble-Alpes (ANR-19-P3IA-0003)

\bibliographystyle{abbrvnat}
\bibliography{biblio}  %%% Uncomment this line and comment out the ``thebibliography'' section below to use the external .bib file (using bibtex) .





% \end{thebibliography}

\end{document}